\documentclass[onecolumn]{aastex63}

\usepackage{amssymb,amsmath,amsfonts,amsbsy}
\usepackage{color}
\usepackage{bbm}
\usepackage{enumerate}
\usepackage{graphicx}
\usepackage{mathtools}
\usepackage{mathrsfs}

\usepackage{pifont}% http://ctan.org/pkg/pifont

\newcommand{\Mspace}{{\mathbb M}}
\newcommand{\Rspace}{{\mathbb R}}
\newcommand{\Sspace}{{\mathbb S}}

\usepackage{hyperref}

\newcommand{\orcid}[1]{\href{https://orcid.org/#1}}

\usepackage{textcase}
\usepackage{ulem}

\usepackage{CJKutf8}

\newcommand{\parallelsum}{\mathbin{\|}}

\begin{document}

\title{Minkowski Tensors in Redshift Space -- Beyond the Plane Parallel Approximation}

\author{Stephen Appleby} 
\email{stephen.appleby@apctp.org}
\affiliation{Asia Pacific Center for Theoretical Physics, Pohang, 37673, Korea}
\affiliation{Department of Physics, POSTECH, Pohang 37673, Korea}
\author{Joby P. Kochappan}
\affiliation{Asia Pacific Center for Theoretical Physics, Pohang, 37673, Korea}
\author{Pravabati Chingangbam}
\affiliation{Indian Institute of Astrophysics, Koramangala II Block, Bangalore 560 034, India}
\affiliation{School of Physics, Korea Institute for Advanced Study, 85
Hoegiro, Dongdaemun-gu, Seoul, 02455, Korea}
\author{Changbom Park}
\affiliation{School of Physics, Korea Institute for Advanced Study, 85
Hoegiro, Dongdaemun-gu, Seoul, 02455, Korea}

\begin{abstract}
The Minkowski tensors (MTs) can be used to probe anisotropic signals in a field, and are well suited for measuring the redshift space distortion (RSD) signal in large scale structure catalogs. We consider how the linear RSD signal can be extracted from a field without resorting to the plane parallel approximation. A spherically redshift space distorted field is both anisotropic and inhomogeneous. We derive expressions for the two point correlation functions that elucidate the inhomogeneity, and then explain how the breakdown of homogeneity impacts the volume and ensemble averages of the tensor Minkowski functionals. We construct the ensemble average of these quantities in curvilinear coordinates and show that the ensemble and volume averages can be approximately equated, but this depends on our choice of definition of the volume average of a tensor and the radial distance between the observer and field. We then extract the tensor Minkowski functionals from spherically redshift space distorted, Gaussian random fields and gravitationally evolved  dark matter density fields at $z=0$ to test if we can successfully measure the Kaiser RSD signal. For the dark matter field we find a significant, $\sim 10\%$ anomalous signal in the MT component parallel to the line of sight that is present even on large scales $R_{\rm G} \gtrsim 15 \, {\rm Mpc}$, in addition to the Kaiser effect. This is due to the line of sight component of the MT being significantly contaminated by the Finger of God effect, which can be approximately modelled by an additional damping term in the cumulants.
\end{abstract}

%\maketitle

\section{Introduction} 

The tensor Minkowski functionals are a rank-$p$ generalisation of the scalar Minkowski functionals \citep{nla.cat-vn2176896, McMullen:1997,Alesker1999,2002LNP...600..238B,HugSchSch07,1367-2630-15-8-083028,JMI:JMI3331,Beisbart:2001gk,Hadwiger, nla.cat-vn1821482}. Being tensors, they are sensitive to directionally dependent signals in data and have found application in a number of disciplines such as material science \citep{PhysRevE.77.051805,Becker2003ComplexDS,Olszowka2006}. The scalar Minkowski functionals and associated morphological statistics have a long and storied history within cosmology \citep{Gott:1989yj,1991ApJ...378..457P,Mecke:1994ax,Schmalzing:1997aj,Schmalzing:1997uc,1989ApJ...345..618M,1992ApJ...387....1P,2001ApJ...553...33P,Park:2009ja,doi:10.1111/j.1365-2966.2010.18015.x,Sahni:1998cr,Bharadwaj:1999jm,vandeWeygaert:2011ip, Park:2013dga,vandeWeygaert:2011hyr,Shivshankar:2015aza,Pranav:2016gwr,Pranav:2018lox,Pranav:2018pnu,Feldbrugge:2019tal,Wilding:2020oza,Munshi:2020tzm}, but the tensors are less widely adopted. They were initially introduced in \citep{Beisbart:2001vb,Beisbart:2001gk,2002LNP...600..238B} to provide a measure of sub-structure of galaxy clusters and spiral galaxies. In the mathematics literature they are defined for structures on flat Euclidean space. In two dimensions, the definition of the translation invariant rank-2 Minkowski tensors were generalised to structures on the two-sphere in \cite{Chingangbam:2017uqv}. More recently, they have been applied to cosmological scale fields \citep{Chingangbam:2017uqv,Ganesan:2016jdk,Appleby:2017uvb,Appleby:2018tzk,Rahman:2021azv} --  Cosmic Microwave Background temperature and polarisation data \cite{Ganesan:2016jdk,K.:2018wpn,Joby:2021,Goyal:2020,Goyal:2021}, and the fields of the epoch of reionization \citep{Kapahtia:2017qrg,Kapahtia:2019ksk,Kapahtia:2021}. In addition, the authors have written a series of papers on the application of the Minkowski tensors to the low redshift matter density field as traced by galaxies \citep{Appleby:2017uvb,Appleby:2018tzk}. The ensemble average of the MTs measured from isotropic and anisotropic, Gaussian random fields were considered in \cite{Chingangbam:2017uqv,Appleby:2018tzk,Chingangbam:2021kov}. Anisotropic random fields were subsequently explored further in \citet{Klatt_2022}, including higher rank statistics. Numerical algorithms with which to extract the MTs from two-dimensional fields can be found in \cite{JMI:JMI3331,Appleby:2017uvb,Schaller2020}. 

In real space, galaxies are assumed to be distributed in a statistically isotropic and homogeneous manner. The cosmic web is locally anisotropic, with filaments feeding matter into nodes, and extended structures aligning on two-dimensional walls. In this picture, isotropy of the matter distribution means that there is no globally preferred direction within the filamentary large scale structure, when averaging over a volume that is large compared to the typical scale of the structures. This statistical isotropy is an axiom within cosmology, motivated by observations of the Cosmic Microwave Background. 

Even if the large scale distribution of the matter field in real space is isotropic, the observed distribution of galaxies is contaminated by their peculiar velocity along the line of sight. This phenomenon was first described in early pioneering works \citep{1987MNRAS.227....1K}, and is referred to as the redshift space distortion (RSD) effect. The RSD effect perturbs the apparent position of galaxies in redshift space only along the line of sight, and hence has rotational symmetry around the central observer. However, it leads to a global alignment of structures in the excursion sets of the density field along the line of sight. This alignment of structures in the field is what we refer to as anisotropy in the context of Minkowski tensors. A significant body of literature has subsequently been devoted to understanding the effect of RSD on the two-point statistics \citep{Hamilton:1997zq,PhysRevD.70.083007,2013PhR...530...87W} and other quantities \citep{Matsubara:1995wj,Codis:2013exa}.

There are two phenomena commonly associated with redshift space distortion. On small scales $\lesssim {\cal O} (1\, {\rm Mpc})$ the Finger of God effect describes the scatter of galaxy positions within bound structures due to their stochastic velocity components \citep{1972MNRAS.156P...1J}. In addition,  coherent in-fall into overdensities – and corresponding outflow from underdensities – occurs on all scales. The latter phenomenon, dubbed the Kaiser effect \citep{1987MNRAS.227....1K}, can be described using linear perturbation theory on large scales. The density field in the late Universe is non-Gaussian due to the non-linear nature of gravitational collapse, but by smoothing the field on sufficiently large scales one can treat the field as approximately Gaussian and the RSD effect as approximately Kaiser-ian. The anisotropic effect of redshift space distortion contains information regarding the growth rate of structure, due to the fact that the signal is a measure of the in-fall rate of matter into gravitational potentials. 

This work is a continuation of a series of papers by the authors, in which we consider the impact of redshift space distortion on the tensor Minkowski functionals. In \cite{Appleby:2018tzk}, the authors described a numerical algorithm used to extract the Minkowski functionals and Cartesian tensors from any three-dimensional field. In \cite{Appleby_2019} we constructed the ensemble expectation value of the Minkowski tensors in redshift space, in the linearized, plane-parallel Kaiser limit and for Gaussian random fields. The latter paper used the so-called `distant observer' approximation, making the simplifying assumption that the field is sufficiently remote from the observer and localised in direction, so that each point in the field practically shares a common line-of-sight vector along which the redshift space distortion operator acts. This, in conjunction with periodic boundary conditions, renders the field anisotropic but homogeneous, and the sky flat for computational purposes. In reality, the radial nature of the RSD signal generates an inhomogeneous field. 

The purpose of this work is two-fold. First, we generalise the calculation in \cite{Appleby_2019} to account for the radial nature of the RSD signal. We calculate the ensemble average of the Minkowski tensors in spherical coordinates, for a field that has been subjected to a radial RSD correction. The calculation requires a careful reappraisal of the Cartesian tensor analysis of \cite{Appleby_2019} to account for the vagaries of curvilinear coordinate systems. In addition, a radial signal is inherently inhomogeneous, and this will have consequences for the assumption of ergodicity that is frequently applied to cosmological fields. Second, we use gravitationally evolved, dark matter N-body simulations to construct mildly non-Gaussian density fields by smoothing over large scales $15 \, {\rm Mpc} < R_{\rm G} < 45 \, {\rm Mpc}$. We compare the extracted Minkowski tensor statistics to their Gaussian expectation values, to determine the scale at which the analytic prediction can be used. This analysis serves as a precursor to a forthcoming paper, in which we will extract these statistics from the BOSS galaxy data and infer the growth rate from the RSD signal.

The paper will proceed as follows. We review the definition of the rank-2 Minkowski tensors in Section \ref{sec:theory}, and also provide details on our approach to ensemble averaging. In Section \ref{sec:pp} we re-state the main results of \cite{Appleby_2019}; the ensemble average of the Minkowski tensors in globally plane-parallel redshift space. In Section \ref{sec:sph} we expand the analysis and derive the expectation value of the MTs in a spherical coordinate system for a field with radial anisotropy relative to a central observer. We repeat this analysis in a Cartesian coordinate system in Section \ref{sec:cart}. In Section \ref{sec:num} we extract the Minkowski tensors from dark matter particle snapshot boxes after applying a radial redshift space distortion correction, to test the scale at which the Gaussian limit is approached and the magnitude of the non-Gaussian corrections. We also compare plane parallel and radial anisotropic signals. We discuss our results in Section \ref{sec:dis}.

Throughout this work, in the main body of the text we focus on the particular Minkowski tensor $W^{0,2}_{1}$, because it is computationally simpler and we expect that it will provide superior constraining power \citep{Appleby_2019}. A second linearly independent, translation invariant Minkowski tensor in three dimensions $W^{0,2}_{2}$ has some additional complications because it is a function of the second derivative of the field. For completeness we include a brief analysis of $W^{0,2}_{2}$ in Appendix \ref{sec:appen1}. The rotation of the spherical basis vectors relative to a great arc tangent vector is presented in Appendix \ref{sec:appen3} and finally some useful identities regarding spherical harmonics and Bessel functions are provided in Appendix \ref{sec:appen2}. 

%%%%%%%%%%%%%%%%%%%%%%%%%%%%%%%%%%%%%%%%%%%%%%%%%%%%%%
\section{Translation Invariant Minkowski Tensors in Three-Dimensions}
\label{sec:theory}

The Minkowski Tensors (MTs) have been elucidated in numerous papers, and we direct the reader to \cite{1367-2630-15-8-083028} for details on the quantities used in this work. Briefly, in three dimensions we define an excursion set $Q$ for a field $\delta(x)$ on a manifold $\Mspace$ as

\begin{equation} 
\label{intro:Au:equn}
Q =\  \{x\in \Mspace: \delta(x)\geq \delta_{t}\},
\end{equation}

\noindent where $\delta_{t}$ is a chosen density threshold value. Initially we take the manifold $\Mspace$ to be three-dimensional Euclidean space $\Rspace^3$. We then define two translation invariant, rank-two tensors as 

\begin{eqnarray}\label{eq:eq2} & & W_{1}^{0,2} \equiv \frac{1}{6V} \int_{\partial Q} {\bf \hat{n}}^{2} \textrm{dA} ,\\ 
\label{eq:eq3} & & W_{2}^{0,2} \equiv {1 \over 3\pi V} \int_{\partial Q}  G_{2} {\bf \hat{n}}^{2} \textrm{dA}
\end{eqnarray} 

\noindent where the boundary $\partial Q$ of $Q$ is a two-dimensional iso-field surface defined by  $\delta(x) = \delta_{t}$. The vector ${\bf \hat{n}}$ is the unit normal vector and $G_{2}$ is the mean curvature at each point of the surface $\partial Q$. We define the symmetric tensor product as ${\bf \hat{n}^{2}} = {\bf \hat{n}} \otimes {\bf \hat{n}} = (\hat{n}_{i} \hat{n}_{j} + \hat{n}_{j} \hat{n}_{i})/2$. The vector ${\bf \hat n}$ is an element of the cotangent space at each point on $\Rspace^3$. Since addition is defined only for vectors or tensors that belong to the same vector space, in order to perform these integrals we must transport all normal vectors to a fiducial point, and addition is then carried out in the cotangent space at that point. This is a trivial step when the manifold is flat space. $W_{1}^{0,2}$ and $W_{2}^{0,2}$ are invariant under translation of the coordinates, which ensures that they are independent of the choice of fiducial point on $\Rspace^{3}$. If the manifold is curved, then the integrals defined in the expressions ($\ref{eq:eq2}$), ($\ref{eq:eq3}$) require a fiducial point at which the average is taken to be specified, as well as the choice of transport path. These details will be important later and considered in Section \ref{sec:volav}.

We will measure $W_1^{0,2}$ and $W_2^{0,2}$ from dark matter point distributions, which are smoothed with a Gaussian kernel to generate a continuous matter field with background density $\rho_{m}$ and fluctuations $\delta(x)$. The fluctuations satisfy $\langle \delta \rangle = 0$, where $\langle ...  \rangle$ represents the ensemble average of this random field. When smoothed on large scales, $\delta(x)$ is assumed to be well approximated as a Gaussian random field but on small scales non-Gaussianities are present due to the mode coupling arising from the non-linear nature of gravitational collapse. In this work we are chiefly concerned with the large scale limit of the density field, where Gaussian statistics can be applied. The non-Gaussian corrections require further study and are beyond the scope of this work. For the remainder of the paper, we will focus specifically on the Minkowski tensor $W_{1}^{0,2}$, and consign the more complicated $W_{2}^{0,2}$ statistic to Appendix \ref{sec:appen1}. 

Following \citet{Schmalzing:1997aj,Schmalzing:1997uc}, we perform a surface to volume integral transform and use $\hat{n}_{i} = \delta_{i}/|\nabla \delta|$ to re-write equation (\ref{eq:eq2}) as 
\begin{eqnarray}
 W_1^{0,2}|_{i}{}^{j}  &=& \frac{1}{6V} \int_{V} \textrm{dV}  \, \delta_{D}\left( \delta - \delta_{t} \right) \frac{\delta_{i} \delta^{j}}{\left| \nabla \delta \right|} ,
 \label{eqn:W_delta}
\end{eqnarray}
\noindent where we use shorthand notation for the gradients of the field $\delta_{i} = \nabla_{i}\delta$, and $\delta_D$ is the Dirac delta function. Given that $\delta$ is assumed to be a smooth random field, its derivatives and in particular the vector $\delta_{i}/|\nabla \delta|$ is well defined at all points over the volume $V$. The right hand side of equation ($\ref{eqn:W_delta}$) is the volume average of the rank-$(1,1)$ tensor 

\begin{equation}\label{eq:t1} w_{i}{}^{j} = {1 \over 6 }  {\delta _{i}\delta^{j} \over |\nabla \delta|} \delta_{D}(\delta-\delta_{t}), \end{equation} 

\noindent where the delta function $\delta_{D}(\delta-\delta_{t})$ can be defined in a distributional sense when constructing the ensemble average or approximately discretized when taking the volume average \citep{Schmalzing:1997aj}. We denote the volume average of this tensor as $\bar{w}_{i}{}^{j} \equiv W^{0,2}_{1}|_{i}{}^{j}$. 

%%%%%%%%%%%%%%%%%%%%%%%%%%%%%%%%%%%%%%%%%%%%%%%%%%%%%%%
\subsection{Ensemble Average and Ergodicity} 
\label{sec:ens}

First, we review the steps made in calculating the ensemble average of $w_{i}{}^{j}$, because there are some subtleties that will become important later. The purpose of this subsection is to highlight the assumptions that are made when deriving the ensemble average of $w_{i}{}^{j}$, and then equating this quantity to the volume average that we measure from cosmological data. 

The ensemble average $\langle ... \rangle$ is the linear sum over possible states of the quantity within the brackets, weighted by the probability of that state --

\begin{equation} \label{eq:ens}  \langle  w_{i}{}^{j} \rangle   = \frac{1}{6} \int \Phi(X,\Sigma)  \, \delta_{D} \left( \delta - \delta_{t} \right) \frac{\delta_{i} \delta^{j}}{\left| \nabla \delta \right|} \textrm{dX}  ,
\end{equation} 

\noindent where $X = (\delta,\delta_{i})$ is shorthand for an array of the field and components of its first derivatives and $\Phi(X,\Sigma)$ is the underlying probability distribution function (PDF) for $X$. Here $w_{i}{}^{j}$ is defined at a point on the manifold, so $\Phi(X,\Sigma)$ is the PDF describing the field and its derivatives at a single location. For a Gaussian random field we have $\Phi(X, \Sigma) \propto \exp[-X^{T} \Sigma^{-1} X/2]$, where $\Sigma$ denotes the covariance between the component fields of $X$. When integrating over $X$, all physical information is contained within the inverse covariance matrix $\Sigma^{-1}$ in $\Phi(X, \Sigma)$. To estimate the ensemble average of $w_{i}{}^{j}$, we require the covariance matrix $\Sigma$. 

In cosmological applications, we measure $\bar{w}_{i}{}^{j}$ from a data set and then equate this quantity to the theoretically predicted $\langle w_{i}{}^{j} \rangle$. That is, we invoke ergodicity to impose $\langle w_{i}{}^{j} \rangle \simeq \bar{w}_{i}{}^{j}$. Ergodicity is known to be exact if a field is  homogeneous, Gaussian, the two-point correlation $\zeta$ of $\delta$ satisfies $\zeta(r)|_{r\to \infty} = 0$ and we take the limit $V \to \infty$~(\citet{Adler} p145). In reality, cosmological fields occupy a finite volume and have finite resolution, and ergodicity is never exactly realised. We tacitly interpret the volume average of a quantity over a finite domain as providing an unbiased estimate of the ensemble average, with an associated uncertainty related to the finite sampling of the probability distribution. 

If the covariance $\Sigma$ between the fields $\delta, \delta_{i}$ contains explicit coordinate dependence, then the ensemble average $\langle w_{i}{}^{j} \rangle$ is sensitive to the position ${\bf x}$ on the manifold at which we take this average -- $\Phi=\Phi(X, \Sigma({\bf x}))$. In this case, it is clear that the ensemble average at any given point cannot be equated to the volume average of the same tensor over the entire manifold. Constancy of $\Sigma$ is a consequence of the fields being homogeneous (see e.g. \citep{Adler,Chingangbam:2021kov}), so when the fields are inhomogeneous we cannot invoke ergodicity and generically $\langle w_{i}{}^{j}\rangle \neq \bar{w}_{i}{}^{j}$. In such a situation, the question of whether we can invoke ergodicity -- even approximately -- depends on the physical properties of the field, manifold and coordinate system adopted. In what follows we will present an example for which $\bar{w}_{i}{}^{j} \simeq \langle w_{i}{}^{j} \rangle$ is an excellent approximation despite the field being inhomogeneous, and a second example for which $\bar{w}_{i}{}^{j}$ completely fails to encapsulate the properties of the ensemble average.

For a homogeneous field, $\Sigma$ and hence $\langle w_{i}{}^{j}\rangle$ are constant over the entire manifold and ergodicity is more naturally realised. Ambiguity remains in the definition of the volume average of a tensor, which is discussed further in Section \ref{sec:volav}. 

\section{Review : Plane Parallel Redshift Space Distortions}
\label{sec:pp}

In Section \ref{sec:sph} we will calculate the ensemble average of $w_{i}{}^{j}$ for a Gaussian field that has been subjected to a spherically symmetric redshift space distortion operator, but before doing so we briefly review the plane parallel result derived in  \cite{Appleby_2019}, aided by earlier work on the Minkowski functionals \citep{1970Ap......6..320D,Adler,Gott:1986uz,10.1143/PTP.76.952,Hamilton:1986,Ryden:1988rk,1987ApJ...319....1G,1987ApJ...321....2W,Matsubara:1994wn,Matsubara:1994we,Matsubara:1995dv,Gay:2011wz,2000astro.ph..6269M,10.1111/j.1365-2966.2008.12944.x}\footnote{See \citet{Buchert:2017uup} for a model-independent approach applying Minkowski functionals to the CMB and using general Hermite expansions of the discrepancy functions with respect to the analytical Gaussian predictions.}. 

We take an isotropic and homogeneous Gaussian random field in a periodic box, adopt a Cartesian coordinate system $x,y,z$ with basis vectors ${\bf e}_{x}$, ${\bf e}_{y}$, ${\bf e}_{z}$, and then apply the plane parallel redshift space distortion operator aligned with one of the coordinate axes taken arbitrarily to be ${\bf e}_{z}$. We preserve periodicity in ${\bf e}_{z}$, so that the field is homogeneous but anisotropic. We simply re-state the main results of \cite{Appleby_2019}, and direct the reader to that work for details of the calculation and \citet{Matsubara:1995wj,Codis:2013exa} for a detailed analysis of the RSD effect on the scalar functionals. 

To linear order in the density fluctuation, the relation between the true position of a tracer particle ${\bf x}$ and its redshift space position ${\bf s}$ is given by 

\begin{equation} {\bf s} = {\bf x} + f {\bf e}_{z} \, ({\bf u} . {\bf e}_{z}) \end{equation} 

\noindent where $f=d\ln D/d\ln a$ and $D$ is the linear growth factor, ${\bf u} = {\bf v}/(aHf)$, ${\bf v}$ is the peculiar velocity and $H$ is the Hubble parameter. We have assumed that every tracer particle is subject to a single, parallel line of sight. The density field in redshift space $\tilde{\delta}$ can be related to its real space counterpart $\delta$ according to 

\begin{equation} \label{eq:pp1} \tilde{\delta}({\bf k}) = (1 + f \mu^{2}) \delta ({\bf k}) , \end{equation} 

\noindent where $\mu = {\bf k}. {\bf 
e}_{z}/|k|$ is the cosine of the angle between the line of sight and wavenumber ${\bf k}$. The cumulants of the field $\tilde{\delta}$ and its gradient are given by \citep{Matsubara:1995wj}

\begin{eqnarray} \label{eq:rc0} \langle \tilde{\delta}({\bf x'}) \tilde{\delta}({\bf x}) \rangle|_{{\bf x'} \to {\bf x}}   &=&  \sigma_{0}^{2} \left[ 1 + {2 \over 3}f + {1 \over 5} f^{2} \right] \\
\langle \tilde{\delta}_{x}({\bf x'}) \tilde{\delta}_{x}({\bf x}) \rangle|_{{\bf x'} \to {\bf x}} =  \langle \tilde{\delta}_{y}({\bf x'}) \tilde{\delta}_{y}({\bf x}) \rangle|_{{\bf x'} \to {\bf x}}  &=&  \sigma_{1}^{2} \left[ {1 \over 3} + {2 \over 15}f + {1 \over 35} f^{2}\right] \\
\label{eq:rc3} \langle \tilde{\delta}_{z}({\bf x'}) \tilde{\delta}_{z}({\bf x}) \rangle|_{{\bf x'} \to {\bf x}}   &=&  \sigma_{1}^{2} \left[ {1 \over 3} + {2 \over 5}f + {1 \over 7}f^{2} \right] \\
\label{eq:rc4} \langle \tilde{\delta}({\bf x'}) \tilde{\delta}_{i}({\bf x}) \rangle|_{{\bf x'} \to {\bf x}}   &=&  0
\end{eqnarray}

\noindent where we have defined the $i^{\rm th}$ isotropic cumulant as 

\begin{equation} \sigma_{i}^{2} = {1 \over 2\pi^{2}} \int k^{2i+2} P(k, R_{\rm G})  dk , \end{equation}

\noindent and have introduced a Gaussian-smoothed power spectrum $P(k,R_{\rm G}) = W^{2}(kR_{\rm G})P(k)$ with $W(k R_{\rm G}) \propto e^{-k^{2}R_{\rm G}^{2}/2}$ for some comoving smoothing scale $R_{\rm G}$. The ensemble expectation value of the components of the Minkowski tensor $W^{0,2}_{1}$ in this particular Cartesian coordinate system, assuming the field is Gaussian, are then \cite{Appleby_2019}

\begin{eqnarray}\label{eq:m1} & &  \langle W^{0,2}_{1}|_{xx} \rangle =  {A_{0} \over 4}\left[ {(2\lambda^{2}-1)\cosh^{-1}\left(2\lambda^{2}-1\right) \over (\lambda^{2}-1)^{3/2}} - {2\lambda \over \lambda^{2}-1}  \right] e^{-\nu^{2}/2} , \\
& &  \langle W^{0,2}_{1}|_{yy} \rangle = \langle W^{0,2}_{1}|_{xx} \rangle , \\ 
\label{eq:m2} & & \langle W^{0,2}_{1}|_{zz} \rangle = A_{0}\left({\lambda^{2} \over \lambda^{2}-1}\right) \left( \lambda - {\cosh^{-1} \lambda \over \sqrt{\lambda^{2}-1}}\right) e^{-\nu^{2}/2} , \\
\label{eq:m3} & & \langle W^{0,2}_{1}|_{xy} \rangle = \langle W^{0,2}_{1}|_{xz} \rangle =  \langle W^{0,2}_{1}|_{yz} \rangle =  0 ,
\end{eqnarray} 

\noindent the constant $A_{0}$ is given by

\begin{equation}\label{eq:a0} A_{0} = {\sigma_{1} \over 6\sqrt{3}\pi\sigma_{0}}\sqrt{105 + 42 f + 9 f^{2} \over 105 + 70 f + 21f^{2}} ,
\end{equation} 

\noindent and 

\begin{equation}\label{eq:lam} \lambda^{2} = {35 + 42 f   + 15 f^{2} \over 35 + 14 f  + 3 f^{2}} , \end{equation}

\noindent and we have introduced the normalised threshold $\nu = \delta_{t}/\tilde{\sigma}_{0}$, where $\tilde{\sigma}_{0}^{2} = \langle \tilde{\delta}({\bf x'}) \tilde{\delta}({\bf x}) \rangle|_{{\bf x'} \to {\bf x}}$. The Minkowski tensor is diagonal in this coordinate system, with discrepant values in the directions perpendicular and parallel to the `line of sight' $z$. A coordinate transform will generate off-diagonal terms, but the eigenvalues remain invariant. Modulo a noise component due to finite sampling, the eigenvalues are equal to the diagonal elements of the MT in this coordinate system. The properties of the field dictate the form of the MT ; anisotropy is represented by unequal eigenvalues, and homogeneity is manifested by the constancy of the cumulants ($\ref{eq:rc0}$-$\ref{eq:rc4}$) over the domain on which the field is defined.

\section{Minkowski Tensors -- Spherical Redshift Space Distortion} 
\label{sec:sph}

The plane parallel limit reviewed in the previous section is an approximation where the observed patch of the density field is sufficiently distant from the observer and localised on the sky so that the line of sight can be approximately aligned with one of the Cartesian axes. Now we generalise and calculate the Minkowski tensors without the plane parallel approximation. Since redshift space distortion acts along the line of sight, we choose to work with the spherical coordinate system with the observer at the origin. The radial and angular basis vectors in this system are denoted ${\bf e}_{r}$, ${\bf e}_{\theta}$, ${\bf e}_{\phi}$, and ${\bf e}_{r}$ is aligned with the line of sight. The redshift space distortion operator is spherically symmetric and applied to an otherwise isotropic and homogeneous Gaussian random field. Under the assumption that the average number density of tracer particles is constant over the manifold, the relation between the density field in real ($\delta$) and redshift ($\tilde{\delta})$ space is given by \citep{Hamilton:1997zq} 

\begin{equation}\label{eq:sphd} \tilde{\delta}({\bf r}) = \left[ 1 + f \left(  {\partial^{2} \over \partial r^{2}} + {2 \over r} {\partial \over \partial r}\right) \nabla^{-2} \right] \delta({\bf r}) , \end{equation} 

\noindent to linear order in the fields. Here $f$ is the growth factor which we assume to be constant, neglecting its redshift dependence. The redshift space distortion operator in square brackets is now radial relative to a central observer located at $r=0$. There is no longer a uniformly parallel line of sight vector over the entire manifold -- the line of sight is now aligned with the radial basis vector ${\bf e}_{r}$. The redshift space field is sensitive to this vector, because tracer particles that are used to define $\tilde{\delta}$ are perturbed according to the component of their velocity parallel to the corresponding line of sight direction. The radial nature of the signal renders the redshift space distorted field inhomogeneous, and the two-point correlation function of $\tilde{\delta}$ is no longer solely a function of the separation between two tracer particles, but now depends on the triangle formed by the observer and the two points. Translation invariance is broken, but residual rotational symmetry around the observer and azimuthal symmetry about the line of sight persist.

\subsection{Ensemble Average $\langle w_{i}{}^{j} \rangle$}
\label{sec:ensav}

The goal of this section is to derive the ensemble average of the tensor $w_{i}{}^{j}$ for the field $\tilde{\delta}$ defined in equation ($\ref{eq:sphd}$), in a spherical coordinate system. The first step is to derive the cumulants $\langle \tilde{\delta}^{2} \rangle$, $\langle \tilde{\delta} \tilde{\delta}_{i} \rangle$ and $\langle \tilde{\delta}_{i}\tilde{\delta}^{j} \rangle$. The variance of the field $\langle \tilde{\delta}^{2} \rangle$ is a scalar quantity and hence invariant under coordinate transformations, but $\langle \tilde{\delta}_{i}\tilde{\delta}^{j}\rangle$ is a rank-$(1,1)$ tensor and $\langle \tilde{\delta}\tilde{\delta}_{i}\rangle$ is a vector, both of which transform non-trivially. Spherical redshift space two-point statistics have been extensively studied in the literature, and we direct the reader to \citet{1992ApJ...385L...5H,1996MNRAS.278...73H, 1996ApJ...462...25Z,1998ApJ...498L...1S, Szapudi:2004gh,Shaw:2008aa,Bonvin:2011bg, Raccanelli:2013gja, 10.1093/mnras/stu2491,Reimberg_2016,Paul:2022xfx} and references therein for details. 

Starting with the scalar cumulant, following \citet{Castorina:2017inr} we define the density field in terms of angular coefficients as 

\begin{equation}  \tilde{\delta}({\bf r}) = \sum_{\ell m} a_{\ell m}(r) Y^{*}_{\ell m}(\hat{r}) . \end{equation} 

\noindent Then the two-point function is given by 

\begin{eqnarray} \langle \tilde{\delta}({\bf r'}) \tilde{\delta}({\bf r}) \rangle = \zeta({\bf r}, {\bf r'}) &=& \sum_{\ell m} \langle a_{\ell m}(r) a_{\ell m}(r') \rangle  Y_{\ell m}(\hat{r}) Y^{*}_{\ell m}(\hat{r}')  \\
&=& \sum_{\ell} {2\ell + 1 \over 4\pi} C_{\ell}(r,r') {\cal L}_{\ell} (\hat{r} . \hat{r}') , \end{eqnarray}

\noindent where 

\begin{equation} C_{\ell}(r,r') = {2 \over \pi} \int dk k^{2} P(k, R_{\rm G}) \left[ j_{\ell}(kr) - f \left( j''_{\ell}(kr) + {2 \over kr}j'_{\ell}(kr) \right) \right] \left[ j_{\ell}(kr') - f \left( j''_{\ell}(kr') + {2 \over kr'}j'_{\ell}(kr') \right) \right] ,\end{equation}

\noindent where primes on the spherical Bessel function $j_{\ell}$ denote differentiation with respect to the argument $kr$ or $kr'$ and ${\cal L}_{\ell}$ are Legendre polynomials. The cumulant is defined as the field correlation in the limit ${\bf r} \to {\bf r}'$, which is 

\begin{eqnarray}\nonumber  \tilde{\sigma}_{0}^{2} \equiv \langle \tilde{\delta}({\bf r'}) \tilde{\delta}({\bf r}) \rangle|_{{\bf r'} \to {\bf r}}   &=& {1 \over 2 \pi^{2}} \int dk k^{2} P(k, R_{\rm G}) \sum_{\ell=0}^{\infty} (2\ell+1) \left[  j_{\ell}^{2}(kr) - 2 f j_{\ell}(kr)j''_{\ell}(kr) + f^{2}\left[j''_{\ell}(kr)\right]^{2} +  \vphantom{\frac{1}{2}} \right. \\ \nonumber & &  \left. - {4 f \over kr} j_{\ell}(kr) j'_{\ell}(kr) + {4 f^{2} \over kr } j''_{\ell}(kr) j'_{\ell}(kr)  + {4 f^{2} \over k^{2}r^{2}} \left[j'_{\ell}(kr)\right]^{2}  \right] ,\\
\label{eq:sig0} &=& {1 \over 2 \pi^{2}} \int dk k^{2} P(k, R_{\rm G}) \left[ 1 + {2 \over 3}f + {1 \over 5} f^{2} \right] + {4 f^{2} \over 3r^{2}} {1 \over 2\pi^{2}} \int dk P(k, R_{\rm G}) .
\end{eqnarray}

\noindent The first term on the right hand side of ($\ref{eq:sig0}$) is the cumulant in the plane parallel limit. The second term is divergent as $r \to 0$ but falls off at large distances from the central observer. The divergent behaviour at $r=0$ is not physical, and can be subtracted via a suitable correction to the space distortion operator in ($\ref{eq:sphd}$). Practically, cosmological data will always occupy a domain excluding the observer and for computational purposes we will excise the $r=0$ point from the manifold in redshift space. Hence the manifold on which the RSD field $\tilde{\delta}$ is defined is not $\mathbb{R}^{3}$, but rather $\mathbb{S}^{2} \times \mathbb{R}_{> 0}$. 

Similarly the radial and angular derivative cumulants can be calculated -- 

\begin{eqnarray} \nonumber  \langle \tilde{\delta}_{r}({\bf r'}) \tilde{\delta}^{r}({\bf r})  \rangle|_{{\bf r'} \to {\bf r}}  &=& {1 \over 2 \pi^{2}} \int dk k^{4} P(k, R_{\rm G}) \sum_{\ell=0}^{\infty} (2\ell+1) \left[ [j'_{\ell}(kr)]^{2} - 2 f j'_{\ell}(kr)j'''_{\ell}(kr) + f^{2}[j'''_{\ell}(kr)]^{2} \vphantom{\frac{1}{2}} \right. \\ 
\nonumber & & + \left.   {4 f^{2} \over  k^{2} r^{2}} j''_{\ell}(kr) j''_{\ell}(kr) + {4 f\over  k^{2}r^{2}} j'_{\ell}(kr) j'_{\ell}(kr) - {4 f^{2} \over k^{2}r^{2}} j'_{\ell}(kr) j'''_{\ell}(kr) + {4 f^{2} \over k^{4} r^{4}} j'_{\ell}(kr) j'_{\ell}(kr) \right] \\
\nonumber  &=& {1 \over 2 \pi^{2}} \int dk k^{4} P(k, R_{\rm G}) \left[ {1 \over 3} + {2 \over 5}f + {1 \over 7}f^{2} \right] + {1 \over 2\pi^{2}r^{2}} \int dk k^{2} P(k, R_{\rm G}) \left[ {4 f \over 3 } + {8 f^{2} \over 5 } \right] + \\
& & +{1 \over 2\pi^{2}r^{4}} \int dk P(k, R_{\rm G}) {4 f^{2} \over 3} 
 \label{eq:sigr} \end{eqnarray} 

\noindent and 

\begin{eqnarray} \nonumber  \langle \tilde{\delta}_{\phi}({\bf r'}) \tilde{\delta}^{\phi}({\bf r})  \rangle|_{{\bf r'} \to {\bf r}}    &=& {1 \over 4\pi^{2}r^{2}} \int dk k^{2}P(k, R_{\rm G}) \sum_{\ell=0}^{\infty} \ell (\ell+1) (2\ell +1) \left[ j_{\ell}^{2}(kr) - 2 f j_{\ell}(kr)j''_{\ell}(kr) + f^{2}j''_{\ell}(kr)j''_{\ell}(kr) + \vphantom{\frac{1}{2}} \right. \\ 
\nonumber  & &  \left. - {4 f \over kr} j_{\ell}(kr) j'_{\ell}(kr) + {4 f^{2} \over kr } j''_{\ell}(kr) j'_{\ell}(kr)  + {4 f^{2} \over k^{2}r^{2}} j'_{\ell}(kr) j'_{\ell}(kr)   \right] \\
\nonumber  &=& {1 \over 2\pi^{2}} \int dk k^{4}P(k, R_{\rm G})  \left[ {1 \over 3} + {2 \over 15}f + {1 \over 35} f^{2}\right] + {1  \over 2\pi^{2}r^{2}}    \int dk k^{2}P(k, R_{\rm G}) \left[-{4 \over 3} f + {8 \over 15} f^{2}\right] + \\ 
& & +  {1 \over 2\pi^{2}r^{4}} \int dk P(k, R_{\rm G}) {4 f^{2} \over 3} \label{eq:sigphi}  \\
\nonumber  \langle \tilde{\delta}_{\theta}({\bf r'}) \tilde{\delta}^{\theta}({\bf r})  \rangle|_{{\bf r'} \to {\bf r}}   &=& {1 \over 2\pi^{2}} \int dk k^{4}P(k, R_{\rm G})  \left[ {1 \over 3} + {2 \over 15}f + {1 \over 35} f^{2}\right] + {1  \over 2\pi^{2}r^{2}}    \int dk k^{2}P(k, R_{\rm G}) \left[-{4 \over 3} f + {8 \over 15} f^{2}\right] +  \\
 & & +{1 \over 2\pi^{2}r^{4}} \int dk P(k, R_{\rm G}) {4 f^{2} \over 3}  \label{eq:sigthe}\end{eqnarray}

\noindent  The cross covariance terms are zero in this coordinate system -- for example

\begin{eqnarray} & &  \langle \tilde{\delta}_{r}({\bf r'}) \tilde{\delta}^{\phi}({\bf r})  \rangle|_{{\bf r'} \to {\bf r}}  = {2 \over \pi} \int dk k^{3}P(k, R_{\rm G}) \sum_{\ell=0}^{\infty}  j_{\ell}(kr) j'_{\ell}(kr) \sum_{m=-\ell}^{\ell} (im) Y_{\ell m}(\hat{r}) Y^{*}_{\ell m}(\hat{r}) = 0 . \end{eqnarray}

\noindent Similarly 

\begin{equation} \langle \tilde{\delta}_{\theta}({\bf r'}) \tilde{\delta}^{\phi}({\bf r})  \rangle|_{{\bf r'} \to {\bf r}} = \langle \tilde{\delta}_{r}({\bf r'}) \tilde{\delta}^{\theta}({\bf r})  \rangle|_{{\bf r'} \to {\bf r}} = 0 .\end{equation} 

\noindent Hence in this coordinate system, the gradient cumulant tensor $\langle \tilde{\delta}_{i}\tilde{\delta}^{j} \rangle$ is diagonal. There is an additional correlation not present for a homogeneous field --  the vector $\langle \tilde{\delta} \tilde{\delta}_{i}\rangle$ has a single non-zero component 

\begin{eqnarray} \nonumber  \langle \tilde{\delta}({\bf r'}) \tilde{\delta}_{r}({\bf r})  \rangle|_{{\bf r'} \to {\bf r}}  &=& {1 \over 2 \pi^{2}} \int dk k^{3} P(k, R_{\rm G})   \sum_{\ell} (2\ell+1) \left[ j_{\ell}(kr) - f(z) \left( j''_{\ell}(kr) + {2 \over kr} j'_{\ell}(kr) \right)  \right] \\
 \nonumber & & \hspace{10mm} \times \left[ j'_{\ell}(kr) - f(z) \left( j'''_{\ell}(kr) + {2 \over kr} j''_{\ell}(kr) - {2 \over k^{2}r^{2}}j'_{\ell}(kr) \right)  \right]  \\
 &=& - {4 f^{2} \over 3 r^{3}} {1 \over 2\pi^{2}} \int dk  P(k, R_{\rm G})
\end{eqnarray}

\begin{figure}[htb]
  \centering 
  \includegraphics[width=0.45\textwidth]{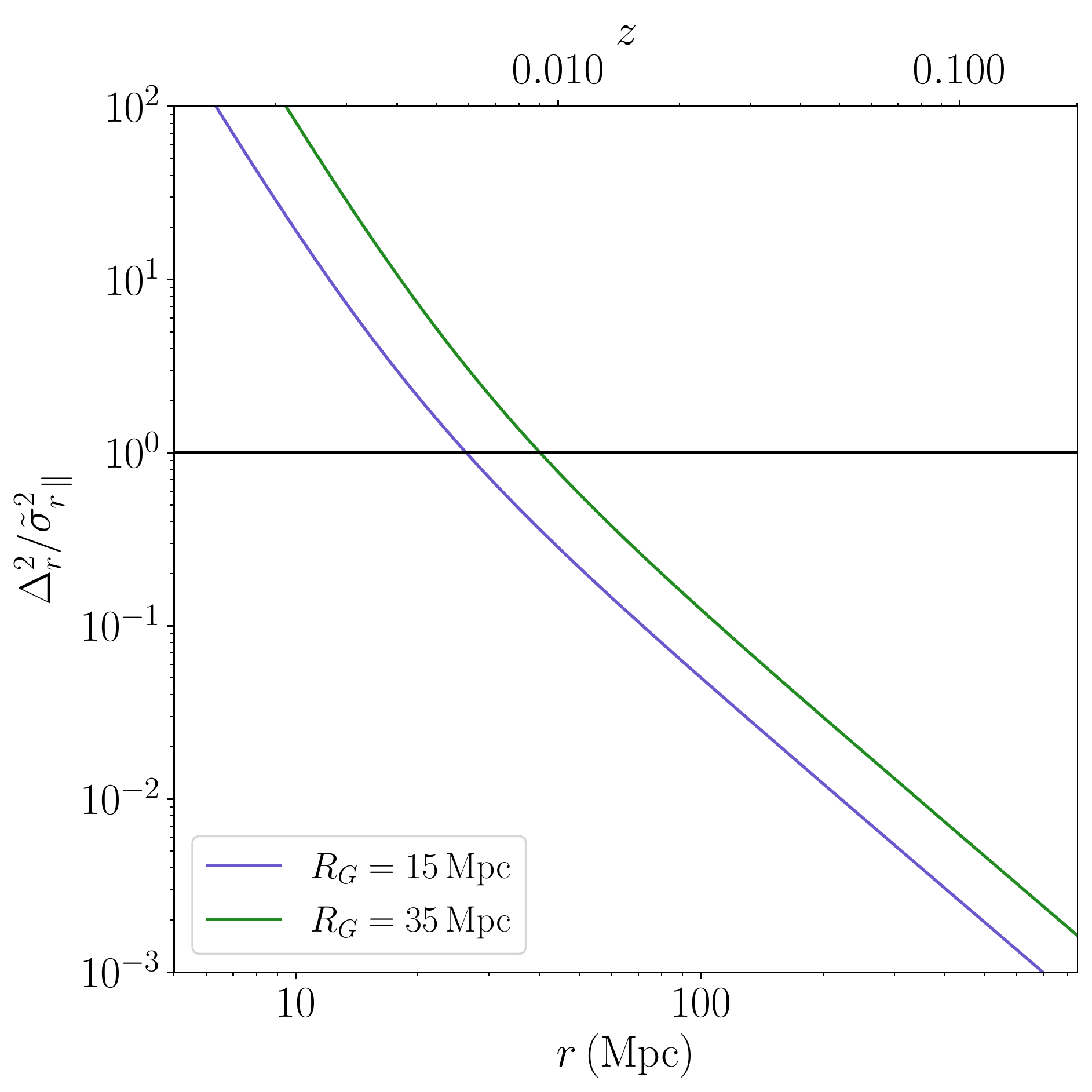} \\
  \caption{Fractional correction to the radial cumulant $\tilde{\sigma}_{r}^{2}$ due to the inhomogeneous contribution $\Delta_{r}^{2}(r)$, relative to its plane parallel limit $\tilde{\sigma}^{2}_{r,\parallelsum}$. Blue/green lines correspond to smoothing scales $R_{\rm G} = 15, 35 \, {\rm Mpc}$ respectively.}
  \label{fig:1}
\end{figure}

\noindent There are two crucial differences between this scenario and the previous plane parallel calculation in \cite{Appleby_2019} -- the cumulants are now explicitly functions of the position on the manifold at which they are estimated and they are no longer defined over $\Rspace^{3}$ since we excise the $r=0$ point. Both are consequences of the inhomogeneous nature of the redshift space distortion signal. In each of the cumulants ($\ref{eq:sig0}$-$\ref{eq:sigthe}$), the first term on the right hand sides correspond to the plane parallel limit, and the remaining terms are corrections that are fractionally suppressed by $\sigma_{0}^{2}/(\sigma_{1}^{2} r^{2})$ and $\sigma_{-1}^{2}/(\sigma_{1}^{2}r^{4})$ at large distances from the observer. Similarly the vector $\langle \tilde{\delta} \tilde{\delta}_{i} \rangle$ has asymptotic behaviour $\langle \tilde{\delta}\tilde{\delta}_{i}\rangle \to 0$ as $\sigma_{-1}^{2}/(\sigma_{0}\sigma_{1}r^{3}) \to 0$. Hence at large distances from the observer the cumulants approach their constant, plane parallel limits. 

To quantify the departure of the cumulants from the plane parallel limit, we numerically evaluate ($\ref{eq:sigr}$) for a typical cold dark matter density field in the linearized limit. Taking cosmological parameters from Table \ref{tab:1}, we generate a linear $\Lambda$CDM matter power spectrum $P(k, R_{\rm G})$ at $z=0$ and use this and $f \simeq \Omega_{\rm m}^{\gamma}$, $\gamma = 6/11$ to numerically reconstruct the plane parallel limit $\tilde{\sigma}_{r, \parallelsum}^{2}$ and radial-dependent correction $\Delta_{r}^{2}$ to the cumulant $\langle \tilde{\delta}_{r}({\bf r}) \tilde{\delta}^{r}({\bf r})  \rangle = \tilde{\sigma}_{r, \parallelsum}^{2} + \Delta_{r}^{2}$, defined as --

\begin{eqnarray} & & \tilde{\sigma}_{r, \parallelsum}^{2} \equiv {1 \over 2 \pi^{2}} \int dk k^{4} P(k, R_{\rm G}) \left[ {1 \over 3} + {2 \over 5}f + {1 \over 7}f^{2} \right] \label{eqn:delta_pp}  \\
& & \Delta_{r}^{2}(r) \equiv  {1 \over 2\pi^{2}r^{2}} \int dk k^{2} P(k, R_{\rm G}) \left[ {4 f \over 3 } + {8 f^{2} \over 5 } \right] + {1 \over 2\pi^{2}r^{4}} \int dk P(k, R_{\rm G}) {4 f^{2} \over 3} \label{eqn:delta_a}
\end{eqnarray} 

\noindent In Figure \ref{fig:1} we present the dimensionless fraction $\Delta_{r}^{2}/\tilde{\sigma}_{r,\parallelsum}^{2}$ as a function of comoving distance $r$ from an observer at $r=0$, and the corresponding redshift (top axis) using the standard $\Lambda$CDM distance-redshift relation with parameters given in Table \ref{tab:1}. We select Gaussian smoothing scales $R_{\rm G} = 15, 35 \, {\rm Mpc}$ (blue, green lines). We only present $\langle \tilde{\delta}_{r}({\bf r}) \tilde{\delta}^{r}({\bf r})  \rangle$, as this is representative of the other cumulants. 

The figure shows that the coordinate dependent corrections to the cumulant are negligible for $r \gg R_{\rm G}$, and that for cosmological density fields that occupy a redshift domain $z > 0.05$ the radial cumulant is practically equal to its plane parallel limit $\langle \tilde{\delta}_{r}({\bf r}) \tilde{\delta}^{r}({\bf r})  \rangle \simeq \tilde{\sigma}_{r,\parallelsum}^{2}$. Conversely, for $r \lesssim R_{G}$ the $\Delta_{r}^{2}$ term is the dominant contribution to the cumulant, which is strongly position dependent. In this regime the cumulants grow without bound as $r \to 0$, so there is always a region for which the field $\tilde{\delta}$ cannot be considered perturbatively small. However, the region $r \lesssim R_{G}$ is not typically utilised in any cosmological scale density field reconstruction, and the plane parallel limit of the cumulants is very accurate for our purposes.

After calculating the cumulants $\langle\tilde{\delta}_{i} \tilde{\delta}^{j} \rangle$, $\langle \tilde{\delta} \tilde{\delta}_{i}  \rangle$, $\langle \tilde{\delta}^{2}  \rangle$, we can now estimate the ensemble average $\langle w_{i}{}^{j} \rangle$ -- 

\begin{equation}\label{eq:fens} \langle w_{i}{}^{j} \rangle = {1 \over 6}  \int \Phi(X,\Sigma(r)) {\tilde{\delta}_{i}\tilde{\delta}^{j} \over |\nabla \tilde{\delta}|} \delta_{D}(\tilde{\delta}-\delta_{t}) dX 
\end{equation}

\noindent where $\Phi(X,\Sigma(r))$ is the probability distribution of the variables $X$. The array $X$ denotes any combination of the stochastic fields ($\tilde{\delta},\tilde{\delta}_r,\tilde{\delta}_{\theta},\tilde{\delta}_{\phi}$) to which $w_{i}{}^{j}$ is sensitive. $\Sigma$ is a square matrix whose dimension is given by the number of components of $X$. 

We use the fact that $\tilde{\delta}_{\theta}$ and $\tilde{\delta}_{\phi}$ are uncorrelated with $\tilde{\delta}$ and $\tilde{\delta}_{r}$ and one another, and their variances are equal as given by equations ($\ref{eq:sigphi},\ref{eq:sigthe}$). Furthermore, if the density field is statistically isotropic on the two-sphere it suffices to calculate $\langle w_{\theta}{}^{\theta} + w_{\phi}{}^{\phi} \rangle$, and then halve this value to obtain the individual elements. To estimate $\langle w_{\theta}{}^{\theta} + w_{\phi}{}^{\phi} \rangle$ and $\langle w_{r}{}^{r} \rangle$ we can use the variables $X = (\tilde{\delta}, \tilde{\delta}_{r}, y)$ where $y=\sqrt{\tilde{\delta}_{\theta}\tilde{\delta}^{\theta}+ \tilde{\delta}_{\phi}\tilde{\delta}^{\phi}}$, where $\tilde{\delta}_{\phi}\tilde{\delta}^{\phi}$ and $\tilde{\delta}_{\theta}\tilde{\delta}^{\theta}$ are given by equations (\ref{eq:sigphi}) and (\ref{eq:sigthe}) respectively.  The quantity $y$ is Rayleigh distributed and uncorrelated with $\tilde{\delta}$ and $\tilde{\delta}_{r}$. The fields $\tilde{\delta}$ and $\tilde{\delta}_{r}$ are Gaussian random variables with non-zero correlations

\begin{equation} \hat{\Sigma}(r) \equiv  \left( \begin{tabular}{cc}
       $\langle \tilde{\delta}^{2}  \rangle$  & $\langle \tilde{\delta} \tilde{\delta}_{r} \rangle$  \\
    $\langle \tilde{\delta} \tilde{\delta}_{r} \rangle$  &  $\langle \tilde{\delta}_{r} \tilde{\delta}^{r} \rangle$
    \end{tabular} \right)
\end{equation}

\noindent Each term in $\hat{\Sigma}$ is non-zero and a function of $r$, but in the limit $r \gg \sigma_{0}/\sigma_{1}$ and $r \gg \sqrt{\sigma_{1}/\sigma_{-1}}$, $\Sigma$ approaches a diagonal form with constant components -- the plane parallel limit of \cite{Appleby_2019}. In the same limit the Rayleigh distribution for $y$ becomes independent of the radial coordinate. Defining ${\bf d} = (\tilde{\delta}, \tilde{\delta}_{r})$, $X = (\textbf{d}, y)$ and the probability distribution is 

\begin{equation}\label{eq:pdf} \Phi({\bf d},y,\Sigma (r)) = {y \over \sigma_{y}^{2}\sqrt{(2\pi)^{2}|\hat{\Sigma}|}} \exp\left[- {1 \over 2} {\bf d}^{T} \hat{\Sigma}^{-1} {\bf d} - {y^{2} \over 2\sigma_{y}^{2}}  \right]  
\end{equation}

\noindent where $\sigma_{y}^{2} = \langle \tilde{\delta}_{\theta}\tilde{\delta}^{\theta}\rangle = \langle \tilde{\delta}_{\phi}\tilde{\delta}^{\phi} \rangle$. Although we cannot perform the integral in ($\ref{eq:fens}$) analytically, $\langle w_{i}{}^{j} \rangle$ can be numerically estimated for any $r$. In the regime $r^{2} \gg \sigma_{0}^{2}/\sigma_{1}^{2}$ and $r^{4} \gg \sigma_{-1}^{2}/\sigma_{1}^{2}$ we can use the plane parallel limit calculated in \citet{Appleby_2019} as an excellent approximation.

\begin{figure}[htb]
  \centering 
 \includegraphics[width=0.65\textwidth]{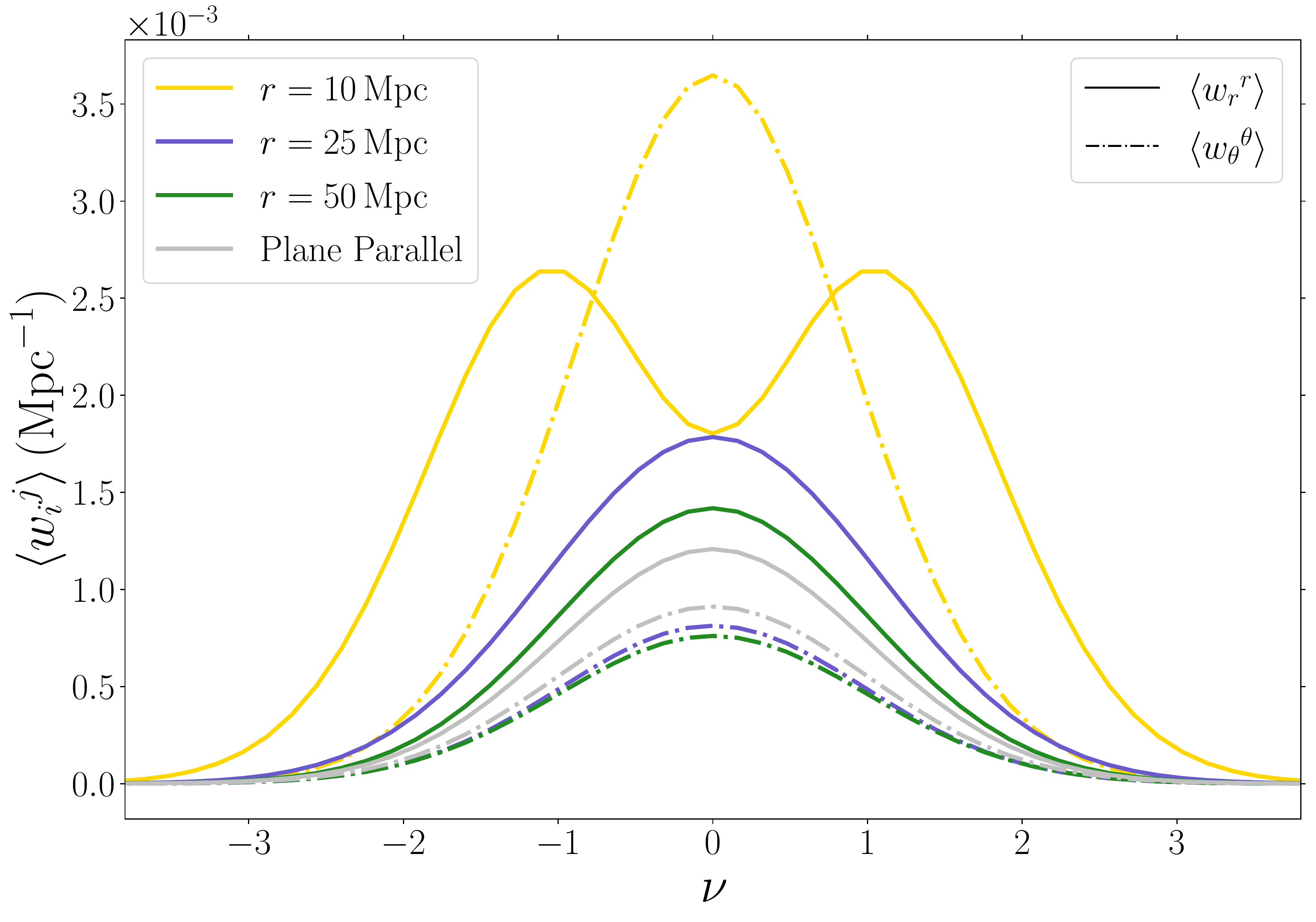} 
  \caption{The ensemble average ($\ref{eq:fens}$) for the spherically redshift space distorted field, evaluated at distance $r=10, 25, 50 \, {\rm Mpc}$ from the central observer (Yellow/Blue/Green lines). We also present an example with very large separation from the observer $r = 10^{3} \, {\rm Mpc}$, which we label as `Plane Parallel' (Grey lines).}
  \label{fig:ens}
\end{figure}

In Figure \ref{fig:ens} we present the ensemble average ($\ref{eq:fens}$) using the probability distribution ($\ref{eq:pdf}$) for fixed $R_{\rm G} = 20 \, {\rm Mpc}$, using the radially dependent cumulants in $\hat{\Sigma}(r)$ and $\sigma_{y}^{2}(r)$. The yellow/blue/green curves correspond to the value of the ensemble average at $r=10, 25, 50 \, {\rm Mpc}$ respectively, and the solid/dashed curves are the $(r,r)$ and $(\theta,\theta)$ components. The $(\phi,\phi)$ components are always equal to the $(\theta,\theta)$ element and so are not plotted. The grey lines correspond to the plane parallel limit of the ensemble average, obtained by taking $r$ to be some arbitrarily high value $r = 10^{3} \, {\rm Mpc}$. For $r < R_{\rm G}$, the ensemble average significantly departs from the standard Gaussian expectation value (cf Yellow, blue curves). This is due to the $r$ dependent terms in the cumulants dominating for $r < R_{\rm G}$, and also the shape change in the $(r,r)$ component is due to the cross correlation contribution $\langle \tilde{\delta}\tilde{\delta}_{r} \rangle \neq 0$. For $r > R_{\rm G}$, the components approach the plane parallel limit (cf. green curves). 

The shape of the $\langle w_{i}{}^{j} \rangle$ curves depend on the correlation properties of the field. When $\langle \delta \delta_{i} \rangle = 0$, the components of $\langle w_{i}{}^{j} \rangle$ possess the well-known $\nu$ functional dependence $e^{-\nu^{2}/2}$. Any correlation between the field and its gradient will modify the shape of these statistics, even for a Gaussian random field. Practically, it would not be feasible to extract the extremely non-standard $\nu$ dependence presented in Figure \ref{fig:ens} for $r < R_{\rm G}$ from large scale structure catalogs, because we measure the volume average $\bar{w}_{i}{}^{j}$ and for $r < R_{\rm G}$ the volume is insufficient to obtain the fair sample required to estimate $\langle w_{i}{}^{j} \rangle$. Still, we can potentially probe small perturbations to the shape of the Minkowski functionals and tensors arising from the $\langle \delta \delta_{i}\rangle$ field correlation. 

\subsection{Volume Average $\bar{w}_{i}{}^{j}$}
\label{sec:volav}

Next we consider what is actually extracted from cosmological data -- the volume average of $w_{i}{}^{j}$. We assume that the continuous field $\tilde{\delta}$ has been sampled at a finite set of points, specifically we take $\tilde{\delta}$ evaluated on a Cartesian grid in a cubic volume. The volume of the cube is $L^{3} \, {\rm Mpc}^{3}$ and each pixel occupies volume $\Delta^{3} \, {\rm Mpc}^{3}$. We denote a discretized field with subscript $\{...\}$ brackets to denote pixel dependence, so $\tilde{\delta}_{\{m,n,p\}}$ is the value of the field in the $m$, $n$, $p$ pixel in $(x,y,z)$ coordinates. We define the Cartesian basis vectors as ${\bf e}_{x}$, ${\bf e}_{y}$, ${\bf e}_{z}$, and spherical basis vectors ${\bf e}_{r}$, ${\bf e}_{\theta}$, ${\bf e}_{\phi}$ in a coordinate system with respect to an observer located at the center of the cube. At each grid point we construct the gradient of the field in Cartesian coordinates using a second order accurate finite difference scheme. Then $w_{i}{}^{j}$ at each point on the grid is given by 

\begin{equation} w_{i}{}^{j}{}_{\{m,n,p\}} = {1 \over 6 }  {\tilde{\delta}_{i}{}_{\{m,n,p\}}\tilde{\delta}^{j}{}_{\{m,n,p\}} \over |\nabla \tilde{\delta}_{\{m,n,p\}}|} \delta_{D}(\tilde{\delta}_{\{m,n,p\}}-\delta_{t}) ,
\end{equation} 

\noindent where the Dirac delta function is also discretized \citep{Schmalzing:1997aj}

\begin{equation}\delta_{D}(\tilde{\delta}_{\{m,n,p\}} - \delta_{t}) =  \begin{cases}
                     \epsilon^{-1} & {\rm if} \quad  |\tilde{\delta}_{\{m,n,p\}}-\delta_{t}| < \epsilon/ 2  \\
                     0 & {\rm Otherwise} .
                 \end{cases} 
\end{equation} 

\noindent $\epsilon$ is a small parameter that we fix as $\epsilon = 10^{-2}$ in what follows. There is a discretization error that comes with this approximation \citep{Lim_2012, Chingangbam:2017uqv}, but we neglect this subtlety. The function $\delta_{D}(\tilde{\delta}_{\{m,n,p\}} - \delta_{t})$ selects a subset of pixels of roughly equal field value which are the points on $\Mspace$ at which we sample the vector field $\tilde{\delta}_{i}{}_{\{m,n,p\}}$ for each threshold $\delta_{t}$. Since the gradient of the field $\tilde{\delta}_{i}$ is approximately uncorrelated with $\tilde{\delta}$ point-wise on the manifold, this sampling generates an unbiased estimate of the underlying vector field $\tilde{\delta}_{i}$ for every $\delta_{t}$. The only caveat is that in spherical redshift space, $\tilde{\delta}_{r}$ is weakly correlated with $\tilde{\delta}$ but the correlation is negligible for $r^{4} \gg \sigma_{-1}^{2}/\sigma_{1}^{2}$. The quantity $w_{i}{}^{j}{}_{\{m,n,p\}}$ is a tensor evaluated at a particular point on the manifold (specified by the ${}_{\{m,n,p\}}$ pixel), and $\bar{w}_{i}{}^{j}$ is their volume average. 

The concept of a volume average of non-Cartesian tensors defined at different points on a manifold is ambiguous. To proceed, we should define a fiducial pixel $\{a,b,c\}$ at which we take the volume average, and a choice of path by which we transport each $w_{i}{}^{j}{}_{\{m,n,p\}}$ to $\{a,b,c\}$. We write the volume average as 

\begin{equation}\label{eq:barf} \bar{w}_{i}{}^{j} (\gamma, a,b,c) = {1 \over 6V} \sum_{m,n,p} \Delta^{3} \delta_{D}(\tilde{\delta}_{\{m,n,p\}}-\delta_{t}) {{}^{\gamma}\tilde{\delta}_{i}{}_{\{m,n,p\}}{}^{\gamma}\tilde{\delta}^{j}{}_{\{m,n,p\}} \over |\nabla \tilde{\delta}_{\{m,n,p\}}|}
\end{equation}

\noindent where the $\gamma$ superscript ${}^{\gamma}\tilde{\delta}_{i}$ denotes the transport of $\tilde{\delta}_{i}$ from $\{m,n,p\}$ to $\{a,b,c\}$ along a path $\gamma$ and 

\begin{equation} V = \sum_{m,n,p} \Delta^{3} \end{equation} 

\noindent We do not use all pixels in the cubic volume, but rather $\sum_{m,n,p}$ represents all pixels that lie in some radial range $r_{\rm min} \leq r \leq r_{\rm max}$, where $r_{\rm min} > R_{\rm G}$ and $r_{\rm max} < L/2$ are selected to ensure that we cut pixels close to the central observer and in the vicinity of the boundary of the box. 

The choice of path $\gamma$ is completely arbitrary. However, the manifold on which the field is defined is $\mathbb{S}^{2} \times \mathbb{R}_{> 0}$ which is geodesically incomplete with respect to Euclidean paths. Since we are adopting a spherical coordinate system and anticipate a preferred signal in the radial direction, it behooves us to select a transport that preserves the radial basis vector. A natural choice that achieves this is great arc transport on the two-sphere from the angular location of $\{m,n,p\}$ to $\{a,b,c\}$ followed by a radial translation to the same distance from the central observer. Great arc transport from $\{m,n,p\}$ to $\{a,b,c\}$ rotates the spherical basis vectors ${\bf e}_{r} \to {\bf e}'_{r}$, ${\bf e}_{\theta} \to {\bf e}'_{\theta}$, ${\bf e}_{\phi} \to {\bf e}'_{\phi}$ such that ${\bf e'}_{r} = {\bf e}_{r}$ but ${\bf e'}_{\theta}$ and ${\bf e'}_{\phi}$ become mixed relative to ${\bf e}_{\theta}$, ${\bf e}_{\phi}$\footnote{Parallel transport along geodesics on $\mathbb{S}^{2}$ preserves the orientation of the tangent space relative to the tangent vector of the transport. The mixing described here arises due to the fact that the angle subtending a great arc tangent vector and the basis vectors ${\bf e}_{\theta}$, ${\bf e}_{\phi}$ varies continuously along the path.}. The mixing of spherical components is unimportant, because we are assuming that the field is isotropic on $\mathbb{S}^{2}$. We explicitly present the rotation of the spherical basis vectors -- relative to great arc tangents -- in Appendix \ref{sec:appen3}.

To perform this transport for all pixels that satisfy $\delta_{D}(\tilde{\delta}_{\{m,n,p\}} - \delta_{t}) \neq 0$, we define ${\bf \hat{T}_1}$ and ${\bf \hat{T}_2}$ as unit vectors pointing to pixels $\{m,n,p\}$ and $\{a,b,c\}$ from the `observer' at $r=0$, and rotate the vector $\tilde{\delta}_{i \, \{m,n,p\}}$ by angle $\cos\theta = {\bf \hat{T}_1} \cdot {\bf \hat{T}_2}$ about the axis defined by ${\bf \hat{T}_1} \times {\bf \hat{T}_2}$. Such a rotation can be used to describe great arc transport. The second stage of transport, along ${\bf e}_{r}$, is trivial and undertaken implicitly. Finally the transported, Cartesian gradient $\tilde{\delta}'_{i \, \{m,n,p\}}$, now defined at $\{a,b,c\}$, is converted into the spherical basis via a coordinate transformation. Note that we used a Cartesian basis to define $\tilde{\delta}_{i}$ and performed a coordinate transformation as a final step, but one could instead define $\tilde{\delta}_{i}$ in a spherical basis then rotate from $\{m,n,p\}$ to $\{a,b,c\}$. The final result will not depend on the ordering of these operations. 

If we used Euclidean paths to transport $\tilde{\delta}_{i}$ to a common point on the manifold (ignoring the geodesic incompleteness), then we would obtain a completely different result. In this case, all three spherical basis vectors ${\bf e}_{r}$, ${\bf e}_{\theta}$, ${\bf e}_{\phi}$ would mix, and $\bar{w}_{i}{}^{j}$ would depend entirely on the volume over which the field is defined. 

The fact that the choice of transport affects the volume average is troubling, because the ensemble average is defined at a point on the manifold and requires no notion of transport. However, we expect that our choice is appropriate for the very specific physical scenario considered in this work. With our path selection, the radial basis vector is preserved and although the angular derivatives become mixed, we are working with a field that is isotropic on $\Sspace^{2}$. 
Other choices of path could be used instead -- for example transport along lines of latitude and longitude. This choice is not angle preserving -- lines of latitude are not generally geodesics. Ultimately there is no unique path definition, but for a field that is isotropic on $\Sspace^{2}$ these details are not important. Also the point on $\Sspace^{2}$ at which we take the average will not impact the volume average for an idealised field that is isotropic on $\Sspace^{2}$. Anisotropic fields on $\Sspace^{2}$ will be considered elsewhere, as many of these subtleties are likely to become problematic in the absence of this symmetry.

We would like to equate the volume and ensemble averages of $w_{i}{}^{j}$, defined in equations ($\ref{eq:barf}$) and ($\ref{eq:fens}$) respectively\footnote{Since we measure $\bar{w}_{i}{}^{j} \equiv W^{0,2}_{1}$ from a cosmological density field, we should not compare the measurement to the ensemble average of the Minkowski tensor $\langle W^{0,2}_{1} \rangle$ but rather $\langle w_{i}{}^{j} \rangle$. When the field is statistically homogeneous, we can write $\langle w_{i}{}^{j} \rangle = \langle W^{0,2}_{1} \rangle$ and there is no distinction to be made.}. As justification, we can appeal to the weak law of large numbers -- for a sequence of identically distributed variables $X_{n}$ we define an average

\begin{equation} \label{eq:sm} \bar{X} = {1 \over N} \sum_{n=1}^{N} X_{n} .\end{equation}

\noindent Then if the covariance between variables $(X_{n},X_{n+m})$ asymptotes to zero as $m \to \infty$, the sample mean $\bar{X}$ in ($\ref{eq:sm}$) approaches the underlying expectation value $E(X)$ in the limit $N \to \infty$. In our example, the summed variables are the combination on the right hand side of equation ($\ref{eq:barf}$). As we take the volume $V \to \infty$, we expect that the pixels will provide a fair sample and the correlation functions of $\tilde{\delta}$ and its gradient satisfy $\zeta(r) \to 0$ as $r \to \infty$. This suggests that the ensemble and sample averages will converge, but in realistic scenarios we deal with finite volumes, and furthermore the fields $\tilde{\delta}$, $\tilde{\delta}_{i}$ are non-Gaussian in the low redshift universe. It is not clear that the sample and ensemble averages converge when higher point correlations are present, and if so how quickly they do as the volume increases \citep{10.1046/j.1365-8711.2003.06130.x}. With our choice of transport, we do not expect the volume average to be sensitive to the point on the sphere at which we take the average, and we will take density fields located at $r \gg R_{\rm G}$ so the radial dependence of the cumulants should be irrelevant. We therefore expect that for this particular physical scenario, our choice of coordinate system and transport will allow us to use the approximation $\bar{w}_{i}{}^{j} \simeq \langle w_{i}{}^{j} \rangle$. We confirm this numerically in Section \ref{sec:num}. However, before moving on to the numerics we present a counter example in Section \ref{sec:cart} for which the notion of ergodicity (even approximate) fails completely.

%%%%%%%%%%%%%%%%%%%%%%%%%%%%%%%%%%%%%%%%%%%%%%%%%%%%%%%%%%%%%%%%%
\section{Spherical Redshift Space, Cartesian Coordinate System}
\label{sec:cart}

In this section, we calculate the cumulants of the spherically redshift space distorted field in Cartesian coordinates $(x,y,z)$, following the methodology of \citet{Castorina:2017inr}. The calculation is extremely tedious, and we simply state some important steps and the results in the main body of the text. The density field in redshift space $\tilde{\delta}({\bf r})$ can be written in terms of its real-space counterpart $\delta({\bf r})$ via

\begin{equation}
    \tilde{\delta}({\bf r}) = \left(1+\frac{f}{3}\right)\int \frac{d^3k}{2\pi^3}\mathcal{L}_0(\hat{k}\cdot\hat{r}) e^{i{\bf k}\cdot{\bf r}}\delta({\bf k}) +  \frac{2f}{r}\int \frac{d^3k}{2\pi^3}\frac{\mathcal{L}_1(\hat{k}\cdot\hat{r})}{ik} e^{i{\bf k}\cdot{\bf r}}\delta({\bf k}) + \frac{2f}{3}\int \frac{d^3k}{2\pi^3}\mathcal{L}_2(\hat{k}\cdot\hat{r}) e^{i{\bf k}\cdot{\bf r}}\delta({\bf k}),
    \label{eqn:del_rsd_exp}
\end{equation}

\noindent where $\mathcal{L}_{\ell}$s are the Legendre polynomials.

We express $\textbf{k}$ as $k_x {\bf e}_x + k_y {\bf e}_y + k_z {\bf e}_z$ in Cartesian coordinates. The differentiation of the first term on the right hand side in equation (\ref{eqn:del_rsd_exp}) with respect to $x$ gives,

\begin{eqnarray}
    \nonumber \left(1+\frac{f}{3}\right)\int \frac{d^3k}{2\pi^3} \frac{\partial}{\partial x}\mathcal{L}_0(\hat{k}\cdot\hat{r}) e^{i{\bf k}\cdot{\bf r}}\delta({\bf k}) &=& \left(1+\frac{f}{3}\right)\int \frac{d^3k}{2\pi^3}\delta({\bf k}) \frac{\partial}{\partial x} e^{i{\bf k}\cdot{\bf r}} \\
     &=& \left(1+\frac{f}{3}\right)\int \frac{d^3k}{2\pi^3} (ik\sin\theta\cos\phi) e^{i{\bf k}\cdot{\bf r}} \delta({\bf k}).
\end{eqnarray}

We treat the other terms in the right hand side of equation (\ref{eqn:del_rsd_exp}) in a similar way and substitute the results into $\left< \frac{\partial}{\partial x}\tilde{\delta}({\bf r}) \frac{\partial}{\partial x'}\tilde{\delta}({\bf r'}) \right>\bigg\rvert_{{\bf r'} \to {\bf r}}$. We then use the relation (\ref{app:b2}) and ${\cal L}({\bf \hat{x}}.{\bf \hat{x}})=1$, along with the result 

\begin{equation}
    \left<\delta({\bf k_1})\delta({\bf k_2})\right> = \left( 2\pi \right)^3 \delta_D^{(3)}({\bf k_1}+{\bf k_2})P(k) ,
\end{equation}

\noindent to get

\begin{eqnarray}
 \nonumber   \left< \frac{\partial}{\partial x}\tilde{\delta}({\bf r}) \frac{\partial}{\partial x'}\tilde{\delta}({\bf r'}) \right>\bigg\rvert_{{\bf r'} \to {\bf r}} &=& \frac{4f^2}{3r^4}\int \frac{\textrm{d}k P(k, R_{\rm G})}{2\pi^2}+\left[ \frac{4f}{3r^4}(2x^2-r^2) +\frac{4f^2}{5r^4}(r^2+x^2) \right] \int \frac{k^2\textrm{d}kP(k, R_{\rm G})}{2\pi^2} \\
     & & +\left[ \frac{1}{3} + \frac{2f}{15r^2}(r^2+2x^2) +\frac{f^2}{35r^2}(r^2+4x^2) \right]\int \frac{k^4\textrm{d}kP(k, R_{\rm G})}{2\pi^2} .
\end{eqnarray}

\noindent Similarly,

\begin{eqnarray}
 \nonumber   \left< \frac{\partial}{\partial y}\tilde{\delta}({\bf r}) \frac{\partial}{\partial y'}\tilde{\delta}({\bf r'}) \right>\bigg\rvert_{{\bf r'} \to {\bf r}} &=& \frac{4f^2}{3r^4}\int \frac{\textrm{d}kP(k, R_{\rm G})}{2\pi^2}+\left[ \frac{4f}{3r^4}(2y^2-r^2) +\frac{4f^2}{5r^4}(r^2+y^2) \right] \int \frac{k^2\textrm{d}kP(k, R_{\rm G})}{2\pi^2} \\
     & & +\left[ \frac{1}{3} + \frac{2f}{15r^2}(r^2+2y^2) +\frac{f^2}{35r^2}(r^2+4y^2) \right]\int \frac{k^4\textrm{d}kP(k, R_{\rm G})}{2\pi^2}  ,
\end{eqnarray}

\begin{eqnarray}
 \nonumber   \left< \frac{\partial}{\partial z}\tilde{\delta}({\bf r}) \frac{\partial}{\partial z'}\tilde{\delta}({\bf r'}) \right>\bigg\rvert_{{\bf r'} \to {\bf r}} &=& \frac{4f^2}{3r^4}\int \frac{\textrm{d}kP(k, R_{\rm G})}{2\pi^2}+\left[ \frac{4f}{3r^4}(2z^2-r^2) +\frac{4f^2}{5r^4}(r^2+z^2) \right] \int \frac{k^2\textrm{d}kP(k, R_{\rm G})}{2\pi^2} \\
     & & +\left[ \frac{1}{3} + \frac{2f}{15r^2}(r^2+2z^2) +\frac{f^2}{35r^2}(r^2+4z^2) \right]\int \frac{k^4\textrm{d}kP(k, R_{\rm G})}{2\pi^2}  ,
\end{eqnarray}

\begin{eqnarray}
    \left< \frac{\partial}{\partial x}\tilde{\delta}({\bf r}) \frac{\partial}{\partial y'}\tilde{\delta}({\bf r'}) \right>\bigg\rvert_{{\bf r'} \to {\bf r}} &=& \left( \frac{8f}{3} + \frac{4f^2}{5} \right)\frac{xy}{r^4}\int \frac{k^2\textrm{d}kP(k, R_{\rm G})}{2\pi^2} + \left( \frac{4f}{15}+\frac{4f^2}{35} \right)\frac{xy}{r^2}\int \frac{k^4\textrm{d}kP(k, R_{\rm G})}{2\pi^2} ,
\end{eqnarray}

\begin{eqnarray}
    \left< \frac{\partial}{\partial x}\tilde{\delta}({\bf r}) \frac{\partial}{\partial z'}\tilde{\delta}({\bf r'}) \right>\bigg\rvert_{{\bf r'} \to {\bf r}} &=& \left( \frac{8f}{3} + \frac{4f^2}{5} \right)\frac{xz}{r^4}\int \frac{k^2\textrm{d}kP(k, R_{\rm G})}{2\pi^2} + \left( \frac{4f}{15}+\frac{4f^2}{35} \right)\frac{xz}{r^2}\int \frac{k^4\textrm{d}kP(k, R_{\rm G})}{2\pi^2} .
\end{eqnarray}

\begin{figure}[htb]
\vspace{.5cm}
  \centering
  $\langle \delta_{x}\delta^{x} \rangle/\sigma_{1}^{2}$ \hskip 4.3cm $\langle \delta_y\delta^y \rangle/\sigma_{1}^{2}$    \hskip 4.3cm     $\langle \delta_y\delta^y \rangle/\sigma_{1}^{2}$   \\
 \includegraphics[width=0.32\textwidth]{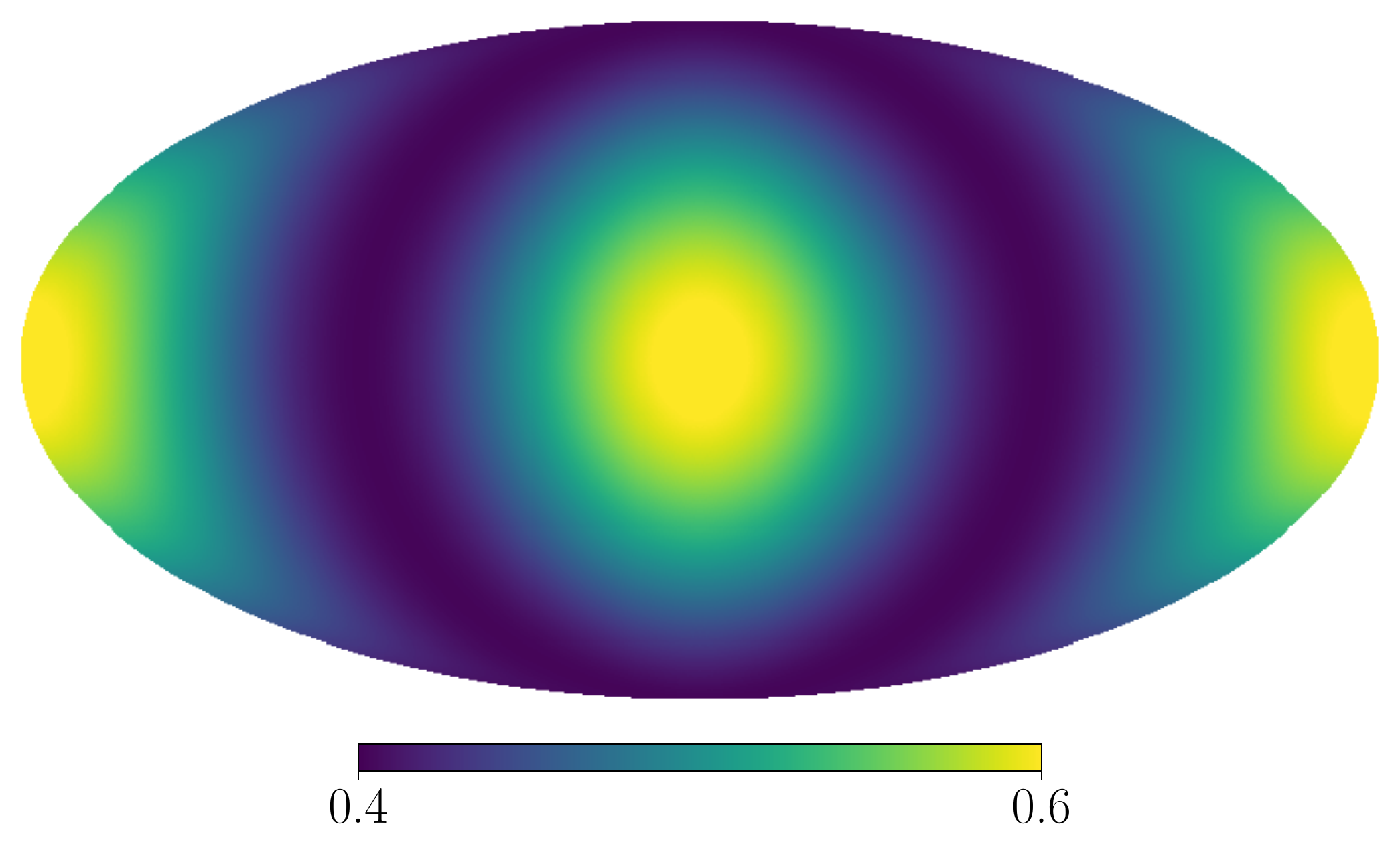} 
 \includegraphics[width=0.32\textwidth]{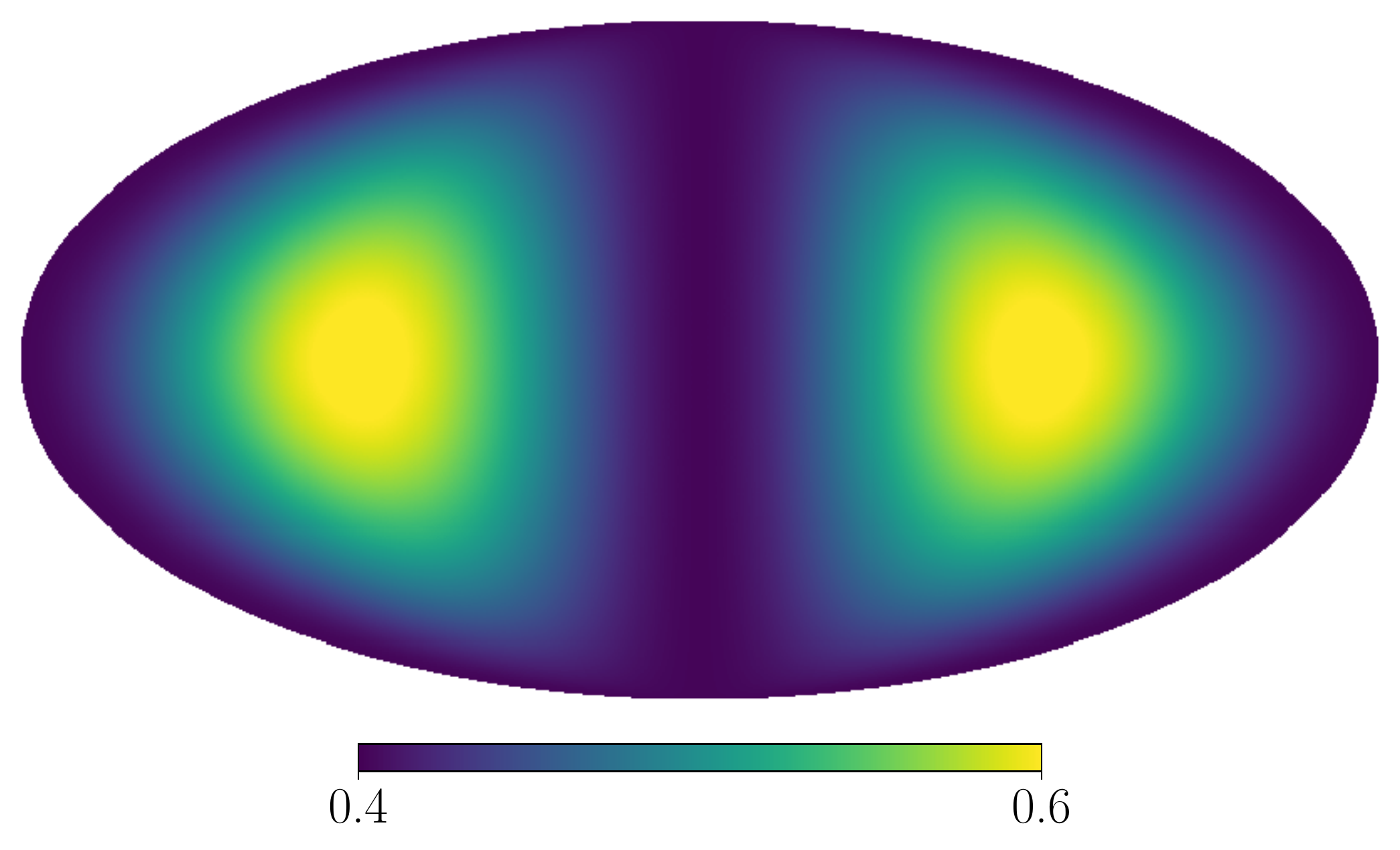}
 \includegraphics[width=0.32\textwidth]{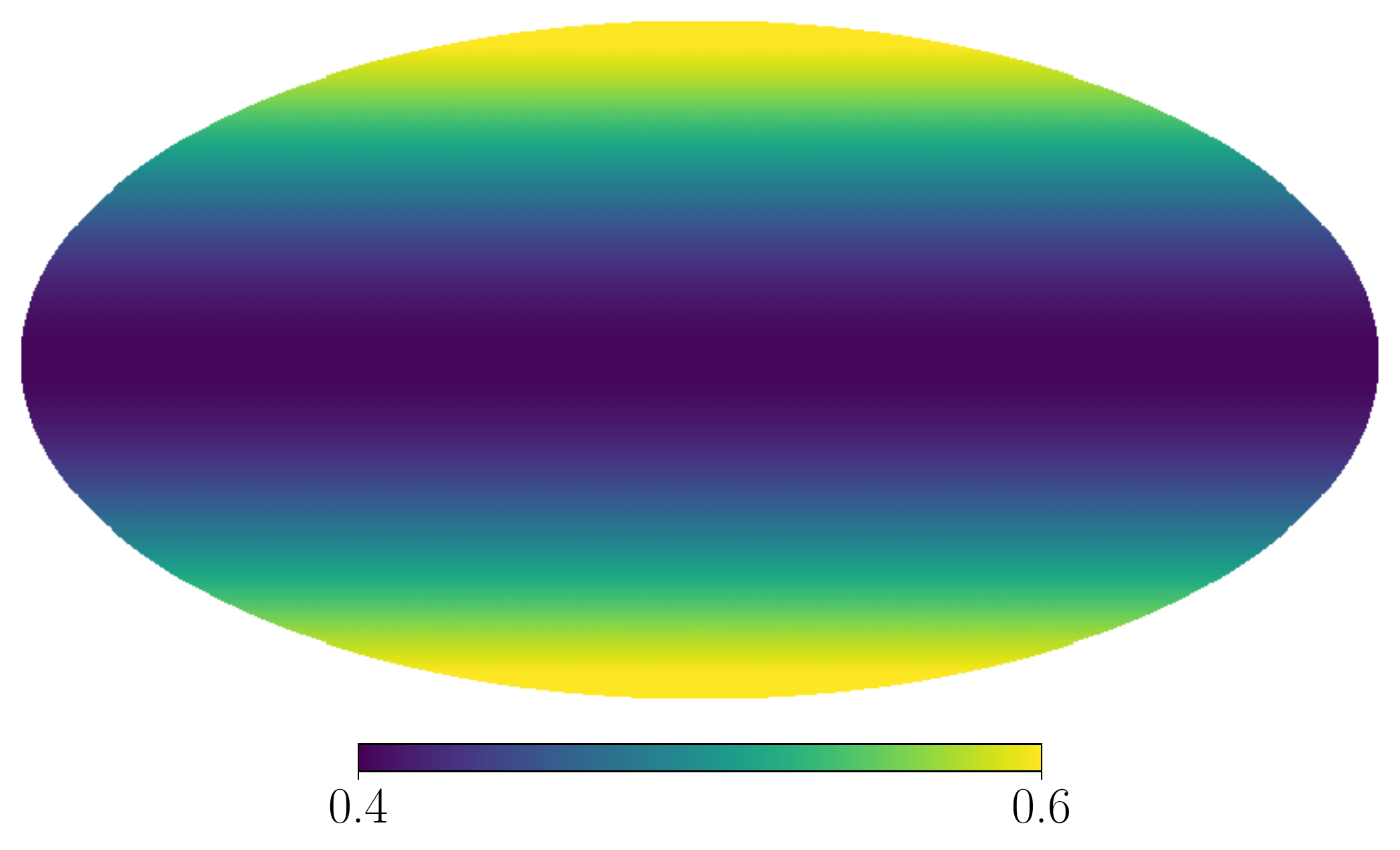} \\
    $\langle \delta_x\delta^y \rangle/\sigma_{1}^{2}$ \hskip 4.3cm $\langle \delta_x\delta^z \rangle/\sigma_{1}^{2}$  \hskip 4.3cm $\langle \delta_y\delta^z \rangle/\sigma_{1}^{2}$ \\
 \includegraphics[width=0.32\textwidth]{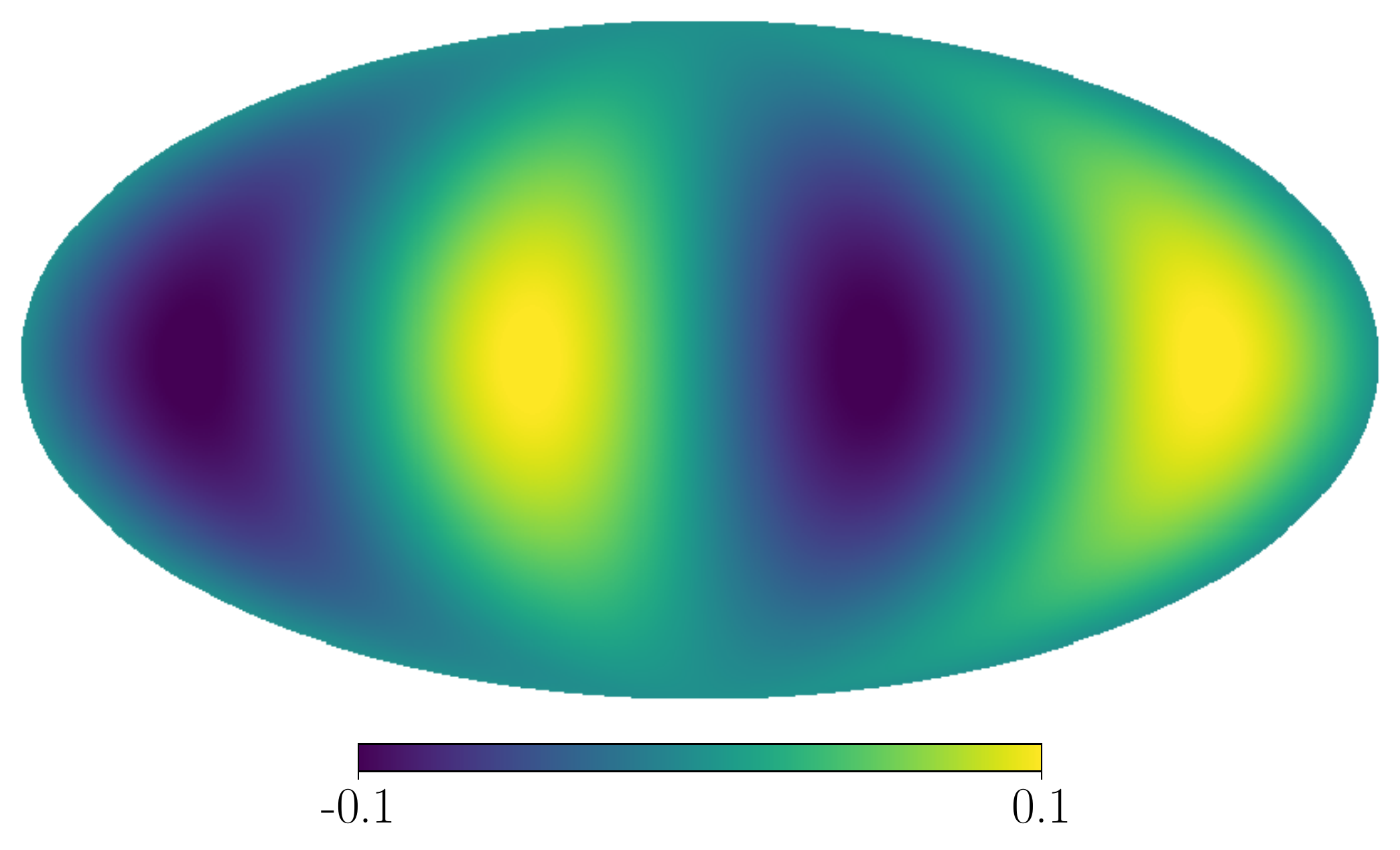} 
 \includegraphics[width=0.32\textwidth]{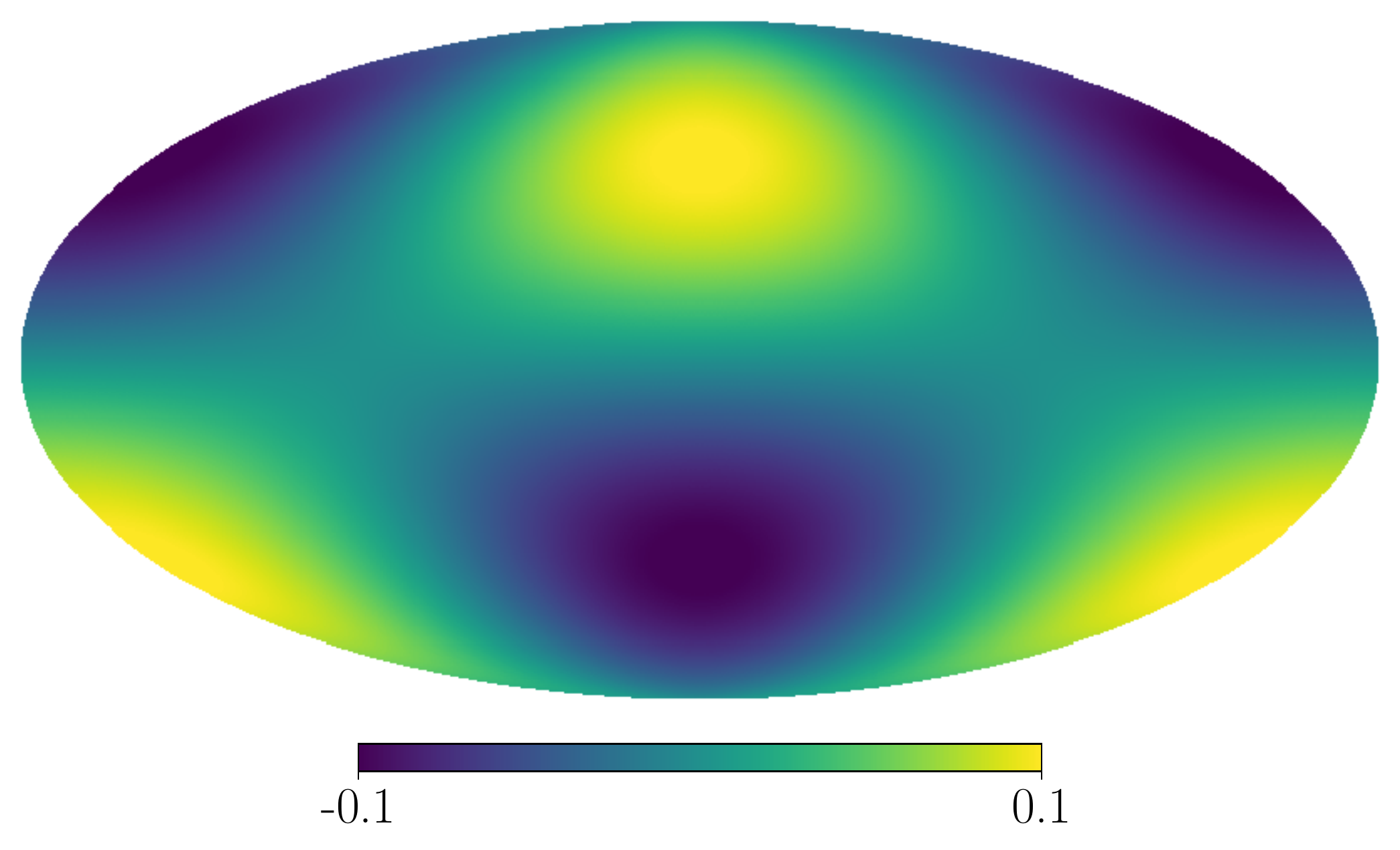} 
 \includegraphics[width=0.32\textwidth]{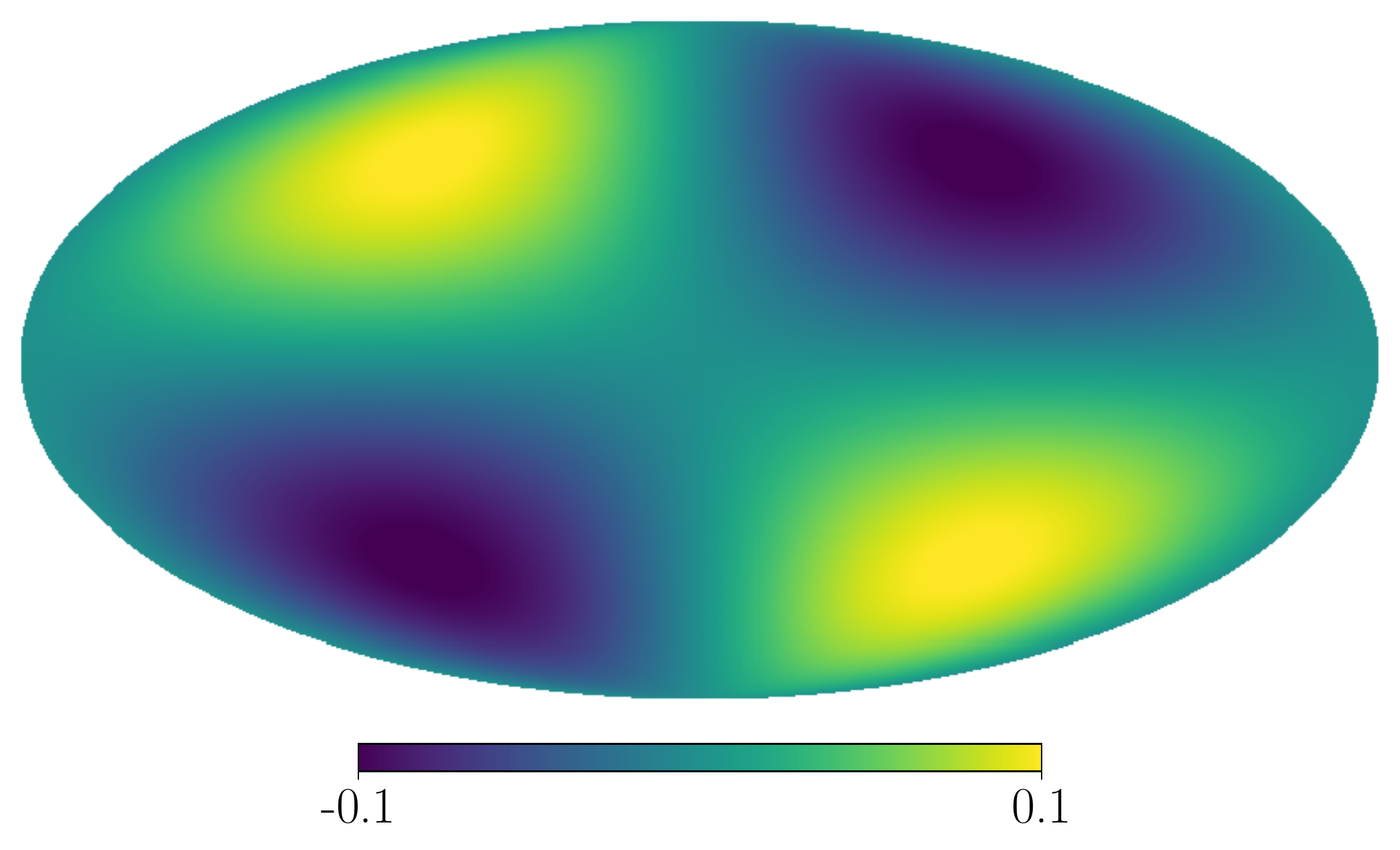} 
  \caption{Cartesian cumulants $\langle \delta_{i}\delta^{j} \rangle/\sigma_{1}^{2}$ projected onto the two-sphere for fixed radial distance $r=200 \, {\rm Mpc}$ from the central observer. The top row panels display the ($x$,$x$), ($y$,$y$) and ($z$,$z$) components from left to right, while the bottom row panels display the off-diagonal components ($x$,$y$), ($x$,$z$) and ($y$,$z$) components, from left to right.}
  \label{fig:4a}
\end{figure}

\noindent In this coordinate system, the cumulant tensor $\langle \tilde{\delta}_{i}\tilde{\delta}^{j} \rangle$ is not diagonal. We visualize the Cartesian cumulants in Figure \ref{fig:4a}. We smooth the power spectrum with a Gaussian kernel with scale $R_{\rm G} = 20 \, {\rm Mpc}$, select a fixed radial distance $r=200 \, {\rm Mpc}$ from the central observer and present Mollweide projections of the dimensionless quantity $\langle \tilde{\delta}_{i}\tilde{\delta}^{j} \rangle/\sigma_{1}^{2}$ on the sphere. The  
top row panels show the diagonal $(x,x)$, $(y,y)$ and $(z,z)$ components (left to right), while the bottom row panels show $(x,y)$, $(x,z)$ and $(y,z)$ components (left to right). The diagonal elements present a series of dipoles on the sphere, and the off-diagonal elements are quadrupolar. All elements are generically non-zero and vary significantly with spatial position. This is in contrast with the cumulants in a spherical coordinate system, which are isotropic on the sphere and vary only weakly with $r$. 

The spatial dependence of $\langle \tilde{\delta}_{i}\tilde{\delta}^{j} \rangle$ means that the vector  $\tilde{\delta}_{i}$ located at different points on $\Rspace^{3}$ are not equally likely to be observed in a realisation. Given $\langle \tilde{\delta}^{2} \rangle$, $\langle \tilde{\delta}\tilde{\delta}_{j}\rangle$ and $\langle \tilde{\delta}_{i}\tilde{\delta}^{j}\rangle$, we can construct the ensemble average $\langle w_{i}{}^{j}\rangle$ in this coordinate system

\begin{equation}\label{eq:hhm} \langle w_{i}{}^{j} \rangle = {1 \over 6}  \int \Phi(X,x,y,z) {\tilde{\delta}_{i}\tilde{\delta}^{j} \over |\nabla \tilde{\delta}|} \delta_{D}(\tilde{\delta}-\delta_{t}) dX , 
\end{equation}

\noindent where $\Sigma(x,y,z)$ is given by

\begin{equation} \Sigma(x,y,z) =  \left( \begin{tabular}{cc}
       $\langle \tilde{\delta}^{2}  \rangle$  & $\langle \tilde{\delta} \tilde{\delta}_{i} \rangle$  \\
    $\langle \tilde{\delta} \tilde{\delta}^{j} \rangle$  &  $\langle \tilde{\delta}_{i} \tilde{\delta}^{j} \rangle$ 
    \end{tabular} \right) .
\end{equation}

It is clear that the volume average $\bar{w}_{i}{}^{j}$ will not generically be representative of the ensemble average $\langle w_{i}{}^{j} \rangle$ in this coordinate system, due to the coordinate dependence of $\langle w_{i}{}^{j} \rangle$. For example, taking the all-sky spatial average of $w_{i}{}^{j}$ extracted from a field with the particular cumulant pattern in Figure \ref{fig:4a} will yield an isotropic result $\bar{w}_{i}{}^{j} \propto \delta_{i}{}^{j}$ -- we confirm this in the following section\footnote{Simply adding Cartesian components of $w_{i}{}^{j}$ at different points of the manifold to obtain $\bar{w}_{i}{}^{j}$ implicitly assumes Euclidean path transport, but neglects the geodesic incompleteness of the manifold. Regardless, we do not use the Cartesian coordinate system other than to provide an example for which ergodicity fails.}. The volume average in this particular case would incorrectly identify the field as isotropic, because the spatial dependence of the signal in this coordinate system would be washed out by the averaging. The volume and ensemble averages cannot be equated even approximately in this example. This conclusion is not in contradiction with the plane parallel limit, because here we are considering an all-sky average. If we instead took a small patch on the sky and aligned the Cartesian coordinate system with one axis pointing to the patch, then the plane parallel limit could be approximately realised. 

The underlying point is that for tensors and inhomogeneous fields, the volume average can reasonably approximate the ensemble average or completely misrepresent it, depending on the properties of the field and choice of coordinate system, volume and transport path.

\section{Numerical Extraction of Minkowski Tensors in Spherical Redshift Space} 
\label{sec:num}

We now confirm numerically some of the results of the previous sections, and furthermore study the conditions under which we can faithfully extract the Kaiser signal from a redshift space distorted, non-Gaussian matter field in the low redshift Universe. The matter density field is assumed to be Gaussian in the early Universe, but the non-linear nature of gravitational collapse couples Fourier modes. This is a scale dependent statement, and by smoothing the late time density field over sufficiently large scales, the standard model of cosmology posits that the density field is perturbatively non-Gaussian. We attempt to extract the Kaiser redshift space distortion signal from the large-scale-averaged density field.

In this work we do not pursue the computational challenges that come with real data, such as galaxy bias, shot noise, complex survey geometries and Malmquist bias -- these issues will be considered elsewhere. When galaxies are scattered radially, the relative volume difference along the line of sight can introduce a spurious radial gradient in the mean density, which must be carefully subtracted. Neglecting these subtleties, we focus specifically on two questions -- can we use the volume average constructed in Section \ref{sec:volav} as an unbiased estimate of the ensemble average derived in Section \ref{sec:ensav}, and over what scales must we smooth the non-Gaussian dark matter field to reproduce the Gaussian limit of these statistics? We also compare the MTs extracted from plane parallel and spherical redshift space distorted fields and confirm that they are indistinguishable for fields occupying cosmological volumes. 

To perform these tests, we use two data sets -- initially Gaussian random fields and then dark matter particle distributions that have been gravitationally evolved to $z=0$. 

%%%%%%%%%%%%%%%%%%%%%%%%%%%%%%%%%%%%%%%%%%%%%%%%%%%%%%%%%%%%%%
\subsection{Gaussian Random Fields} 
\label{sec:grf}

For Gaussian random fields, we start by generating an isotropic and homogeneous field $\delta$ in a periodic cube of side length $L = 1490 \, {\rm Mpc}$ ( $= 1000 \, h^{-1} {\rm Mpc}$), using an input linear $\Lambda$CDM matter power spectrum $P(k,R_{\rm G})$ at $z=0$ with cosmological parameters given in Table \ref{tab:1}. We smooth the field with Gaussian kernel $W(k R_{\rm G}) \propto e^{-k^{2}R_{\rm G}^{2}/2}$. The field is sampled on a Cartesian grid with $N_{\rm p} = 512$ pixels per side, with resolution $\Delta = 2.9 \, {\rm Mpc}$. We then create plane parallel and spherical redshift space distorted fields. For the plane parallel case, we apply the standard operator (cf. equation ($\ref{eq:pp1}$)) to $\delta$, in Fourier space, using $f = \Omega_{\rm m}^{6/11}$ and aligning the RSD correction with the ${\bf e}_{z}$ axis of the box. 

To construct a spherically redshift space distorted field, we generate a second isotropic field $\Omega \equiv  \nabla^{-2}\delta$ on the grid, and construct the gradient $\nabla_{i} \Omega$ in the Cartesian coordinate system. Then we infer the radial derivative $\partial_{r}\Omega$ using a standard transformation (we provide our angle conventions explicitly in equation (\ref{eq:other1})). We repeat this procedure on $\partial_{r}\Omega$ to obtain the second derivative $\partial_{rr}\Omega$, and finally define the spherically redshift space distorted density field as 

\begin{equation} \tilde{\delta}_{\{m,n,p\}} =  \delta_{\{m,n,p\}} + f \left( \partial_{rr}\Omega_{\{m,n,p\}} + {2 \over r}\partial_{r}\Omega_{\{m,n,p\}} \right)   .
\end{equation} 

This field is masked such that $\tilde{\delta}_{\{m,n,p\}}$ is assigned zero value and not used in our analysis if the pixel $\{m,n,p\}$ is such that it's radial distance from the `observer' at the center of the box, lies outside the range $100<r \le 630 $ in Mpc units. 
We use `all-sky' data, taking the complete $4\pi r^2$ area on $\mathbb{S}^{2}$ relative to the central observer. 

\begin{table}[tb]
\begin{center}
 \begin{tabular}{||c  c ||}
 \hline
 Parameter \, & Fiducial Value \\ [0.5ex] 
 \hline\hline
 $\Omega_{\rm m}$ & $0.318$   \\ 
 $h$ & $0.671$   \\
 $w_{\rm de}$ & $-1$ \\
 $n_{\rm s}$ & $0.962$   \\
 $\sigma_{8}$ & $0.834$ \\ 
  \hline 
\end{tabular}
\caption{\label{tab:1}Fiducial cosmological parameters used in this work, selected to match the fiducial cosmology of the Quijote simulations \cite{Villaescusa-Navarro:2019bje}. }
\end{center} 
\end{table}

For each dataset, we calculate the mean $\bar{\delta}$ and variance $\tilde{\sigma}_{0}^{2}$ of the unmasked pixels, and define the zero mean, unit variance field $(\tilde{\delta}_{\{m,n,p\}} - \bar{\delta})/\tilde{\sigma}_{0}$. The volume average $\bar{w}_{i}{}^{j}$ is calculated for each of the three datasets -- isotropic, plane parallel and spherical redshift space distorted. For the isotropic and plane parallel fields, we use the entire box with periodic boundary conditions, and $\bar{w}_{i}{}^{j}$ is defined in the Cartesian coordinate system of the box. From the Cartesian lattice we use a simple second order accurate finite difference scheme to construct the gradients $\delta_{i}$ and $\tilde{\delta}_{i}$, and since we use Euclidean paths to collect tensors in $\Rspace^{3}$ we can simply take a sum of $w_{i}{}^{j}_{\{m,n,p\}}$ pixels without any explicit transport transformation. Hence the volume averages are 

\begin{eqnarray} {}^{\rm re}\bar{w}_{i}{}^{j} &=& {1 \over 6V} \sum_{m,n,p} \Delta^{3} \delta_{D}(\delta_{\{m,n,p\}}-\nu) {\delta_{i}{}_{\{m,n,p\}} \delta^{j}{}_{\{m,n,p\}} \over |\nabla \delta_{\{m,n,p\}}|}  ,\\
{}^{\rm pp}\bar{w}_{i}{}^{j} &=& {1 \over 6V} \sum_{m,n,p} \Delta^{3} \delta_{D}(\tilde{\delta}_{\{m,n,p\}}-\nu) {\tilde{\delta}_{i}{}_{\{m,n,p\}}\tilde{\delta}^{j}{}_{\{m,n,p\}} \over |\nabla \tilde{\delta}_{\{m,n,p\}}|} ,
\end{eqnarray}

\noindent where the superscripts denote `real space' (re) and `plane parallel' (pp) and $\nu$ is the root mean square normalised threshold $\nu = \delta_{t}/\sigma_{0}$ or $\nu = \delta_{t}/\tilde{\sigma}_{0}$,  respectively.  

For the spherically distorted field, we follow the procedure outlined in Section \ref{sec:volav} - we randomly select an unmasked pixel $\{a,b,c\}$ as the fiducial point at which we take the spatial average, with unit vector pointing to the pixel denoted ${\bf \hat{T}_2}$. Then for each pixel selected by the discretized delta function $\delta_{D}(d_{\{m,n,p\}}-\nu) \neq 0$, where $d$ is either $\tilde\delta$ or $\delta$, we define the unit vector pointing to this pixel as ${\bf \hat{T}_1}$, and use ${\bf \hat{T}_1}$ and ${\bf \hat{T}_2}$ to construct a unit quaternion $q$ which is used to rotate the Cartesian gradient vector $\tilde{\delta}'_{i} = q \tilde{\delta}_{i} q^{*}$, reflecting its change of orientation when transported from ${\bf \hat{T}_1}$ to ${\bf \hat{T}_2}$. The components of the quaternion are given in Appendix \ref{sec:appen2}. At $\{a,b,c\}$, the rotated Cartesian gradient is transformed to the spherical coordinate basis ${\bf e}_{r}$, ${\bf e}_{\theta}$, ${\bf e}_{\phi}$. Note that there is no unique rotation/great arc transport for pixels at antipodal points on the sphere to $\{a,b,c\}$; for these we select a random rotation axis in the plane perpendicular to ${\bf \hat{T}_2}$ (we have confirmed that different choices do not affect our numerical results). The volume average for the spherically redshift-space distorted case (superscript ${}^{\rm sp}$) is 

\begin{equation}  {}^{\rm sp}\bar{w}_{i}{}^{j} = {1 \over 6V} \sum_{m,n,p} \Delta^{3} \delta_{D}(\tilde{\delta}_{\{m,n,p\}}-\nu) {{}^{\gamma}\tilde{\delta}_{i}{}_{\{m,n,p\}} {}^{\gamma}\tilde{\delta}^{j}{}_{\{m,n,p\}} \over |\nabla \tilde{\delta}_{\{m,n,p\}}|} ,
\end{equation}

\noindent where ${}^{\gamma}$ denotes great arc transport, and the tensor is defined in a spherical basis. We measure $\bar{w}_{i}{}^{j}$ over $N_{\nu} = 51$ threshold values $\nu$ equi-spaced over the range $-3.8 \leq \nu \leq 3.8$,  for $N_{\rm real} = 50$ realisations of a Gaussian random field. We repeat the measurements for fields smoothed with scale $R_{\rm G}$ over the range $15 \, {\rm Mpc} \leq R_{\rm G} \leq 45 \, {\rm Mpc}$. 

Before presenting the numerical results, we discuss a way to check the Gaussian nature of a random field. For a general weakly non-Gaussian field we can expand the components of the Minkowski tensors as a series of Hermite polynomials\footnote{$H_{n}(\nu)$ are the probabilist's Hermite polynomials, the first few of which are given by $H_{0}(\nu)=1$, $H_{1}(\nu) = \nu$, $H_{2}(\nu) = \nu^{2} - 1$.}, as follows,
\begin{equation}
\bar{w}_{i}{}^{j} = e^{-\nu^2/2}\left(A|_i^j H_0(\nu) + a_1|_i^j H_1(\nu) + a_2|_i^j H_2(\nu)+\ldots\right).
\end{equation}
This expansion is equivalent to Matsubara's perturbative expansion for the scalar Minkowski functionals \citep{2003ApJ...584....1M}, albeit the expansion coefficients are assigned to each Hermite polynomial and not to powers of the variance. The coefficients contain information of the generalized skewness, kurtosis and higher moments of the field.  The coefficients $A|_i^j, a_1|_i^j, a_2|_i^j$ can be computed using the orthogonality properties of the Hermite polynomials, as,
\begin{eqnarray} A|_{i}{}^{j} &=& {1 \over \sqrt{2\pi}} \int_{-\nu_{{\rm max}}}^{\nu_{\rm max}} \bar{w}_{i}{}^{j}(\nu) H_{0}(\nu) d\nu ,
\label{eqn:Aij}\\ 
a_{1}|_{i}{}^{j} &=& {1 \over \sqrt{2\pi}} \int_{-\nu_{{\rm max}}}^{\nu_{{\rm max}}} \bar{w}_{i}{}^{j}(\nu) H_{1}(\nu) d\nu ,
\label{eqn:a1ij}\\
a_{2}|_{i}{}^{j} &=& {1 \over 2\sqrt{2\pi}} \int_{-\nu_{{\rm max}}}^{\nu_{{\rm max}}} \bar{w}_{i}{}^{j}(\nu) H_{2}(\nu) d\nu,
\label{eqn:a2ij}
\end{eqnarray} 
where $\nu_{\rm max} \to \infty$. For our analysis we take $\nu_{{\rm max}} = 3.8$, after checking that our results are not sensitive to reasonable variations of this value\footnote{The Hermite polynomials are exactly orthogonal only in the limit $\nu_{{\rm max}}\to \infty$. However, since $\bar{w}_{i}{}^{j}$ is exponentially damped at large thresholds, it suffices to choose finite $\nu_{{\rm max}}$. Taking $\nu_{{\rm max}}$ to be too large in a finite volume dataset can generate biased values of the Hermite polynomial coefficients (cf Appendix A-4 of \citet{Appleby:2021lfq}).}.  For Gaussian random fields, the coefficients of the higher order terms in the expansion $a_1,a_2$ etc should be consistent with zero in real and redshift space, so we refer to the coefficient of $H_{0}(\nu)$ as the `amplitude' of the MT components. 

Using the above way of representing weakly non-Gaussian random fields, in the Gaussian and plane parallel limits we have \citep{Appleby_2019}

\begin{equation} {}^{\rm pp}\bar{w}_{i}{}^{j} = {}^{\rm pp}A_{G}|_{i}{}^{j} H_{0}(\nu) e^{-\nu^{2}/2} 
\end{equation} 

\noindent with

\begin{eqnarray}\label{eq:amp1} & & {}^{\rm pp}A_{G}|_{x}{}^{x} = {A_{0} \over 4}\left[ {(2\lambda^{2}-1)\cosh^{-1}\left(2\lambda^{2}-1\right) \over (\lambda^{2}-1)^{3/2}} - {2\lambda \over \lambda^{2}-1}  \right]  , \\
\label{eq:amp2} & & {}^{\rm pp}A_{G}|_{y}{}^{y} = {}^{\rm pp}A_{G}|_{x}{}^{x}  , \\
\label{eq:amp3} & & {}^{\rm pp}A_{G}|_{z}{}^{z} =  A_{0}\left({\lambda^{2} \over \lambda^{2}-1}\right) \left( \lambda - {\cosh^{-1} \lambda \over \sqrt{\lambda^{2}-1}}\right) ,
\end{eqnarray}

\noindent where $A_{0}$ and $\lambda$ are defined in equations ($\ref{eq:a0}$) and ($\ref{eq:lam}$). In real space we have \citep{Appleby:2018tzk}

\begin{equation}\label{eq:Agau} {}^{\rm re}A_{\rm G}|_{x}{}^{x} = {}^{\rm re}A_{\rm G}|_{y}{}^{y} = {}^{\rm re}A_{\rm G}|_{z}{}^{z} = {\sigma_{1} \over 9\sqrt{3} \pi \sigma_{0}} .
\end{equation}

\noindent For the Minkowski functional $W_{1}$, the coefficient $a_{1}$ of $H_{1}(\nu)$ is one of two terms induced as a leading order non-Gaussian correction and $a_{2}$ is one of several higher order contributions. Hence we use these terms as proxies to study the non-Gaussian corrections of the MTs that are induced by gravitational collapse. As mentioned above, for the Gaussian random fields considered in this subsection, $a_{1}$ and $a_{2}$ should be consistent with zero in real, plane parallel and spherical redshift space. The redshift space distortion operator does not change the Gaussian nature of the field. We check that the numerically computed values of $a_{1}$ and $a_{2}$ are consistent with zero in our calculations, when measuring the Minkowski tensor of Gaussian random fields.

\begin{figure}[htb]
  \centering 
 \includegraphics[width=0.45\textwidth]{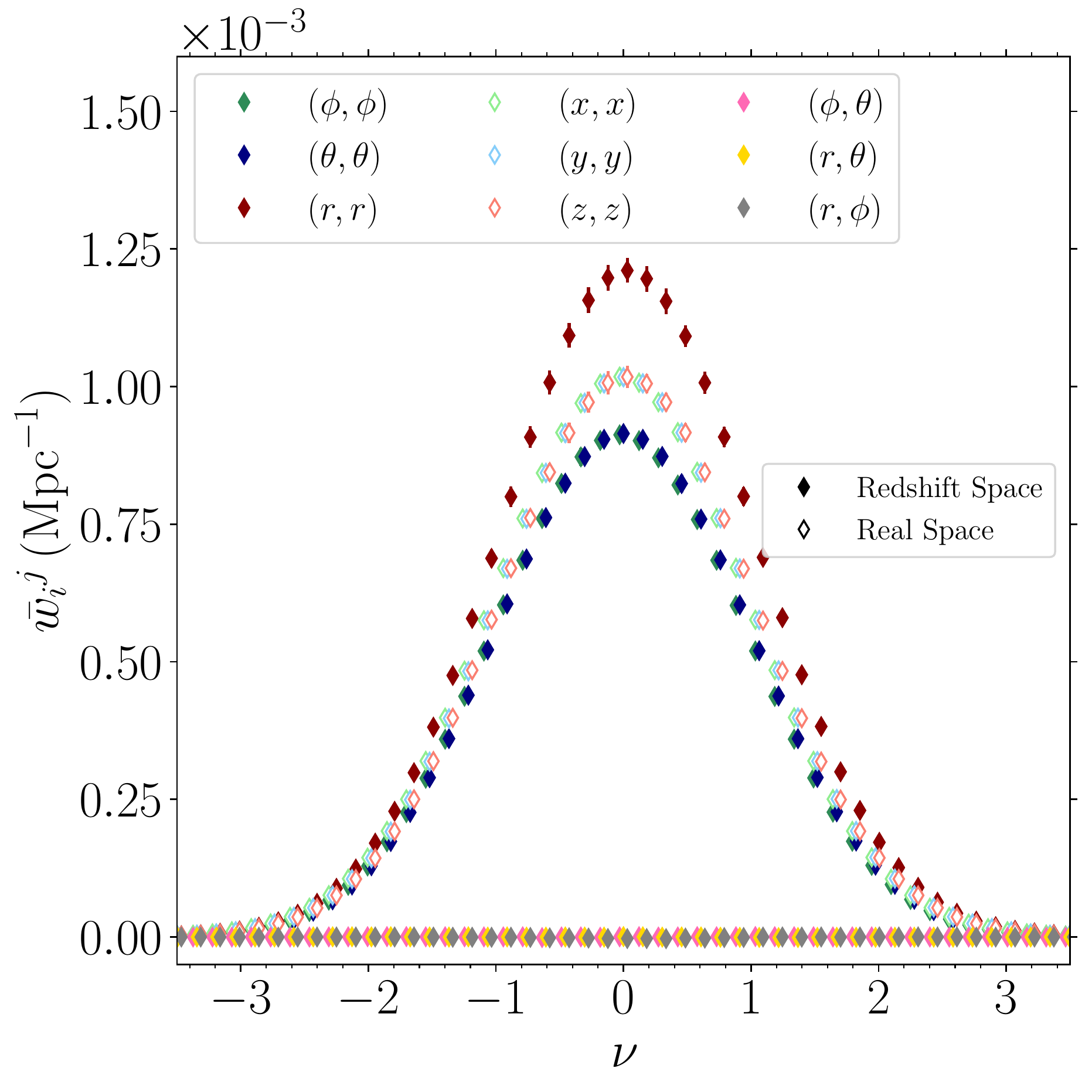} 
 \includegraphics[width=0.45\textwidth]{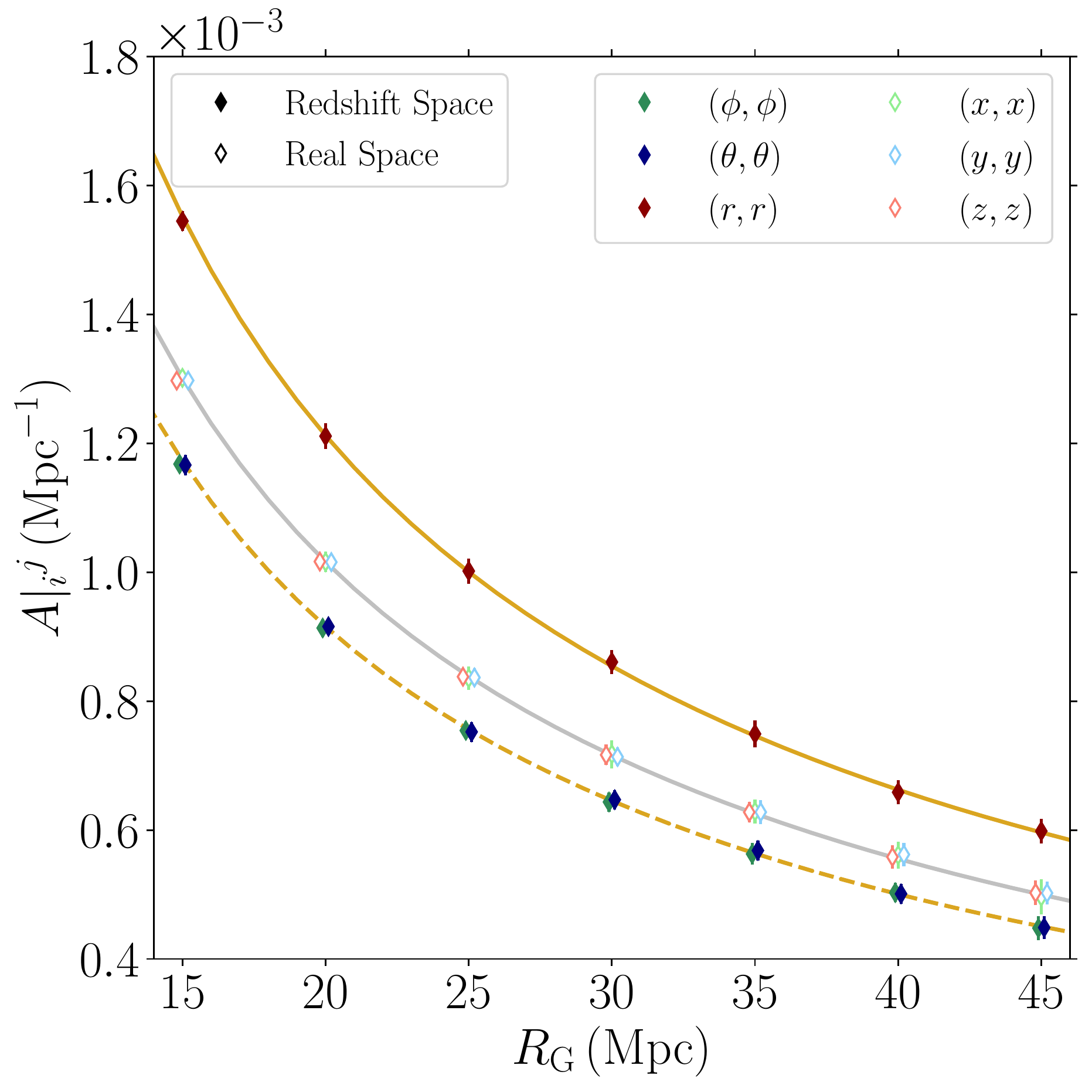} \\
 \includegraphics[width=0.45\textwidth]{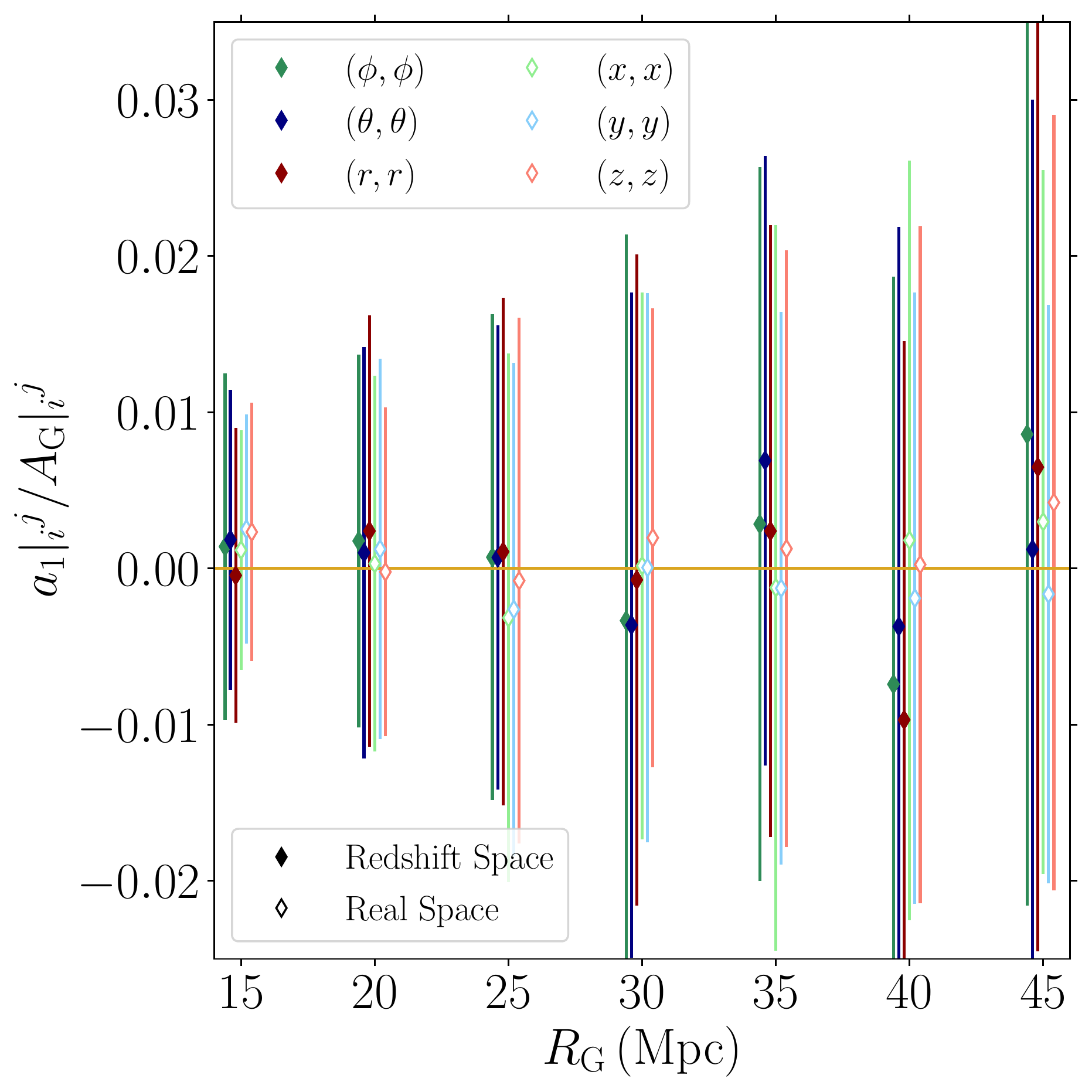} 
 \includegraphics[width=0.45\textwidth]{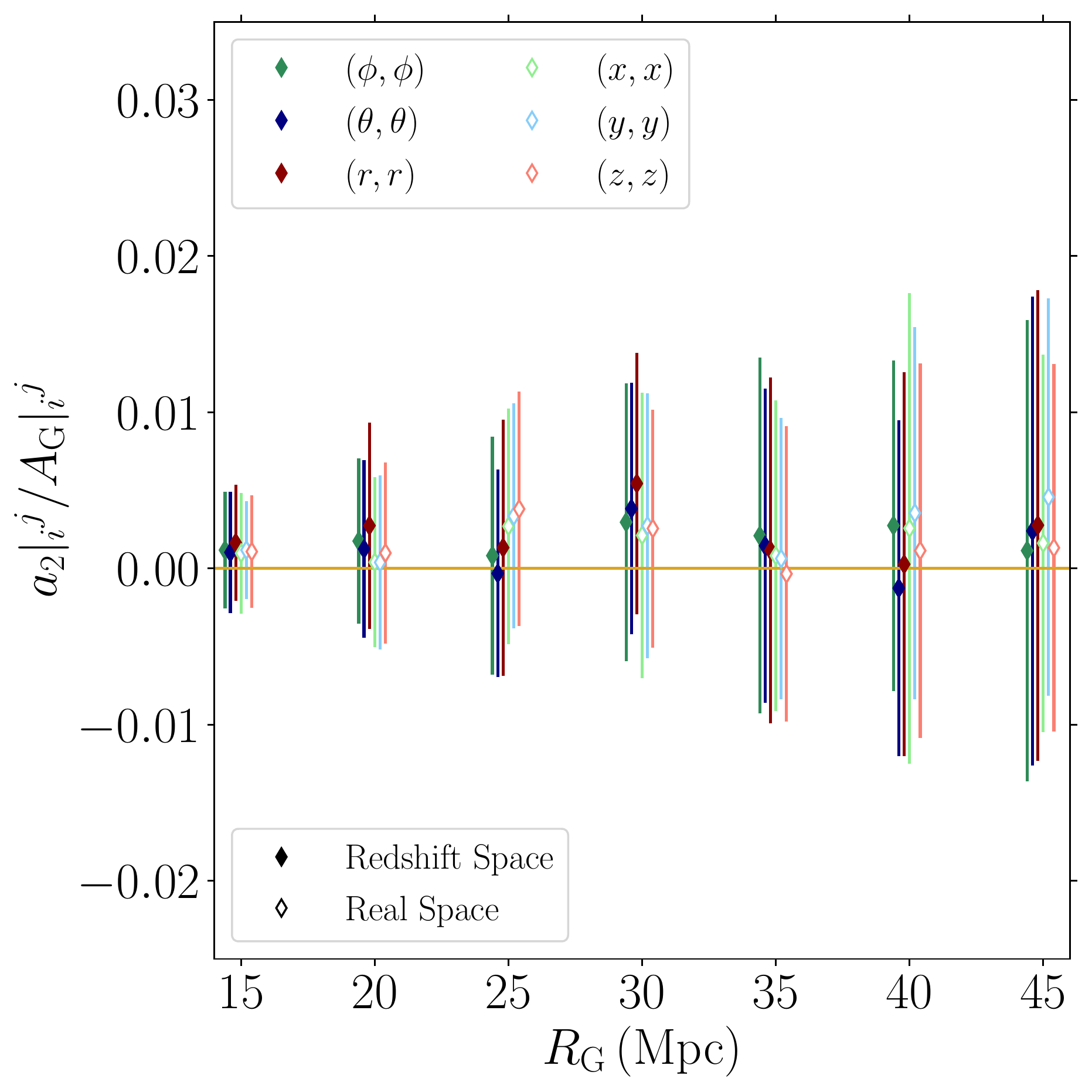} 
  \caption{{\em Top left panel}: Components of the the Minkowski tensor extracted from a GRF in real space (open diamonds), and spherical redshift space (filled diamonds). The off-diagonal elements for the spherical redshift space field are also presented and consistent with zero. The field has been smoothed with scale $R_{G} = 20 \, {\rm Mpc}$.{\em Top right panel}: The amplitude of $\bar{w}_{i}{}^{j}$ as a function of smoothing scale of the field. The solid/dashed gold lines are the plane parallel expectation values in redshift space and the solid silver line is the isotropic expectation value. {\em Bottom panels}: The coefficients of the $H_{1}(\nu)$ (left), $H_{2}(\nu)$ (right) Hermite polynomials. They are consistent with zero for a Gaussian field, in both real and redshift space.}
  \label{fig:grf}
\end{figure}

In Figure \ref{fig:grf} (top left panel) we present the diagonal and off-diagonal components of ${}^{\rm re}\bar{w}_{i}{}^{j}$ and ${}^{\rm sp}\bar{w}_{i}{}^{j}$ extracted from the fields smoothed with $R_{\rm G} = 20 \, {\rm Mpc}$. The points/error bars correspond to the mean and root-mean-square (rms) of the realisations, hence we are presenting the ensemble average of the volume average. The filled/open diamonds are measurements in spherical redshift and real space respectively. The diagonal components in real space are equal, modulo a noise component (cf light green/blue/red open diamonds). The real-space volume average satisfies ${}^{\rm re}\bar{w}_{i}{}^{j} \propto \delta_{i}{}^{j}$ in every coordinate system. In redshift space, the radial component of $\bar{w}_{i}{}^{j}$ is significantly larger than the angular components -- this is the Kaiser signal. The off-diagonal components of  ${}^{\rm re}\bar{w}_{i}{}^{j}$, ${}^{\rm pp}\bar{w}_{i}{}^{j}$ and ${}^{\rm sp}\bar{w}_{i}{}^{j}$ are all consistent with zero.

In Figure \ref{fig:grf} we present the values of $A|_{i}{}^{j}$ (top right panel), $a_{1}|_{i}{}^{j}$ (bottom left panel) and $a_{2}|_{i}{}^{j}$ (bottom right panel) for $\bar{w}_{i}{}^{j}$ extracted from the real and spherical redshift space distorted fields as a function of smoothing scale $R_{\rm G}$. In the top right panel, the solid/dashed gold lines are the corresponding plane parallel Kaiser limits given in equations ($\ref{eq:amp1}-\ref{eq:amp3}$) and the solid silver line is the isotropic expectation value in equation ($\ref{eq:Agau}$).

\noindent The volume averages ${}^{\rm re}\bar{w}_{i}{}^{j}$ and ${}^{\rm sp}\bar{w}_{i}{}^{j}$ extracted from the spherical RSD and real space data sets match the ensemble averages derived in \cite{Appleby:2018tzk, Appleby_2019}. Similarly the coefficients $a_{1}$, $a_{2}$ are consistent with zero at all scales probed (cf bottom panels). This is expected - we generated Gaussian random fields and the application of the linear redshift space distortion operator preserves Gaussianity. This provides a check on the ergodicity condition $\langle w_{i}{}^{j} \rangle \simeq \bar{w}_{i}{}^{j}$, and indicates that our definition of the volume average can be used to reproduce the ensemble average. 

\begin{figure}[htb]
  \centering 
 \includegraphics[width=0.32\textwidth]{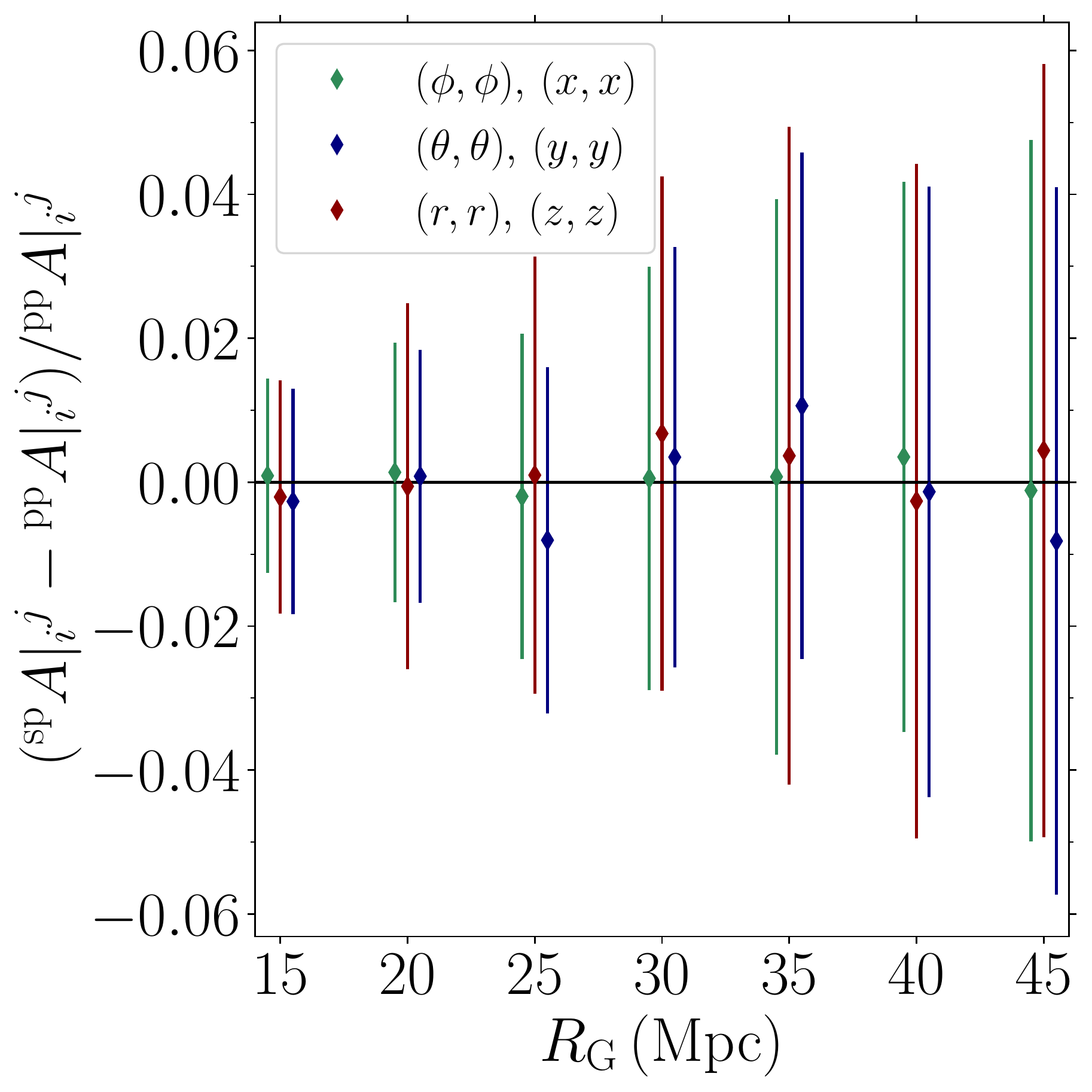} 
\includegraphics[width=0.32\textwidth]{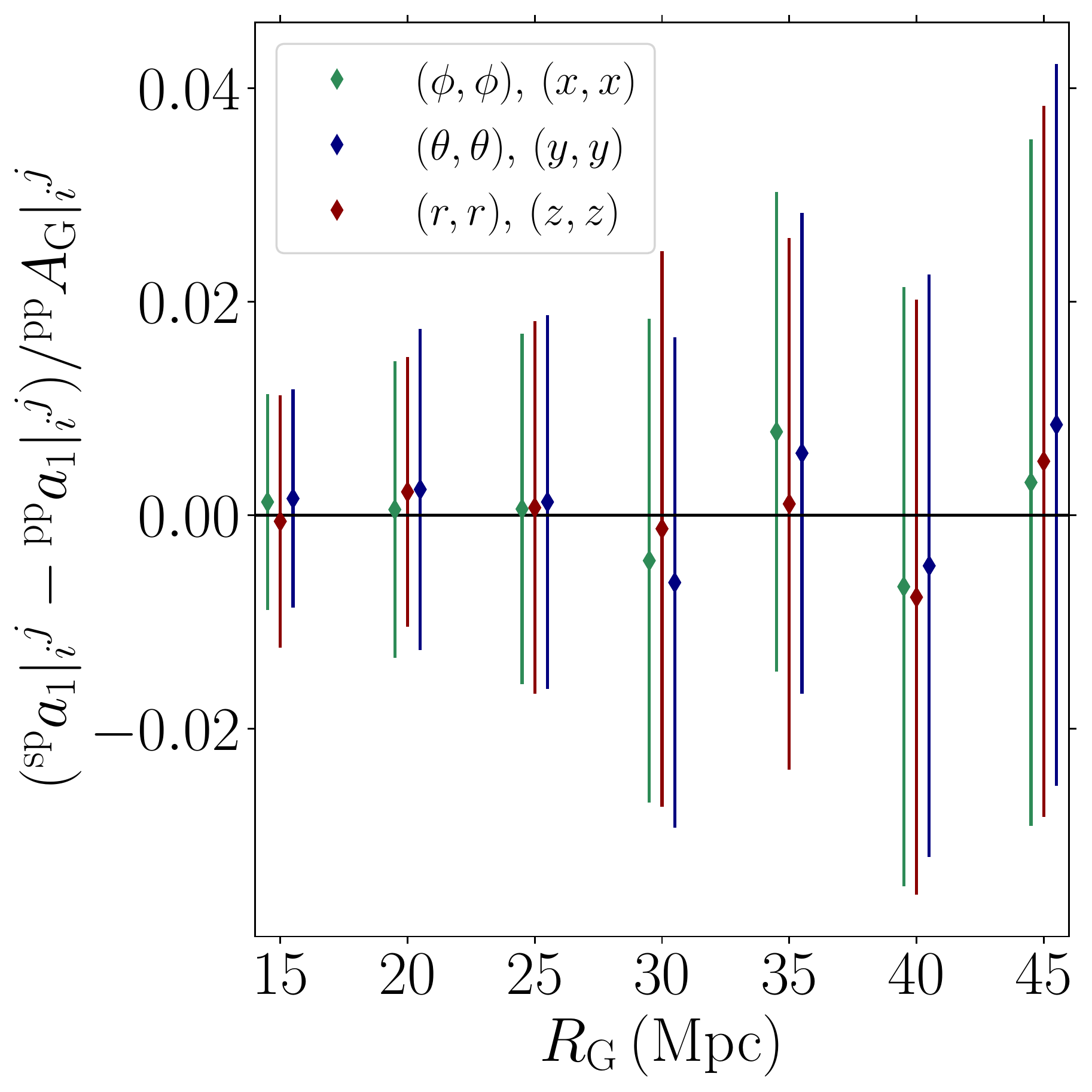} 
 \includegraphics[width=0.32\textwidth]{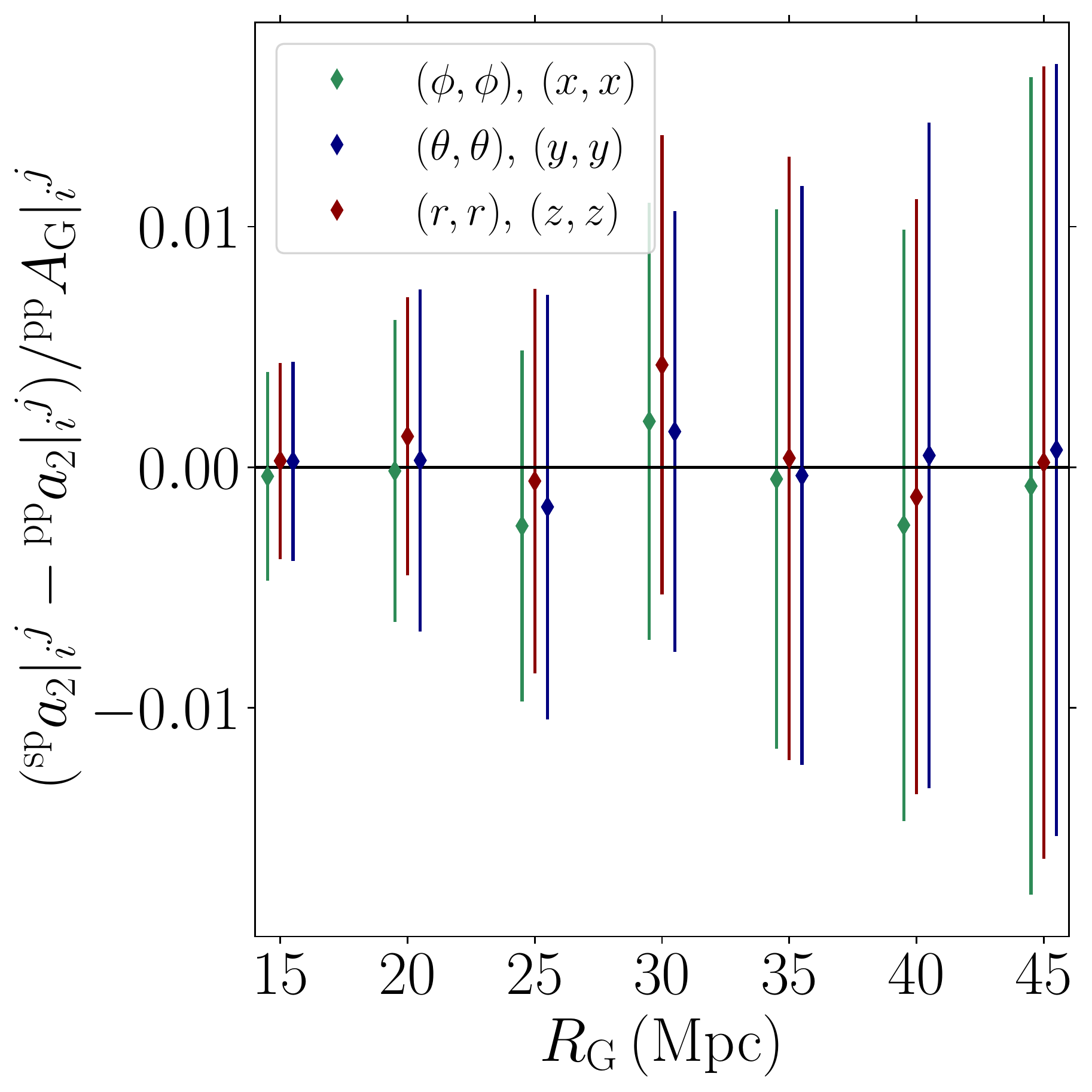} 
  \caption{Fractional difference between $A|_{i}{}^{j}$ [Left], $a_{1}|_{i}{}^{j}$ [Middle] and $a_{2}|_{i}{}^{j}$ [Right] extracted from the spherical and plane parallel redshift space distorted Gaussian random fields. }
  \label{fig:sp_pp_grf}
\end{figure}

Finally, in Figure \ref{fig:sp_pp_grf} we present the fractional differences $({}^{\rm sp}A|_{i}{}^{j} - {}^{\rm pp}A|_{i}{}^{j})/{}^{\rm pp}A|_{i}{}^{j}$ (left panel), $({}^{\rm sp}a_{1}|_{i}{}^{j} - {}^{\rm pp}a_{1}|_{i}{}^{j})/{}^{\rm pp}A_{\rm G}|_{i}{}^{j}$ (central panel) and $({}^{\rm sp}a_{2}|_{i}{}^{j} - {}^{\rm pp}a_{2}|_{i}{}^{j})/{}^{\rm pp}A_{\rm G}|_{i}{}^{j}$ (right panel) as a function of smoothing scale $R_{\rm G}$. These quantities are all consistent with zero at all scales probed, confirming that the plane parallel and spherical redshift space distorted fields are statistically indistinguishable for data that is at cosmological distance $ > 100 \, {\rm Mpc}$ from the observer.

%%%%%%%%%%%%%%%%%%%%%%%%%%%%%%%%%%%%%%%%%%%%%%%%%%%%%%%%%%%%%%%%%%%%%%%%
\subsection{Non-Gaussian Dark Matter Fields} 
\label{sec:ngrf}

To study the gravitationally evolved non-Gaussian dark matter density field, we use $N_{\rm real} = 50$, $z=0$ snapshot boxes from the Quijote simulations \citep{Villaescusa-Navarro:2019bje}). These are a suite of cosmological scale dark matter simulations in which $\sim 44,000$ realisations of $512^{3}$ particles are gravitationally evolved in boxes of size $L = 1490 \, {\rm Mpc}$ ( $= 1000 \, h^{-1} {\rm Mpc}$) from $z=127$ to $z=0$. We take $N_{\rm real} = 50$, $z=0$ snapshot boxes and generate real space density fields by binning the dark matter particles into a regular $512^3$ Cartesian grid of resolution $\Delta = 2.9 \, {\rm Mpc}$ using a cloud-in-cell scheme. Defining the number density field $\delta_{\{i,j,k\}} = (n_{\{i,j,k\}} - \bar{n})/\bar{n}$, where $n_{\{i,j,j\}}$ is the number of particles in the $\{i,j,k\}$ pixel and $\bar{n}$ is the mean number of particles per pixel. We smooth this field with a Gaussian kernel $W(kR_{\rm G}) \propto e^{-k^{2}R_{G}^{2}/2}$ in Fourier space. 

To generate the plane parallel and spherical redshift space distorted fields, we take the real-space positions of the particles ${\bf x}$ and perturb them according to

\begin{eqnarray} & & {\bf s} = {\bf x} +  {\bf e}_{z} ({\bf v}.{\bf e}_{z}) {(1+z) \over H(z)} , \\
& & {\bf s} = {\bf x} + {\bf e}_{r} ({\bf v}.{\bf e}_{r} ) {(1+z) \over H(z)} , \end{eqnarray} 

\noindent respectively, where ${\bf v}$ is the velocity of the particle, ${\bf e}_{z}$ is the unit vector aligned with the $z$ direction of the Cartesian grid and ${\bf e}_{r}$ is the radial basis vector to the particle from an observer at the center of the box. We take redshift zero snapshot boxes, so we fix $z=0$ and $H(z) = H_{0}$.

\begin{figure}[htb]
  \centering 
 \fbox{\large Spherical polar coordinates}\\
  \vspace{.2cm}
 \includegraphics[width=0.99\textwidth]{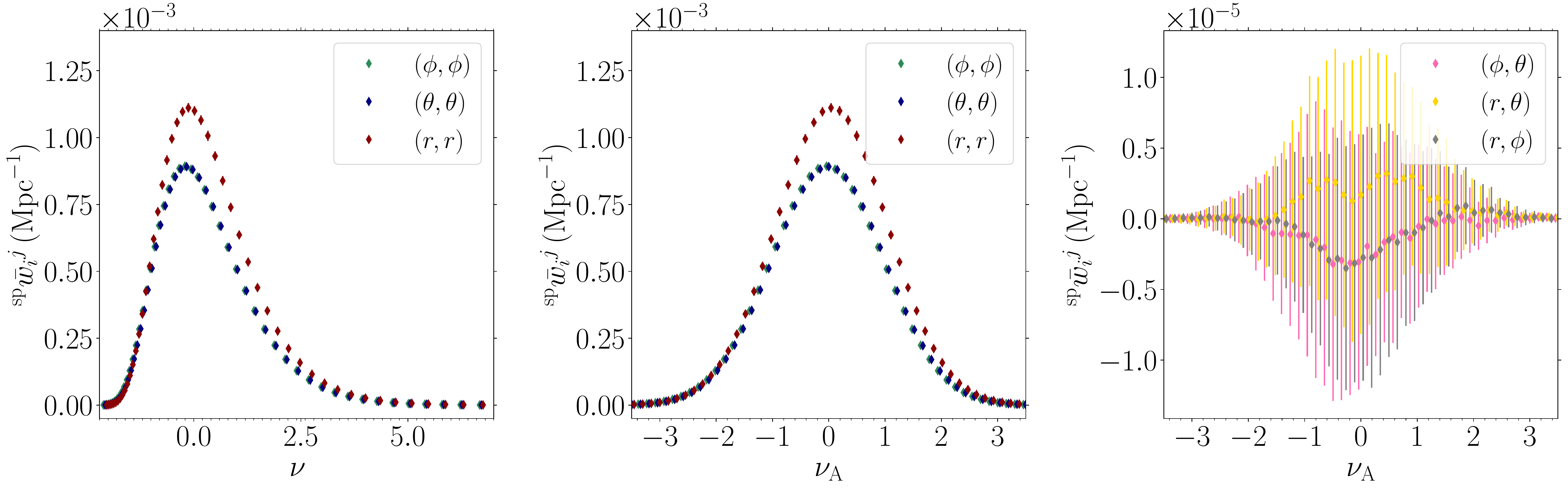}\\
 \fbox{\large Cartesian coordinates}\\
  \vspace{.2cm}
 \includegraphics[width=0.99\textwidth]{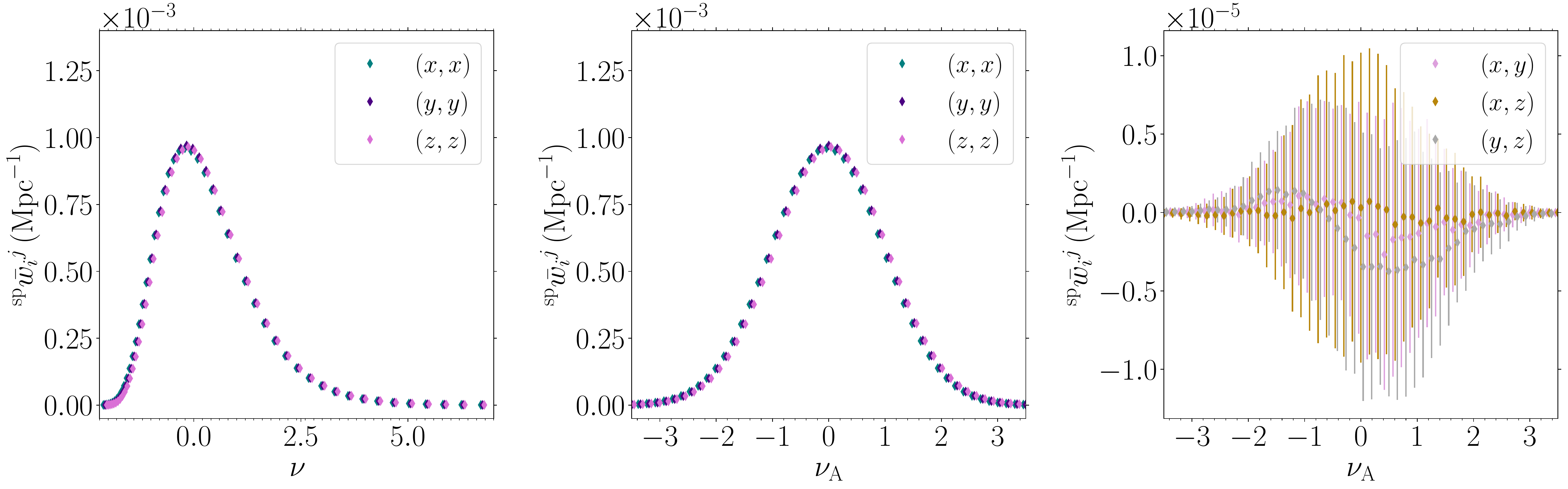} 
  \caption{The Minkowski Tensors measured from $N_{\rm real}$=50 redshift space distorted Quijote dark matter simulations, as a function of the thresholds $\nu$, and the re-scaled thresholds $\nu_A$. The top panels display ${}^{\rm sp}\bar{w}_{i}{}^{j}$ from the spherically redshift space distorted fields in a spherical coordinate system, while the bottom panels show $\bar{w}_{i}{}^{j}$ extracted from the same spherical RSD fields but in a Cartesian coordinate system.}
  \label{fig:2a}
\end{figure}

For the redshift space distorted fields we bin the particles into pixels with the cloud-in-cell scheme according to their redshift space position, using the same $\Delta = 2.90 \, {\rm Mpc}$ Cartesian grid. We apply periodic boundary conditions for the plane parallel corrected box along ${\bf e}_{z}$, which renders the field homogeneous but anisotropic.  The spherical redshift space distortion operator is incompatible with periodicity. So we exclude all pixels that lie at distances $r \leq 50 \, {\rm Mpc}$ and $r \geq 700 \, {\rm Mpc}$ from the central observer in our calculations of  $\bar{w}_{i}{}^{j}$.
The outer boundary of the shell is at least $50 \, {\rm Mpc}$ from the edges of the box, so all particles affected by the periodic boundary are excluded. Finally, we smooth these pixel boxes with Gaussian kernel $W(kR_{\rm G}) \propto e^{-k^{2}R_{G}^{2}/2}$ in Fourier space, and then further exclude all pixels that lie a distance $r \leq 100 \, {\rm Mpc}$ and $r \geq 670 \, {\rm Mpc}$ from the central observer. This last step eliminates pixels that are affected by sampling near the boundary. The end result is a set of three fields from which we extract ${}^{\rm re}\bar{w}_{i}{}^{j}$, ${}^{\rm pp}\bar{w}_{i}{}^{j}$ and ${}^{\rm sp}\bar{w}_{i}{}^{j}$. 

We calculate the mean $\bar{\delta}$ and variance $\tilde{\sigma}_{0}^{2}$ of the unmasked pixels for each field, and define the zero mean, unit variance quantity $(\tilde{\delta}_{\{m,n,p\}} - \bar{\delta})/\tilde{\sigma}_{0}$. The quantities ${}^{\rm re}\bar{w}_{i}{}^{j}$, ${}^{\rm pp}\bar{w}_{i}{}^{j}$ and ${}^{\rm sp}\bar{w}_{i}{}^{j}$ are measured over $n_{\nu} = 301$ values of threshold density $\nu$ from the minimum and maximum values of the field in each simulation. We then re-scale the iso-density threshold $\nu$ to $\nu_{\rm A}$, where $\nu_{\rm A}$ is the threshold for which the excursion set has the same volume fraction as a corresponding Gaussian field:
\begin{equation}\label{eq:afrac} 
f_{\rm A} = {1 \over \sqrt{2\pi}} \int^{\infty}_{\nu_{A}} e^{-t^{2}/2} \, dt , 
\end{equation}
where $f_{\rm A}$ is the fractional volume of the field above $\nu_{\rm A}$. Expressing the MTs as a function of $\nu_{\rm A}$ as opposed to $\nu$ partially Gaussianizes the statistics \citep{1987ApJ...319....1G,1987ApJ...321....2W,1988ApJ...328...50M}. To perform this re-scaling, we use spline interpolation on the $W^{0,2}_{1}$ versus $\nu$ calculated data and construct $W^{0,2}_{1}$ versus $\nu_A$ at  $n_{\nu_{A}}=41$ values equi-spaced over the range $-3.8 < \nu_{\rm A} < 3.8$.

In Figure \ref{fig:2a} we present the components of the Minkowski tensor ${}^{\rm sp}\bar{w}_{i}{}^{j}$  as a function of $\nu$ (top-left panel) and $\nu_{\rm A}$ (top-middle panel) for the fields smoothed with comoving scale $R_{\rm G} = 20 \, {\rm Mpc}$. The off-diagonal components are presented in the top-right panel and are consistent with zero. The same is true for all smoothing scales tested in this work. The top panels represent the components of ${}^{\rm sp}\bar{w}_{i}{}^{j}$ in the spherical basis. 
In the bottom panels of Figure \ref{fig:2a} we present the components of ${}^{\rm sp}\bar{w}_{i}{}^{j}$ in a Cartesian basis, calculated using Euclidean paths to transport tensors to a common location on the manifold. We plot the $(x,x)$, $(y,y)$, $(z,z)$ components as a function of $\nu$ (left), $\nu_{A}$ (middle) and the off-diagonal elements (right panel). The Minkowski tensors in the top and bottom panels are both extracted from the same spherical redshift space distorted density field, only the coordinate systems and choice of transport paths differ. In the bottom panels, we observe that the diagonal elements of the Minkowski tensor are statistically equivalent, and the off-diagonal elements consistent with zero. Hence ${}^{\rm sp}\bar{w}_{i}{}^{j} \propto \delta_{i}{}^{j}$, and the volume average incorrectly infers that the field is isotropic. As discussed in Section \ref{sec:cart}, in a Cartesian basis the spherical redshift space distortion operator generates spatially dependent cumulants, and taking the volume average washes out the anisotropic signal. 

%%%%%%%%%%%%%%%%%%%%%%%%%
\begin{figure}[htb]
  \centering 
 \includegraphics[width=0.45\textwidth]{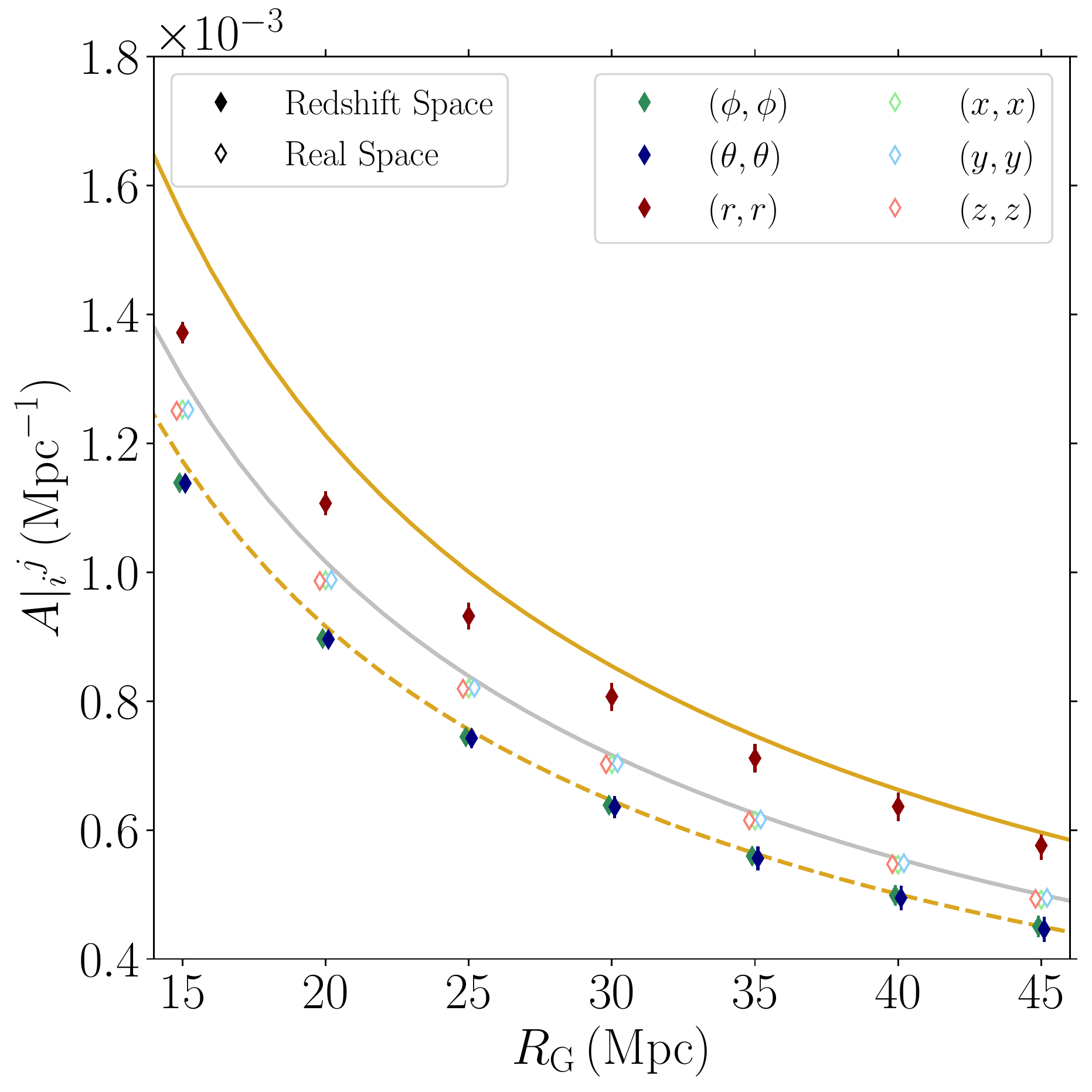} 
\includegraphics[width=0.45\textwidth]{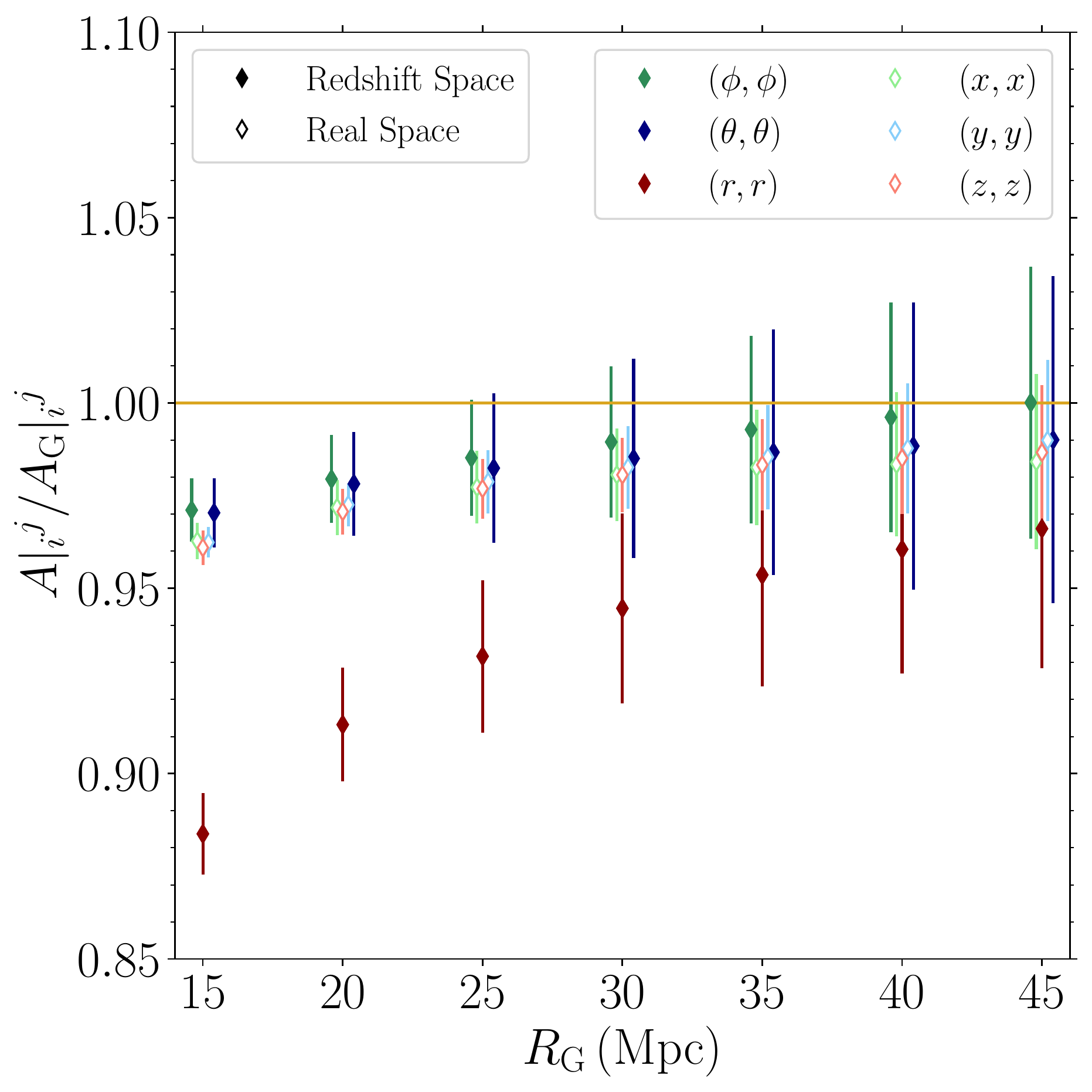} \\ 
 \includegraphics[width=0.45\textwidth]{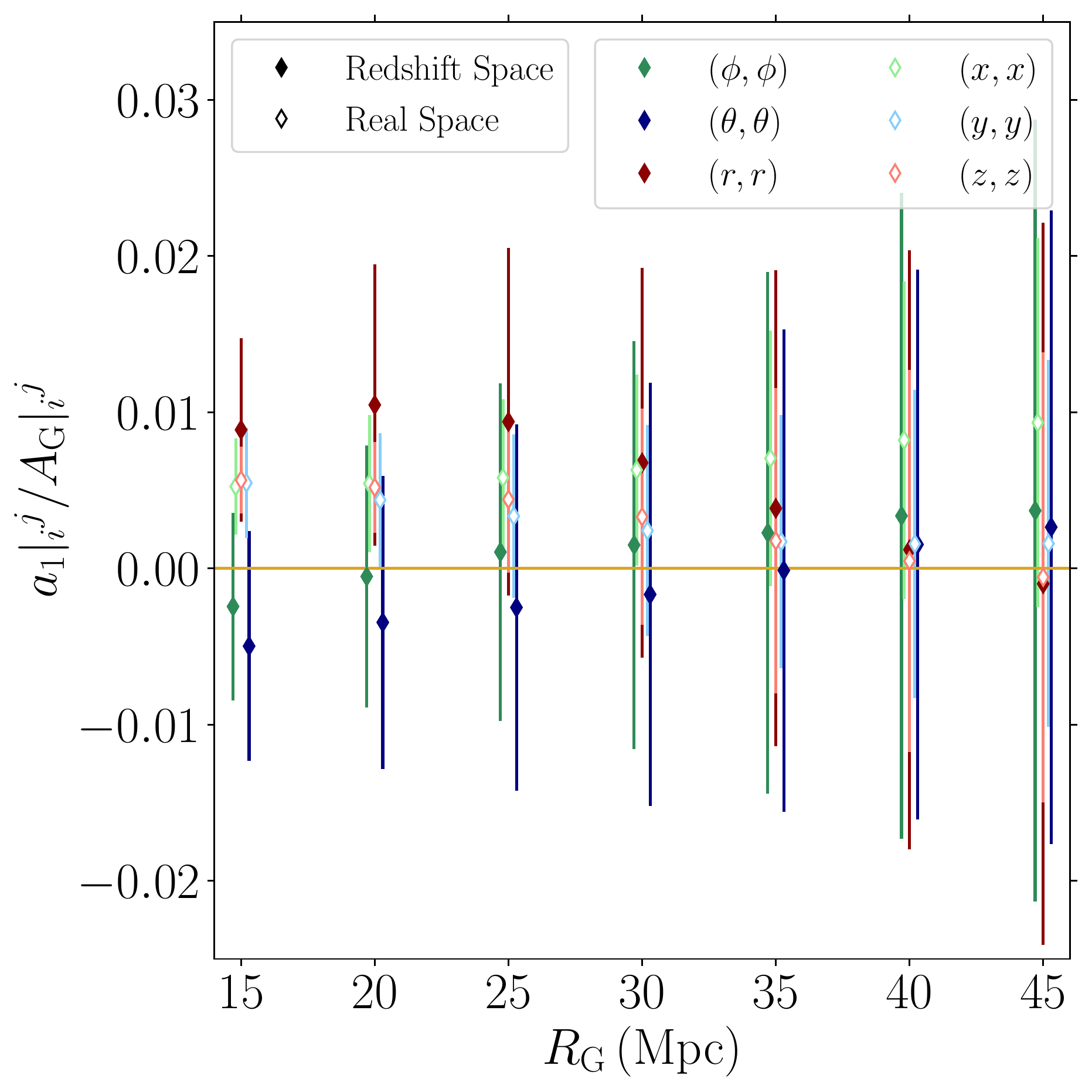} 
\includegraphics[width=0.45\textwidth]{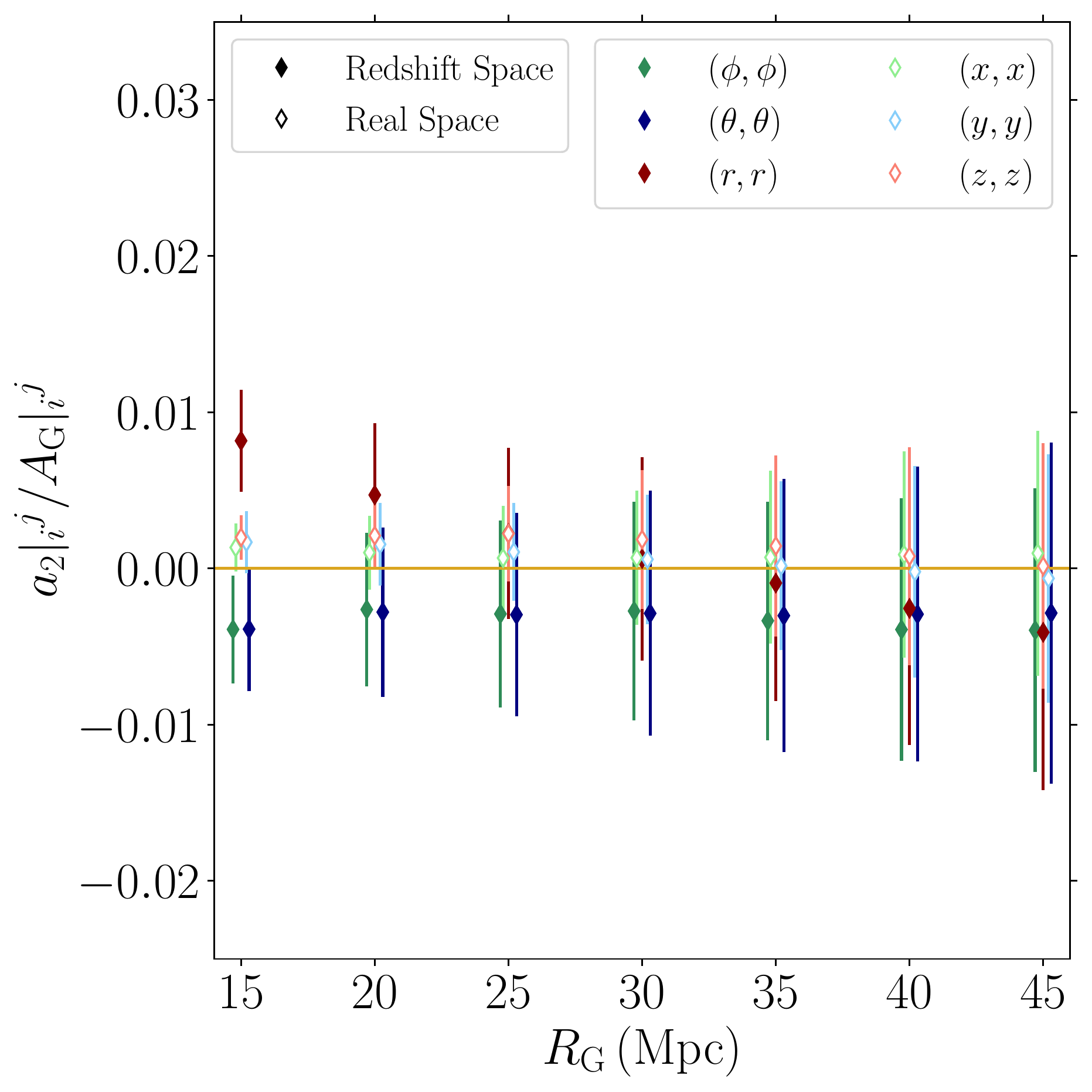} 
  \caption{$A|_{i}{}^{j}$ (top left), $a_{1}|_{i}{}^{j}$ (lower left) and $a_{2}|_{i}{}^{j}$ (lower right) quantities as defined in equations (\ref{eqn:Aij}), (\ref{eqn:a1ij}) and (\ref{eqn:a2ij}) measured from $N_{\rm real}$=50 Quijote simulations at $z$=0, in real and redshift space. The solid/dashed gold lines in the top left panel represent the expectation values of the radial/angular components for a Gaussian field in the plane parallel limit and the silver line is the isotropic Gaussian expectation value. The top right panel shows the fractional difference between the Quijote measurements of $A|_{i}{}^{j}$ and the Gaussian limit of this quantity in real and redshift space.}
  \label{fig:2}
\end{figure}

Next we explore the information contained in the coefficients $A,\, a_1$ and $a_2$ defined in section~\ref{sec:grf}. In the top left panel of Figure \ref{fig:2} we present the components $A|_{i}{}^{j}$ with $(i,j)=(r,r), (\theta, \theta), (\phi,\phi)$ (dark red, blue, green filled diamonds respectively) in redshift space. The points/error bars are the mean and rms values of the $N=50$ snapshot boxes and points that overlap have been slightly perturbed for visual clarity. For comparison we also show the expectation values for ${}^{\rm sp}A_{G}|_{r}{}^{r}$ (solid gold line) and ${}^{\rm sp}A_{G}|_{\theta}{}^{\theta} = {}^{\rm sp}A_{G}|_{\phi}{}^{\phi}$ (dashed gold line) in the limit $r \to \infty$ for a Gaussian random field with a linear $\Lambda$CDM power spectrum and the same cosmological parameters as the Quijote simulations. 
In the top right panel we exhibit the ratio of ${}^{\rm sp}A|_{i}{}^{j}$ extracted from the Quijote simulations and the Gaussian plane parallel expectation values ($\ref{eq:amp1}-\ref{eq:amp3}$). We also present ${}^{\rm re}A|_{i}{}^{j}$  divided by the isotropic expectation value ($\ref{eq:Agau}$), with $(i,j)=(x,x), (y, y), (z,z)$ (light red, blue, green open diamonds respectively) extracted from the corresponding real space snapshot boxes without any velocity correction applied to the particle positions. 

The results for the isotropic field (light open diamonds) present no surprises. The amplitude of each component $(x,x)$, $(y,y)$, $(z,z)$ are statistically indistinguishable, and the Gaussian limit is an excellent approximation at quasi-linear scales $R_{G} \gtrsim 30 \, {\rm Mpc}$ (cf. top panels). Below this scale, the amplitude of the Minkowski tensor components starts to drop relative to the Gaussian expectation (cf top right panel). This is due to the `gravitational smoothing' effect first observed in \citet{1988ApJ...328...50M} for the scalar functionals. The $a_{1}$ component (cf bottom left) is consistent with zero on large scales, but is ${\cal O}(0.01)$ at quasi-linear scales $R_{G} \sim 25 \, {\rm Mpc}$. The $a_{2}$ term (cf bottom right), which we expect to be induced at higher order in a $\sigma_{0}$ expansion of non-Gaussianity, is consistent with zero at all scales probed.

In redshift space (dark filled diamonds), the picture changes considerably. The most striking difference is the strong departure of ${}^{\rm sp}A|_{r}{}^{r}$ from its Gaussian expectation value (cf red filled diamonds, top panels). Even on large scales $R_{\rm G} \sim 40 \, {\rm Mpc}$, the Gaussian, Kaiser formula ($\ref{eq:amp1}$) is not a particularly good approximation. In contrast, the Kaiser approximation ($\ref{eq:amp2},\ref{eq:amp3}$) is excellent for the perpendicular components (green/blue filled diamonds, top panels). It was noted in \cite{Kim:2014axe} that the Gaussian, Kaiser limit is only a good approximation for the scalar Minkowski functionals when the density field is smoothed on very large scales. Our results support this statement, and further show that the radial component of the field is the origin of the breakdown. In addition to the decrease in $A|_{r}{}^{r}$, the non-Gaussian terms $a_{1,2}|_{i}{}^{j}$ are larger for the $(r,r)$ component in redshift space, but remain small at the scales probed. The fact that $a_{2}$ is induced at a statistically significant level on scales $R_{G} \leq 20 \, {\rm Mpc}$ suggests that novel non-Gaussian contributions are induced in redshift space (cf red filled diamonds, lower right panel). 

In \citet{Appleby_2019} it was noted that the ratio of parallel and perpendicular components of the Minkowski tensor would provide a relatively pure measurement of $f$ (or $\beta = f/b$ for biased tracers). However, it is clear that ${}^{\rm sp}A|_{r}{}^{r}$ strays far from the Kaiser limit. The perpendicular components ${}^{\rm sp}A|_{\theta}{}^{\theta}$, ${}^{\rm sp}A|_{\phi}{}^{\phi}$ remain closer to their Gaussian expectation values on small scales, but their values are not sensitive to $f$ alone. Specifically, each individual component of the Minkowski tensors are sensitive to $n_{s}$, $\Omega_{c}h^{2}$ and $f$. Measuring the ratios  ${}^{\rm sp}A|_{\theta}{}^{\theta}/{}^{\rm sp}A|_{r}{}^{r}$, ${}^{\rm sp}A|_{\phi}{}^{\phi}/{}^{\rm sp}A|_{r}{}^{r}$ would potentially break these degeneracies, but only after  we have resolved the origin of the ${}^{\rm sp}A|_{r}{}^{r}$ behaviour. 

\begin{figure}[htb]
  \centering 
 \includegraphics[width=0.32\textwidth]{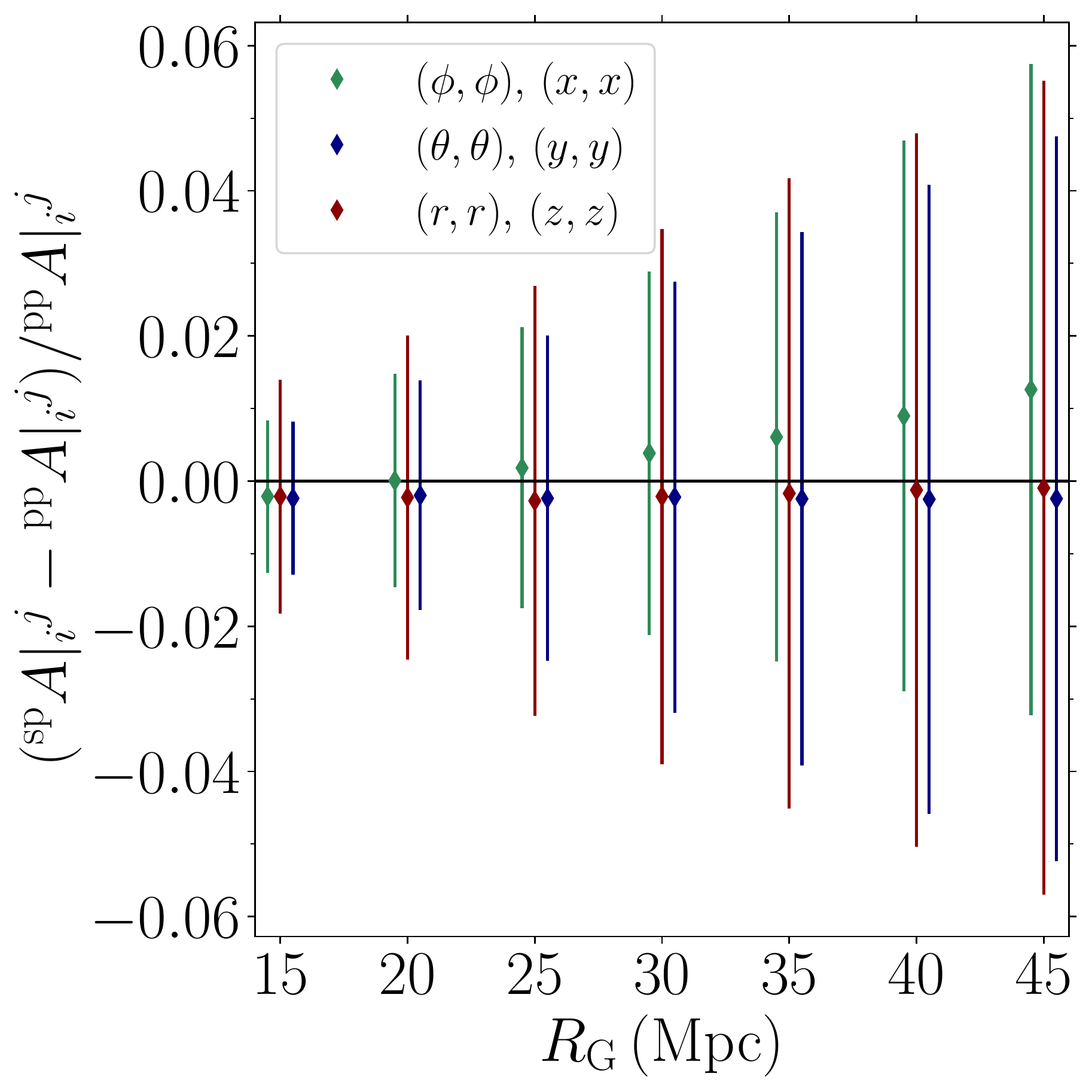} 
\includegraphics[width=0.32\textwidth]{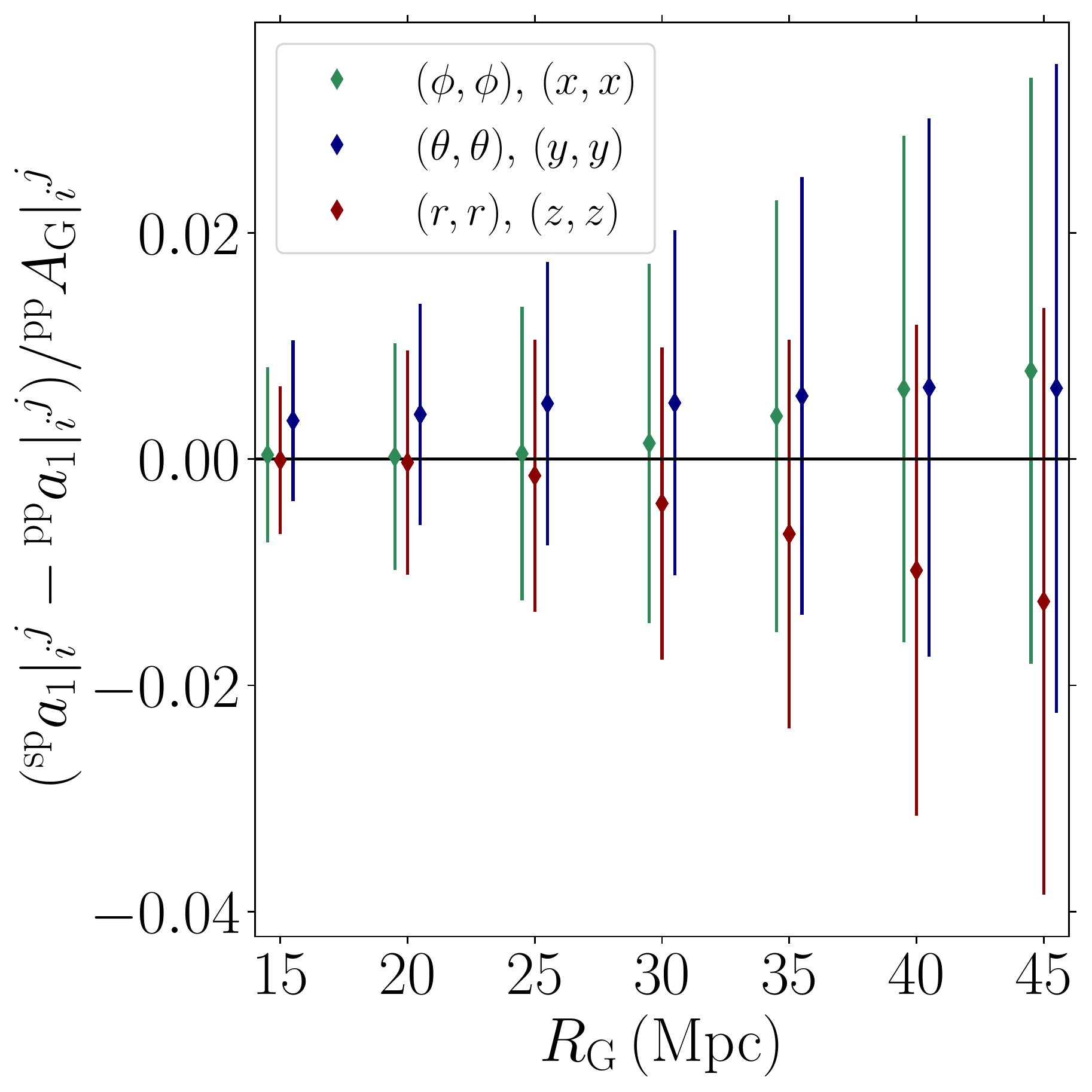} 
 \includegraphics[width=0.32\textwidth]{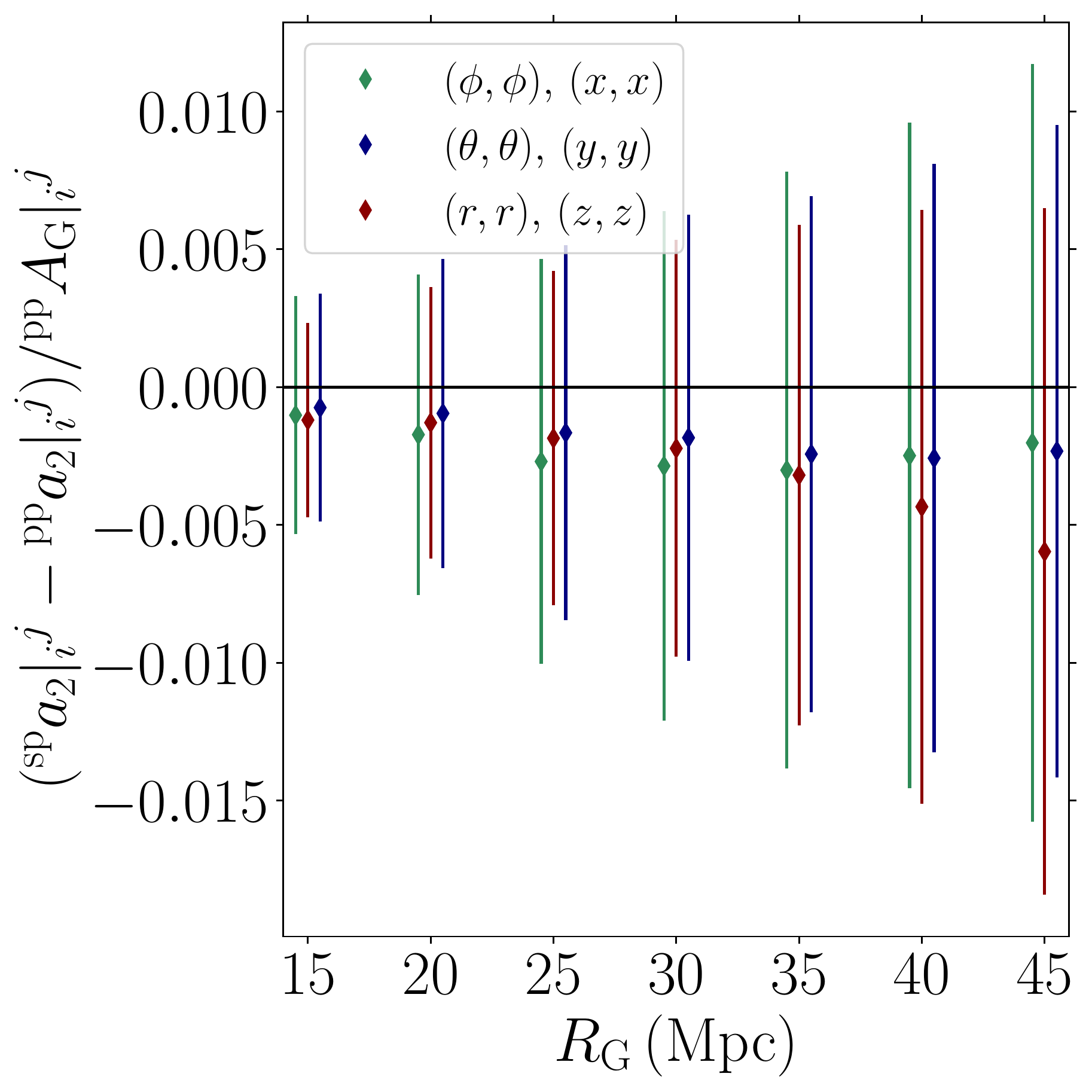} 
  \caption{Fractional difference between $A|_{i}{}^{j}$ (Left), $a_{1}|_{i}{}^{j}$ (middle) and $a_{2}|_{i}{}^{j}$ (right) extracted from the spherical and plane parallel redshift space distorted Quijote dark matter snapshot boxes. }
  \label{fig:sp_pp}
\end{figure}

The large departure of ${}^{\rm sp}A|_{r}{}^{r}$ from the Kaiser limit is not due to the imposition of the spherical redshift space distortion operator. To hightlight this, in Figure \ref{fig:sp_pp} we present the fractional differences $({}^{\rm sp}A|_{i}{}^{j} - {}^{\rm pp}A|_{i}{}^{j})/{}^{\rm pp}A|_{i}{}^{j}$ (left panel), $({}^{\rm sp}a_{1}|_{i}{}^{j} - {}^{\rm pp}a_{1}|_{i}{}^{j})/{}^{\rm pp}A_{\rm G}|_{i}{}^{j}$ (central panel) and $({}^{\rm sp}a_{2}|_{i}{}^{j} - {}^{\rm pp}a_{2}|_{i}{}^{j})/{}^{\rm pp}A_{\rm G}|_{i}{}^{j}$ (right panel). All three fractional differences are consistent with zero over all scales probed in this work, meaning that the spherical and plane parallel redshift space fields possess statistically indistinguishable Minkowski tensor functionals, similar to the Gaussian random fields in the previous subsection.

%%%%%%%%%%%%%%%%%%%%%%%%%%%%%%%%%%%%%%%%%%%%%%%%%%%%%%%%%%%%%%%%%%%%%%%%%%%%%%%%%%%%%
\subsection{Non-Gaussian Effects along the line of sight}

The significant drop in the amplitude of the Minkowski tensor component parallel to the line of sight on small scales observed in the previous subsection can be interpreted as the Finger of God effect, which scatters particle positions over megaparsec scales due to the large peculiar velocity dispersion $\sigma_{\rm pec}$ associated with bound structures \citep{1972MNRAS.156P...1J}. The dominant effect of $\sigma_{\rm pec}$ is an amplitude decrease in $A|_{r}{}^{r}$ which is consistent with an additional, anisotropic damping factor acting on the power spectrum. The Finger of God effect has a long history within theoretical and observational cosmology \cite{1972MNRAS.156P...1J,1994ApJ...431..569P,1995ApJ...448..494F}, and it is well known that its effect on the power spectrum is imprinted even on relatively large scales \citep{Juszkiewicz:1998em,Hikage:2013yja,10.1093/mnras/stu1051,10.1093/mnras/stu1391,Tonegawa:2020wuh,Okumura:2015fga}. Observations of the two-point functions indicate that the Kaiser limit is only accurate on the largest scales \citep{PhysRevD.70.083007,10.1111/j.1365-2966.2010.17581.x,Jennings_2010, Okumura_2010,Kwan_2012,10.1093/mnras/stu2460}.

Our analysis provides two new insights into this phenomenon in the context of the Minkowski statistics. First, the components of the Minkowski tensor perpendicular to the line of sight remain well described by the Kaiser approximation, even on relatively small scales $R_{\rm G }\gtrsim 25 \, {\rm Mpc}$. Second, on ``small scales'' $R_{\rm G} \lesssim 20 \, {\rm Mpc}$ the non-Gaussianity of the components $\bar{w}_{r}{}^{r}$ and $\bar{w}_{\theta}{}^{\theta}$ differ with considerable statistical significance; this can be observed in the $a_{2}|_{i}{}^{j}$ coefficient in Figure \ref{fig:2} (bottom right panel). This indicates that additional non-Gaussian effects are induced in redshift space parallel to the line of sight.

\begin{figure}[htb]
  \centering 
 \includegraphics[width=0.48\textwidth]{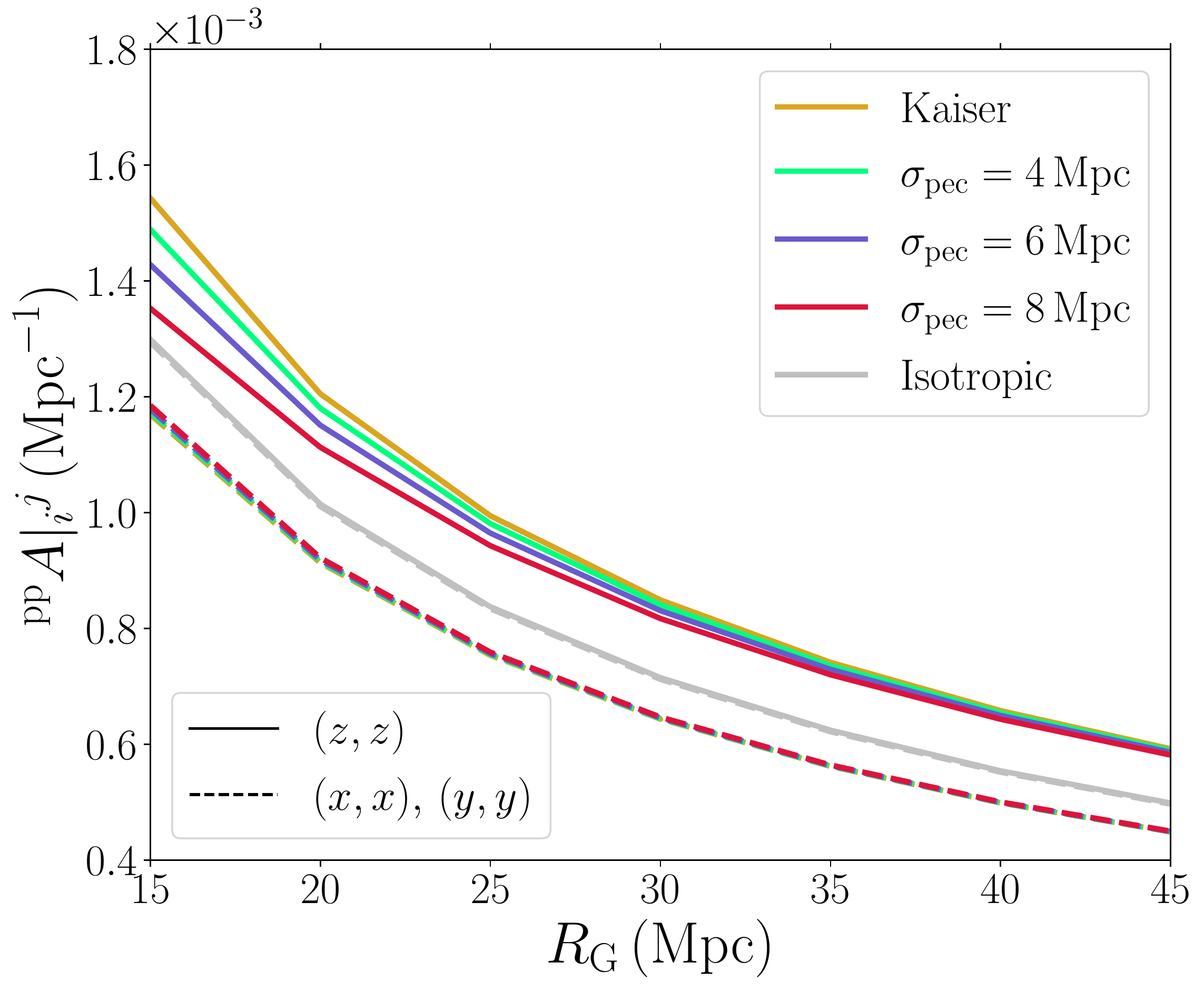} 
\includegraphics[width=0.48\textwidth]{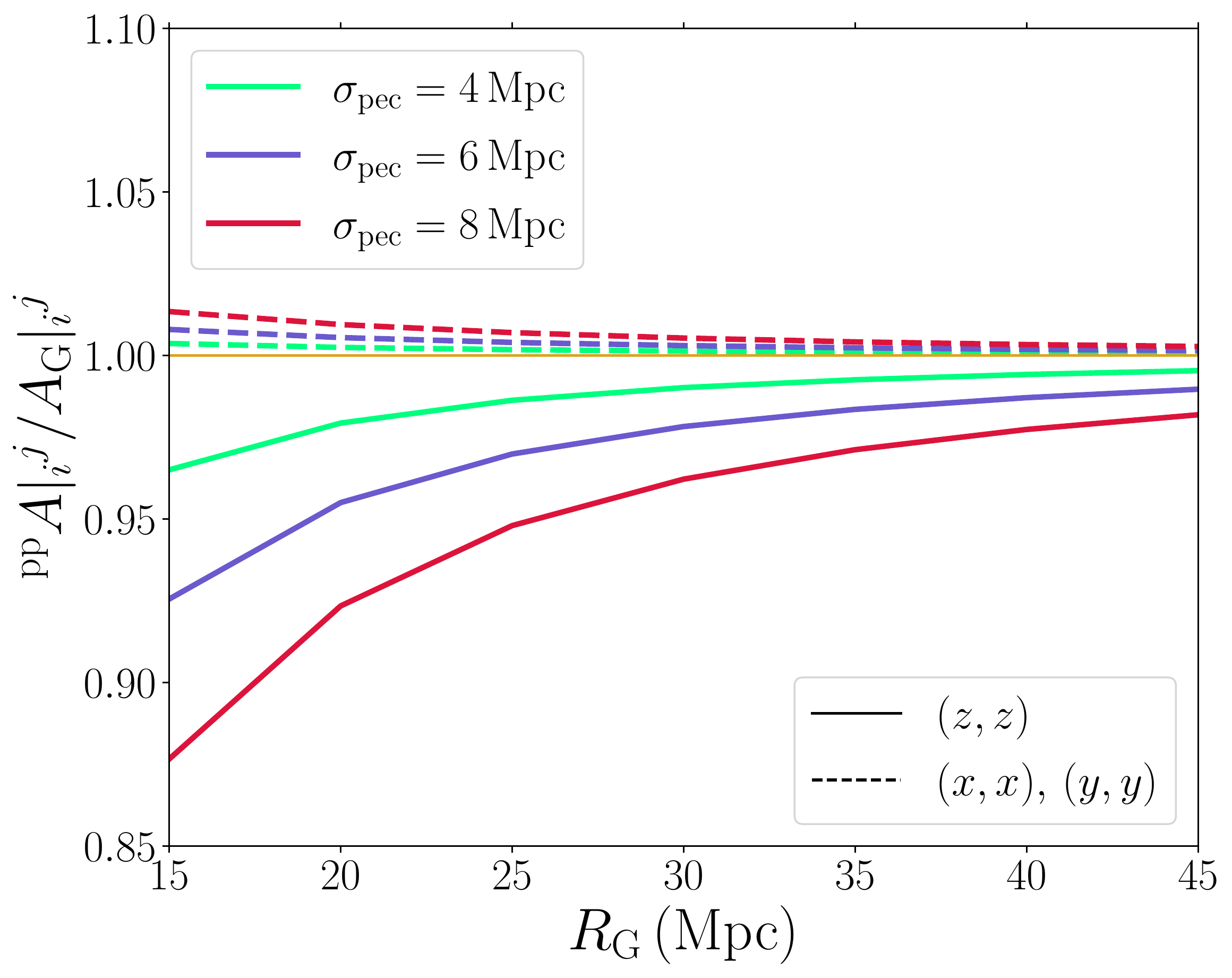} 
  \caption{The effect of an exponential Finger of God damping term on the amplitude of the Minkowski tensor components in plane-parallel redshift space as a function of smoothing scale $R_{\rm G}$.   }
  \label{fig:FoG}
\end{figure}

Regarding the amplitude decrease in the $A|_{r}{}^{r}$ component, we can attempt to model this effect using the standard approach in the literature -- following \citet{1976Ap&SS..45....3P,Peacock:1993xg,1994ApJ...431..569P,Desjacques:2009kt,PhysRevD.70.083007} we introduce an additional damping kernel $P(k,R_{\rm G}) \to P(k,R_{\rm G}) e^{-k_{\parallelsum}^{2}\sigma_{\rm pec}^{2}}$ into the power spectrum that is used in defining the cumulants. Returning to the plane parallel limit, we can write the cumulants in redshift space as 

\begin{eqnarray} & & \tilde{\sigma}_{0}^{2} \equiv \langle \tilde{\delta}({\bf x'}) \tilde{\delta}({\bf x}) \rangle|_{{\bf x'} \to {\bf x}} = {1 \over (2\pi)^{2}} \int_{-1}^{1} d\mu \int_{0}^{\infty} dk k^{2} (1 + f\mu^{2})^{2} P(k,R_{\rm G})e^{-k^{2}R_{\rm G}^{2}} e^{-k^{2}\mu^{2} \sigma^{2}_{\rm pec}} ,  \\
& & \tilde{\sigma}_{z}^{2} \equiv \langle \tilde{\delta}_{z}({\bf x'}) \tilde{\delta}_{z}({\bf x}) \rangle|_{{\bf x'} \to {\bf x}} =  {1 \over (2\pi)^{2}} \int_{-1}^{1} d\mu \int_{0}^{\infty} dk k^{4} \mu^{2} (1 + f\mu^{2})^{2} P(k,R_{\rm G})e^{-k^{2}R_{\rm G}^{2}} e^{-k^{2}\mu^{2} \sigma^{2}_{\rm pec}} , \\
& & \tilde{\sigma}_{x}^{2} \equiv \langle \tilde{\delta}_{x}({\bf x'}) \tilde{\delta}_{x}({\bf x}) \rangle|_{{\bf x'} \to {\bf x}} =  {1 \over 2(2\pi)^{2}} \int_{-1}^{1} d\mu \int_{0}^{\infty} dk k^{4} (1-\mu^{2}) (1 + f\mu^{2})^{2} P(k,R_{\rm G})e^{-k^{2}R_{\rm G}^{2}} e^{-k^{2}\mu^{2} \sigma^{2}_{\rm pec}}  , \\
& & \tilde{\sigma}_{y}^{2} = \tilde{\sigma}_{x}^{2} ,
\end{eqnarray}

\noindent where $\sigma_{\rm pec}$ is a free parameter that describes the velocity dispersion of tracer particles within bound structures and $\mu^{2} = k_{z}^{2}/k^{2}$. If we use these cumulants to derive the ensemble average $\langle {}^{\rm pp}\bar{w}_{i}{}^{j} \rangle$, the additional anisotropic exponential damping term due to the Finger of God effect introduces a significant drop in the $(z, z)$ component, but does not have a large effect on the perpendicular $(x, x)$, $(y, y)$ elements. The amplitudes ${}^{\rm pp}A_{i}{}^{j}$ as a function of $R_{\rm G}$ are presented in the left panel of Figure \ref{fig:FoG}, keeping all parameters fixed and varying $\sigma_{\rm pec} = 0, 4, 6, 8 \, {\rm Mpc}$ (yellow, green, blue, red lines respectively). The components parallel and perpendicular to the line of sight are presented as solid/dashed lines respectively, and we have included the isotropic limit $f = \sigma_{\rm pec} = 0$ (silver lines) and Kaiser limit $\sigma_{\rm pec}=0$ (yellow lines). The right panel exhibits the ratio of the Finger-of-God affected ensemble averages to the Kaiser limit. The large decrease in the parallel cumulant is clearly observed on all scales and the result is in qualitative agreement with the dark matter snapshot results (cf top right panel of Figure \ref{fig:2}). The perpendicular components increase by $\sim 1-2\%$ relative to the Kaiser approximation. We also observe this effect in the dark matter data -- in the top right panel of Figure \ref{fig:2} the $(\theta,\theta)$, $(\phi,\phi)$ components in redshift space are marginally higher than the isotropic components (top right panel, blue/green filled diamonds and light blue/green/red open diamonds respectively). However, in the dark matter snapshot case, all components in real and redshift space have a systematically lower amplitude relative to the Gaussian limit due to the non-Gaussianity of the field (cf top right panel, Figure \ref{fig:2}) which requires further modelling. 

Attempting to simultaneously constrain $n_{s}$, $\Omega_{c}h^{2}$, $f$ and $\sigma_{\rm pec}$ from the Minkowski tensors will yield strong degeneracies. Potentially some of these can be broken since the Finger of God contribution is scale dependent (cf Figure \ref{fig:FoG}), while the Kaiser signal is independent of our choice of $R_{\rm G}$. Hence measuring the MTs at multiple scales will provide simultaneous constraints on $\sigma_{\rm pec}$ and $f$. We must be careful to check for additional, non-Gaussian effects since these will also be scale dependent. A study of perturbative non-Gaussianity in redshift space is beyond the scope of this work and will be conducted elsewhere. 

An alternative approach to mitigating the Finger-of-God effect is to iteratively correct galaxy positions using some higher order prescription \citep{1991ApJ...379....6N,1993ApJ...405..449G,Narayanan_1998,2010ApJ...714..207P}, to reduce the large scatter induced by stochastic velocities within bound structures. This method attempts to reconstruct the galaxy density field in redshift space, but with non-linear effects removed. Many such reconstruction methods rely on the plane parallel approximation, so this approach requires further development to be applied to radial redshift space distortion. A comparison of these different approaches will be a direction of future study.

%%%%%%%%%%%%%%%%%%%%%%%%%%%%%%%%%%%%%%%%%%%%%%%%%%%%%%%%%%%%%%%%%%%%%%%%%%%%%%%%%
\section{Discussion} 
\label{sec:dis}

We have presented an analysis of the rank-two tensor Minkowski functionals for an anisotropic and inhomogeneous Gaussian random field, in particular an isotropic and homogeneous field that has been subjected to the spherically symmetric redshift space distortion operator. Anisotropy here means that the structures in the field share a common alignment along the radial direction, leading to an inequality in the diagonal components of the Minkowski tensors parallel and perpendicular to the line of sight. The inhomogeneity of the field introduces some significant pitfalls -- the ensemble average is now a function of position on the manifold, and the volume average of the statistics will not necessarily be representative of the ensemble average. This statement depends on the coordinate system selected, the volume occupied by the field and also the choice of path transport used to define a volume average of the tensors.

For the spherically redshift space distorted field, there is a singularity in the cumulants at $r=0$ which indicates that this point must be excised from the manifold. This fact, in conjunction with the assumed symmetry properties of the field -- isotropic on $\mathbb{S}^{2}$ -- suggest that spherical coordinates and great arc transport provide a natural framework to measure the Minkowski tensors. We constructed the cumulants of the density field and the gradient in this system and found that they are only weakly coordinate dependent at large distances from a central observer at $r=0$, and furthermore are insensitive to angular position on $\mathbb{S}^{2}$ perpendicular to the line of sight. Similarly the volume average is insensitive to the specifics of how we transport vectors on $\mathbb{S}^{2}$. Of course, we are free to adopt any coordinate system that we want. However, we cannot naively equate volume and ensemble averages when the field is inhomogeneous. We have presented evidence that a spherical coordinate system allows us to extract the Kaiser signal from the components of the volume average ${}^{\rm sp}\bar{w}_{i}{}^{j}$. In contrast, the volume average in Cartesian coordinates does not necessarily replicate the ensemble average due to the non-trivial coordinate dependence of the cumulants. It is important to stress that the volume average of a tensor is generically ambiguous and our choice of coordinates and transport determines the properties of $\bar{w}_{i}{}^{j}$. We can choose a definition that approximately respects the properties of the ensemble average $\langle w_{i}{}^{j} \rangle$, but ergodicity is not exactly realised except in highly idealised scenarios ; Gaussian and isotropic fields, Euclidean manifolds. We have argued that it can be approximately realised for anisotropic and inhomogeneous fields, but only with careful contrivance.

We extracted the Minkowski tensors from Gaussian random fields and gravitationally evolved dark matter snapshot boxes at $z=0$, for three different fields (isotropic, plane parallel and spherical redshift space distorted). We found that the plane parallel and spherical redshift space fields are statistically indistinguishable if the data is sufficiently distant from a central observer at $r=0$. At cosmological distances, the inhomogeneous nature of the cumulants in spherical coordinates is negligible. 

The effect of non-Gaussianity on the MTs is an order $\sim 3\% - 1\%$ effect for the isotropic fields over the range $15 \, {\rm Mpc}  \leq R_{\rm G} \leq 45 \, {\rm Mpc}$, manifesting as a decrease in the amplitude of the diagonal elements, and inducing a non-zero value of the coefficient of $H_{1}(\nu_{\rm A})$ Hermite polynomial that mildly skews the MT as a function of $\nu_{\rm A}$. However, in redshift space the component of the MT parallel to the line of sight for the non-Gaussian dark matter field significantly departs from the Kaiser limit, even for large smoothing scales. The most significant effect is an amplitude decrease that is approximately $\sim 12\%$ on scales $R_{\rm G} \sim 15 \, {\rm Mpc}$. This signal is due to large peculiar velocities along the line of sight from nonlinear regions of the density field, which can scatter particle positions over megaparsec scales. To extract the Kaiser signal from the data, we must model the non-linear velocity component and account for this additional signal. The non-Gaussianity of the redshift space field is also observed in the dark matter data, which indicates that on scales $R_{\rm G} \lesssim 20 \, {\rm Mpc}$, treating the Finger-of-God effect purely in terms of a suppression of the power spectrum is insufficient. Perturbative non-Gaussianity in redshift space is an important area of future study, and the Minkowski tensors are necessary for studying the directional dependence of the non-Gaussian signal. The scalar Minkowski functionals, which are proportional to the trace of these quantities, contain directionally averaged information.

Although we have focused on the radial anisotropy generated by redshift space distortion, even in real space we can expect a radially anisotropic signal. This is due to the fact that we observe tracer particles on the lightcone, and the density field  evolves significantly from the beginning of the matter dominated epoch to the present. At the level of linearized perturbations the evolution can be absorbed into a $z$-dependent galaxy bias, amplitude of the matter power spectrum and the growth rate $f(z)$ in the redshift space distortion signal. In reality, the picture is more complicated on small scales and the Minkowski functionals and tensors will exhibit systematic evolution when measured at different epochs due to non-Gaussianity induced by gravitational collapse. The non-Gaussian evolution can be potentially measured and quantified, and this will be the focus of future work. In this work we have neglected the time dependence of $f(z)$, as this effect is tied to the evolution of the field and hence beyond the scope of our analysis. 

The Minkowski functionals and tensors provide a method to test the fundamental assumptions on which the standard model of cosmology is based. Without the need for {\it a priori} assumptions, the Minkowski functionals provide a measure of the non-Gaussianity of the field as a function of scale, agnostic of the nature of non-Gaussianity. Similarly the eigenvalues of the Minkowski tensors can be used to quantify the isotropy of a field without assuming the presence or absence of this symmetry property. A test of statistical homogeneity is more difficult to engineer, but coordinate dependent cumulants are a smoking gun for inhomogeneous signals. Constructing a test of statistical homogeneity using the tensor transformation properties of the MTs is an interesting direction for future study.

%%%%%%%%%%%%%%%%%%%%%%%%%%%%%%%%%%%%%%%%%%%%%%%%%%%%%%%%%%%%%%%%%%%%%%%%%%%%%

\section*{Acknowledgment}{SA and JK are supported by an appointment to the JRG Program at the APCTP through the Science and Technology Promotion Fund and Lottery Fund of the Korean Government, and were also supported by the Korean Local Governments in Gyeongsangbuk-do Province and Pohang City. This work is supported by Korea Institute for Advanced Study (KIAS) grant funded by the Korea government.}

\bibliography{biblio}{}

\appendix

%%%%%%%%%%%%%%%%%%%%%%%%%%%%%%%%%%%%%%%%%%%%%%%%%%%%%%%%%
\section{Minkowski Tensor $W^{0,2}_{2}$}
\label{sec:appen1}

The second independent, translation invariant Minkowski tensor considered in \citet{Appleby_2019} is given by --

\begin{eqnarray}\label{app1:w22}  W_{2}^{0,2}|_{i}{}^{j} &\equiv&   {1 \over 3\pi V} \int_{\partial Q}  G_{2} \hat{n}_{i} \hat{n}^{j} \textrm{dA} \\
\label{app:w22_2 }&=&  \frac{1}{3\pi V} \int_{V} \textrm{dV}  \, \delta_{D}\left( \delta - \delta_{t} \right) G_{2} \frac{\delta_{i} \delta^{j}}{\left| \nabla \delta \right|} ,
\end{eqnarray} 

\noindent where the scalar quantity $G_{2}$ is the mean curvature at each point of the iso-field surface, and can be written as 

\begin{equation}\label{eq:g2} G_{2} = -{1 \over 2} \nabla . {\bf \hat{n}} = -{1 \over 2} \nabla . \left( {\nabla \delta \over |\nabla \delta|} \right) .
\end{equation}

\noindent Similarly to the $W^{0,2}_{1}$ case in the main body of the text, the quantity $W^{0,2}_{2}|_{i}{}^{j}$ can be interpreted as the volume average of the following tensor

\begin{equation}  v_{i}{}^{j} = {1 \over 3\pi} \delta_{D}\left( \delta - \delta_{t} \right) G_{2} \frac{\delta_{i} \delta^{j}}{\left| \nabla \delta \right|} .
\end{equation}

$G_{2}$ is a function of both first and second derivatives of the field. Hence when constructing the ensemble average $\langle v_{i}{}^{j} \rangle$, we must use a ten-dimensional multivariate probability distribution involving the field and its first and second derivatives -- $X = (\delta, \delta_{i},\delta_{jk})$. If the field is homogeneous and isotropic or plane parallel redshift space distorted, then the dependence of $G_{2}$ on the second derivatives $\delta_{jk}$ does not contribute to the ensemble average, hence $\langle v_{i}{}^{j} \rangle$ reduces to an integral over the joint probability distribution of $\delta$ and $\delta_{i}$. For an inhomogeneous field we cannot assume this remains true and must construct the corresponding full $10 \times 10$ covariance matrix, $\Sigma$, for $\delta$, $\delta_{i}$ and $\delta_{jk}$. For a spherically redshift space distorted field, due to the assumed residual isotropy on the two-sphere, many off-diagonal terms are zero. The expression for $\Sigma$ takes the form

\begin{equation}
\Sigma = 
\begin{pmatrix} 
  \langle \delta^{2} \rangle & \langle \delta\delta_{r} \rangle   &  0 & 0 & \langle \delta\delta_{rr} \rangle & 0 & 0 &  \langle \delta\delta_{\theta\theta} \rangle & \langle \delta\delta_{\phi\phi} \rangle & 0 \\ 
  
    \langle \delta_{r}\delta \rangle & \langle \delta_{r}^{2} \rangle   &  0 & 0 & \langle \delta_{r}\delta_{rr} \rangle & 0 & 0 &  \langle \delta_{r}\delta_{\theta\theta} \rangle & \langle \delta_{r}\delta_{\phi\phi} \rangle & 0 \\ 
    
      0 & 0   &  \langle \delta_{\theta}^{2} \rangle & 0 & 0 & \langle \delta_{\theta}\delta_{r\theta} \rangle & 0 &  0 & \langle \delta_{\theta}\delta_{\phi\phi} \rangle & 0 \\ 
      
  0 & 0   &  0 & \langle \delta_{\phi}^{2} \rangle & \langle \delta_{\phi}\delta_{rr} \rangle & 0 & \langle \delta_{\phi}\delta_{r\phi} \rangle &  0 & 0 & \langle \delta_{\phi}\delta_{\theta\phi} \rangle \\ 
  
   \langle \delta_{rr}\delta \rangle & \langle \delta_{rr}\delta_{r} \rangle   &  0 & 0 & \langle \delta_{rr}^{2} \rangle & 0 & 0 &  \langle \delta_{rr}\delta_{\theta\theta} \rangle & \langle \delta_{rr}\delta_{\phi\phi} \rangle & 0 \\ 
   
   0 & 0  &  \langle \delta_{r\theta}\delta_{\theta} \rangle & 0 & 0 & \langle \delta_{r\theta}^{2} \rangle & 0 &  0 & \langle \delta_{r\theta}\delta_{\phi\phi} \rangle & 0 \\ 
   
   0 & 0   &  0 & \langle \delta_{r\phi}\delta_{\phi} \rangle & 0 & 0 & \langle \delta_{r\phi}^{2} \rangle &  0 & 0 & \langle \delta_{r\phi}\delta_{\theta\phi} \rangle \\ 
   
   \langle \delta_{\theta\theta}\delta \rangle & \langle \delta_{\theta\theta}\delta_{r} \rangle   &  0 & 0 & \langle \delta_{\theta\theta}\delta_{rr} \rangle & 0 & 0 &  \langle \delta_{\theta\theta}^{2} \rangle & \langle \delta_{\theta\theta}\delta_{\phi\phi} \rangle & 0 \\ 
   
   \langle \delta_{\phi\phi}\delta \rangle & \langle \delta_{\phi\phi}\delta_{r} \rangle   &  \langle \delta_{\phi\phi}\delta_{\theta} \rangle & 0 & \langle \delta_{\phi\phi}\delta_{rr} \rangle & \langle \delta_{\phi\phi}\delta_{r\theta} \rangle & 0 &  \langle \delta_{\phi\phi}\delta_{\theta\theta} \rangle & \langle \delta_{\phi\phi}^{2} \rangle & 0 \\ 
   
   0 & 0   &  0 & \langle \delta_{\theta\phi}\delta_{\phi} \rangle & 0 & 0 & \langle \delta_{\theta\phi}\delta_{r\phi} \rangle &  0 & 0 & \langle \delta_{\theta\phi}^{2} \rangle 
\end{pmatrix}
\end{equation} 

\noindent This is the covariance matrix of the partial derivatives. If one uses covariant derivatives as random variables, then different correlations will be present. The $4\times 4$ block in the top left corner has been calculated in the main body of the text. In this appendix we calculate the other terms as follows --  

\begin{eqnarray} 
   \nonumber   \left<\delta_{rr}^2\right>  &=&  \left(\frac{1}{\pi^2}\right)\left(\frac{1}{10} + \frac{f}{7} + \frac{f^2}{18}\right)\int k^6 P(k,R_{\rm G})dk + \left(\frac{1}{\pi^2r^2}\right)\left(\frac{4f}{5} + \frac{6f^2}{7}\right)\int k^4 P(k,R_{\rm G})dk + \left(\frac{8f^2}{3\pi^2r^6}\right)\int P(k,R_{\rm G}) dk \\
   & & \\
    \nonumber   \left<\delta_{\phi\phi}^2\right> &=&  \left(\frac{r^4\sin^4\theta}{\pi^2}\right)\left(\frac{1}{10} + \frac{f}{35} + \frac{f^2}{210}\right)\int k^6P(k,R_{\rm G})dk + \\
    \nonumber  & &   \left(\frac{r^2}{\pi^2}\right)\left[\frac{\sin^2\theta}{6} + \left(\frac{\sin^2\theta}{15}-\frac{6\sin^4\theta}{5}\right)f + \left(\frac{\sin^2\theta}{70} + \frac{12\sin^4\theta}{35}\right)f^2\right]\int k^4P(k,R_{\rm G})dk \\
    & &  +\left(\frac{1}{\pi^2}\right)\left[\frac{-2f\sin^2\theta}{3} + \left(\frac{2\sin^2\theta}{5} + \frac{18\sin^4\theta}{5}\right)f^2\right]\int k^2P(k,R_{\rm G})dk + \frac{2f^2\sin^2\theta}{3\pi^2r^2}\int P(k,R_{\rm G})dk \\
    \nonumber & & \\
    \nonumber \left<\delta_{\theta\theta}^2\right> & = & \left(\frac{r^4}{\pi^2}\right)\left(\frac{1}{10} + \frac{f}{35} + \frac{f^2}{210}\right)\int k^6P(k,R_{\rm G})dk + \left(\frac{r^2}{\pi^2}\right)\left(\frac{1}{6} - \frac{17f}{15} + \frac{5f^2}{14}\right)\int k^4P(k,R_{\rm G})dk \\
    & & +\left(\frac{1}{\pi^2}\right)\left(-\frac{2f}{3} + 4f^2\right)\int k^2P(k,R_{\rm G})dk + \frac{2f^2}{3\pi^2r^2}\int P(k,R_{\rm G})dk \\
    \nonumber & & \\
    \nonumber \left<\delta_{r\phi}^2\right> & = & \left(\frac{r^2\sin^2\theta}{\pi^2}\right)\left(\frac{1}{30} + \frac{f}{35} + \frac{f^2}{126}\right)\int k^6P(k,R_{\rm G})dk + \left(\frac{\sin^2\theta}{\pi^2}\right)\left(\frac{1}{6} + \frac{f}{5} + \frac{3f^2}{10}\right)\int k^4P(k,R_{\rm G})dk \\
    & & +\left(\frac{\sin^2\theta}{\pi^2r^2}\right)\left(\frac{2f}{3} + \frac{4f^2}{5}\right)\int k^2P(k,R_{\rm G})dk + \frac{2f^2\sin^2\theta}{3\pi^2r^4}\int P(k,R_{\rm G})dk \\
    \nonumber & & \\
    \nonumber \left<\delta_{r\theta}^2\right> & = & \left(\frac{r^2}{\pi^2}\right)\left(\frac{1}{30} + \frac{f}{35} + \frac{f^2}{126}\right)\int k^6P(k,R_{\rm G})dk + \left(\frac{1}{\pi^2}\right)\left(\frac{1}{6} + \frac{f}{5} + \frac{3f^2}{10}\right)\int k^4P(k,R_{\rm G})dk \\
    & & +\left(\frac{1}{\pi^2r^2}\right)\left(\frac{2f}{3} + \frac{4f^2}{5}\right)\int k^2P(k,R_{\rm G})dk + \frac{2f^2}{3\pi^2r^4}\int P(k,R_{\rm G})dk \\
    \nonumber & & \\
    \nonumber \left<\delta_{\theta\phi}^2\right> & = & \left(\frac{r^4\sin^2\theta}{\pi^2}\right)\left(\frac{1}{30}+\frac{f}{105}+\frac{f^2}{630}\right)\int k^6 P(k,R_{\rm G})dk \\ 
    \nonumber & & + \left(\frac{r^2}{\pi^2}\right)\left[\left(\frac{1}{6}-\frac{\sin^2\theta}{6}\right)+\left(\frac{1}{15}-\frac{7\sin^2\theta}{15}\right)f + \left(\frac{1}{70}+\frac{\sin^2\theta}{10}\right)f^2\right]\int k^4 P(k,R_{\rm G})dk \\
    \nonumber & & + \left(\frac{1}{\pi^2}\right)\left[\left(-\frac{2}{3}+\frac{2\sin^2\theta}{3}\right)f + \left(\frac{2}{5}+\frac{4\sin^2\theta}{5}\right)f^2\right]\int k^2 P(k,R_{\rm G})dk \\
    & & + \left(\frac{f^2}{\pi^2r^2}\right)\left(\frac{2}{3}-\frac{2\sin^2\theta}{3}\right)\int P(k,R_{\rm G}) dk \\
    \nonumber & & \\
    \nonumber \left<\delta_{rr}\delta_{\phi\phi}\right> & = & \left(\frac{r^2\sin^2\theta}{\pi^2}\right)\left(\frac{1}{30} + \frac{f}{35} + \frac{f^2}{126}\right)\int k^6 P(k,R_{\rm G})dk \\ 
    \nonumber & & + \left(\frac{\sin^2\theta}{\pi^2}\right)\left(\frac{2f}{15} + \frac{2f^2}{7}\right)\int k^4 P(k,R_{\rm G})dk +\left(\frac{\sin^2\theta}{\pi^2r^2}\right)\left(\frac{2f}{3} + \frac{4f^2}{5}\right)\int k^2P(k,R_{\rm G})dk \\
    & & - \left(\frac{4f^2\sin^2\theta}{3\pi^2r^4}\right)\int P(k,R_{\rm G})dk \\
    \nonumber & & \\
    \nonumber \left<\delta_{rr}\delta_{\theta\theta}\right> & = & \left(\frac{r^2}{\pi^2}\right)\left(\frac{1}{30} + \frac{f}{35} + \frac{f^2}{126}\right)\int k^6 P(k,R_{\rm G})dk + \left(\frac{1}{\pi^2}\right)\left(\frac{2f}{15} + \frac{2f^2}{7}\right)\int k^4 P(k,R_{\rm G})dk + \\  & & \left(\frac{1}{\pi^2r^2}\right)\left(\frac{2f}{3} + \frac{4f^2}{5}\right)\int k^2P(k,R_{\rm G})dk
    - \left(\frac{4f^2}{3\pi^2r^4}\right)\int P(k,R_{\rm G})dk \\
    \nonumber & & \\
    \left<\delta_{r\theta}\delta_{\phi\phi}\right> & = & \left(\frac{r\sin 2\theta}{\pi^2}\right)\left(\frac{-1}{12} - \frac{f}{30} - \frac{f^2}{140}\right)\int k^4 P(k,R_{\rm G})dk + \left(\frac{f^2\sin 2\theta}{3\pi^2r^3}\right)\int P(k,R_{\rm G})dk \\
    \nonumber & & 
\end{eqnarray}

\begin{eqnarray} 
    \left<\delta_{r\phi}\delta_{\theta\phi}\right> & = & \left(\frac{r\sin 2\theta}{\pi^2}\right)\left(\frac{1}{12} + \frac{f}{30} + \frac{f^2}{140}\right)\int k^4 P(k,R_{\rm G})dk - \left(\frac{f^2 \sin 2\theta}{3\pi^2r^3}\right)\int P(k,R_{\rm G}) dk \\
    \nonumber & & \\
    \left<\delta\delta_r\right> & = & \left(\frac{-2f^2}{3\pi^2r^3}\right)\int P
    (k)dk \\
    \nonumber & & \\
    \nonumber \left<\delta\delta_{rr}\right> & = & \left(\frac{1}{\pi^2}\right)\left(-\frac{1}{6}-\frac{f}{5}-\frac{f^2}{14}\right)\int k^4 P(k,R_{\rm G})dk + \left(\frac{1}{\pi^2r^2}\right)\left(-\frac{2f}{3}-\frac{4f^2}{5}\right)\int k^2P(k,R_{\rm G})dk \\  
    & & + \left(\frac{4f^2}{3\pi^2r^4}\right)\int P(k,R_{\rm G})dk \\
    \nonumber & & \\
    \left<\delta_r\delta_{rr}\right> & = & \left(\frac{1}{\pi^2r^3}\right)\left(-\frac{2f}{3}-\frac{4f^2}{5}\right)\int k^2 P(k,R_{\rm G})dk - \left(\frac{4f^2}{3\pi^2r^5}\right)\int P(k,R_{\rm G})dk \\
    \nonumber & & \\
    \left<\delta_{r\theta}\delta_{\theta}\right> & = & \left(\frac{r}{\pi^2}\right)\left(\frac{1}{6}+\frac{f}{15} + \frac{f^2}{70}\right)\int k^4 P(k,R_{\rm G})dk - \left(\frac{2f^2}{3\pi^2r^3}\right)\int P(k,R_{\rm G})dk \\
    \nonumber & & \\
    \left<\delta_{r\phi}\delta_{\phi}\right> & = & \left(\frac{r\sin^2\theta}{\pi^2}\right)\left(\frac{1}{6}+\frac{f}{15} + \frac{f^2}{70}\right)\int k^4 P(k,R_{\rm G})dk - \left(\frac{2f^2\sin^2\theta}{3\pi^2r^3}\right)\int P(k,R_{\rm G})dk \\
    \nonumber & & \\
    \nonumber \left<\delta_{\theta\theta}\delta\right> & = & \left(\frac{r^2}{\pi^2}\right)\left(-\frac{1}{6}-\frac{f}{15} - \frac{f^2}{70}\right)\int k^4 P(k,R_{\rm G})dk + \left(\frac{1}{\pi^2}\right)\left(\frac{2f}{3} - \frac{2f^2}{5}\right)\int k^2P(k,R_{\rm G})dk \\
     & & - \left(\frac{2f^2}{3\pi^2r^2}\right)\int P(k,R_{\rm G})dk \\
     \nonumber & & \\
     \nonumber \left<\delta_{\phi\phi}\delta\right> & = & \left(\frac{r^2\sin^2\theta}{\pi^2}\right)\left(-\frac{1}{6}-\frac{f}{15} - \frac{f^2}{70}\right)\int k^4 P(k,R_{\rm G})dk + \left(\frac{\sin^2\theta}{\pi^2}\right)\left(\frac{2f}{3} - \frac{2f^2}{5}\right)\int k^2P(k,R_{\rm G})dk \\
     & & - \left(\frac{2f^2\sin^2\theta}{3\pi^2r^2}\right)\int P(k,R_{\rm G})dk \\
     \nonumber & & \\
     \left<\delta_{\theta\theta}\delta_r\right> & = & \left(\frac{r}{\pi^2}\right)\left(-\frac{1}{6}-\frac{f}{15} - \frac{f^2}{70}\right)\int k^4 P(k,R_{\rm G})dk + \left(\frac{2f^2}{3\pi^2r^3}\right)\int P(k,R_{\rm G})dk \\
     \nonumber & & \\
     \left<\delta_{\phi\phi}\delta_r\right> & = & \left(\frac{r\sin^2\theta}{\pi^2}\right)\left(-\frac{1}{6}-\frac{f}{15} - \frac{f^2}{70}\right)\int k^4 P(k,R_{\rm G})dk + \left(\frac{2f^2\sin^2\theta}{3\pi^2r^3}\right)\int P(k,R_{\rm G})dk \\
     \nonumber & & \\
     \nonumber \left<\delta_{\phi\phi}\delta_{\theta}\right> & = & \left(\frac{r^2\sin 2\theta}{\pi^2}\right)\left(-\frac{1}{12}-\frac{f}{30} - \frac{f^2}{140}\right)\int k^4 P(k,R_{\rm G})dk + \left(\frac{\sin 2\theta}{\pi^2}\right)\left(\frac{f}{3} - \frac{f^2}{5}\right)\int k^2P(k,R_{\rm G})dk \\
     & & - \left(\frac{f^2\sin 2\theta}{3\pi^2r^2}\right)\int P(k,R_{\rm G})dk \\
     \nonumber & & \\
     \nonumber \left<\delta_{\theta\phi}\delta_{\phi}\right> & = & \left(\frac{r^2\sin 2\theta}{\pi^2}\right)\left(\frac{1}{12}+\frac{f}{30} + \frac{f^2}{140}\right)\int k^4 P(k,R_{\rm G})dk + \left(\frac{\sin 2\theta}{\pi^2}\right)\left(-\frac{f}{3} + \frac{f^2}{5}\right)\int k^2P(k,R_{\rm G})dk \\
     & & + \left(\frac{f^2\sin 2\theta}{3\pi^2r^2}\right)\int P(k,R_{\rm G})dk
\end{eqnarray}

These are the correlations between partial derivatives of the field, although covariant derivatives can be used instead. If we take the limit $r \to \infty$, then the covariance matrix reduces to the plane parallel limit. Hence similar to the main body of the text, if the field is sufficiently distant from the observer at $r=0$, the ensemble average $\langle v_{i}{}^{j}\rangle$ is well approximated by the plane parallel result in \citet{Appleby_2019}. More concretely, the dimensionless terms $\sigma_{-1}^{2} /(r^{6} \sigma_{2}^{2})$, $\sigma_{1}^{2}/(r^{2}\sigma_{2}^{2})$, $\sigma_{0}^{2}/(r^{4}\sigma_{2}^{2})$ must all be negligible to satisfy the plane parallel limit. 

The volume average of $v_{i}{}^{j}$ in a spherical basis is given by

\begin{equation}  {}^{\rm sp}\bar{v}_{i}{}^{j} =  {1 \over 3\pi V} \sum_{m,n,p} \Delta^{3} \delta_{D}(\tilde{\delta}_{\{m,n,p\}}-\delta_{t}) G_{2\, \{m,n,p\}} {{}^{\gamma}\tilde{\delta}_{i}{}_{\{m,n,p\}} {}^{\gamma}\tilde{\delta}^{j}{}_{\{m,n,p\}} \over |\nabla \tilde{\delta}_{\{m,n,p\}}|} .
\end{equation}

\noindent This is straightforward to construct, the only complication beyond ${}^{\rm sp}\bar{w}_{i}{}^{j}$ is that we must additionally estimate $G_{2}$ at each pixel using ($\ref{eq:g2}$). Since $G_{2}$ is a scalar quantity, we use Cartesian coordinates and a simple second order accurate finite difference scheme for the first and second derivatives to reconstruct $G_{2\, \{m,n,p\}}$ at each pixel. 

\begin{figure}[htb]
  \centering 
 \includegraphics[width=0.45\textwidth]{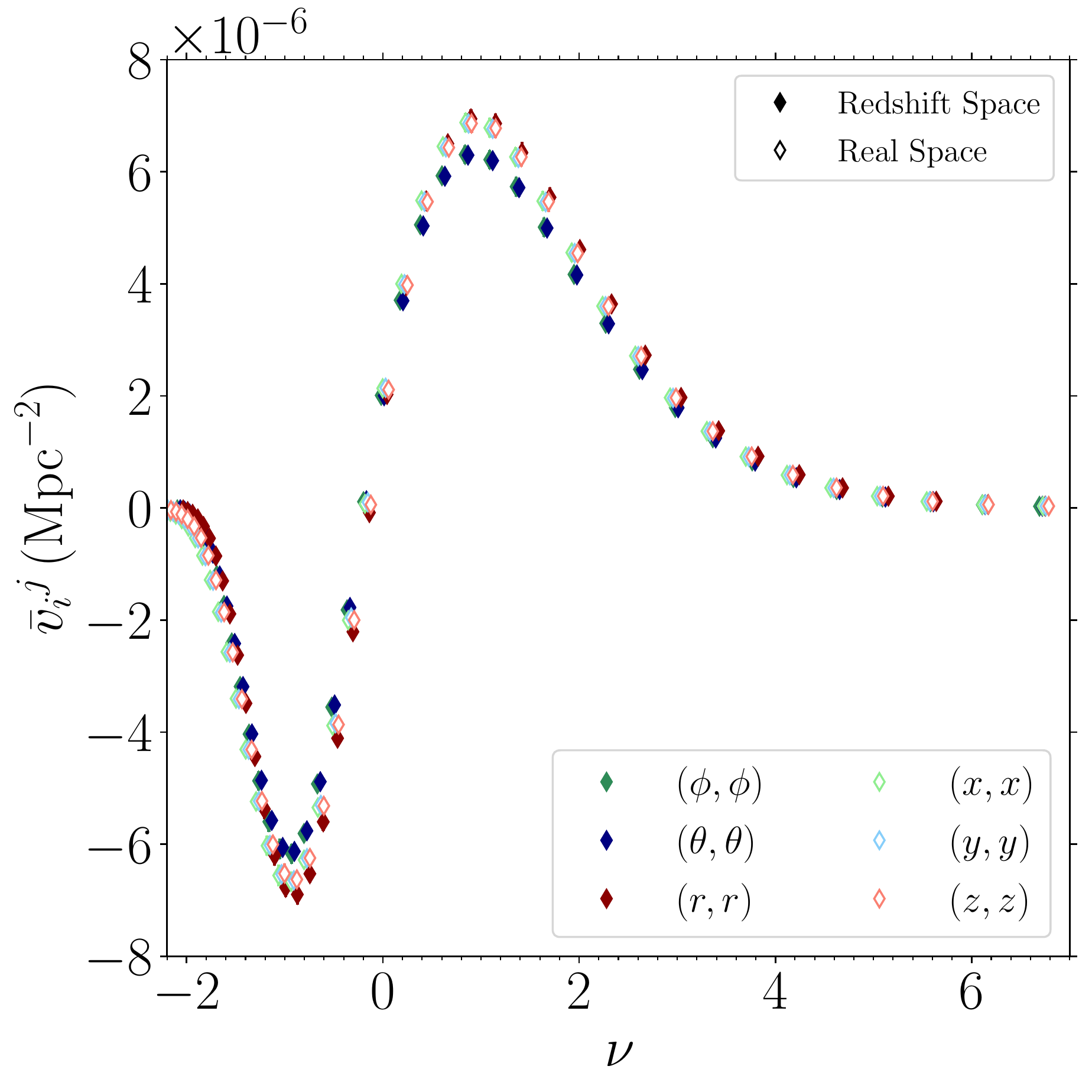} 
 \includegraphics[width=0.45\textwidth]{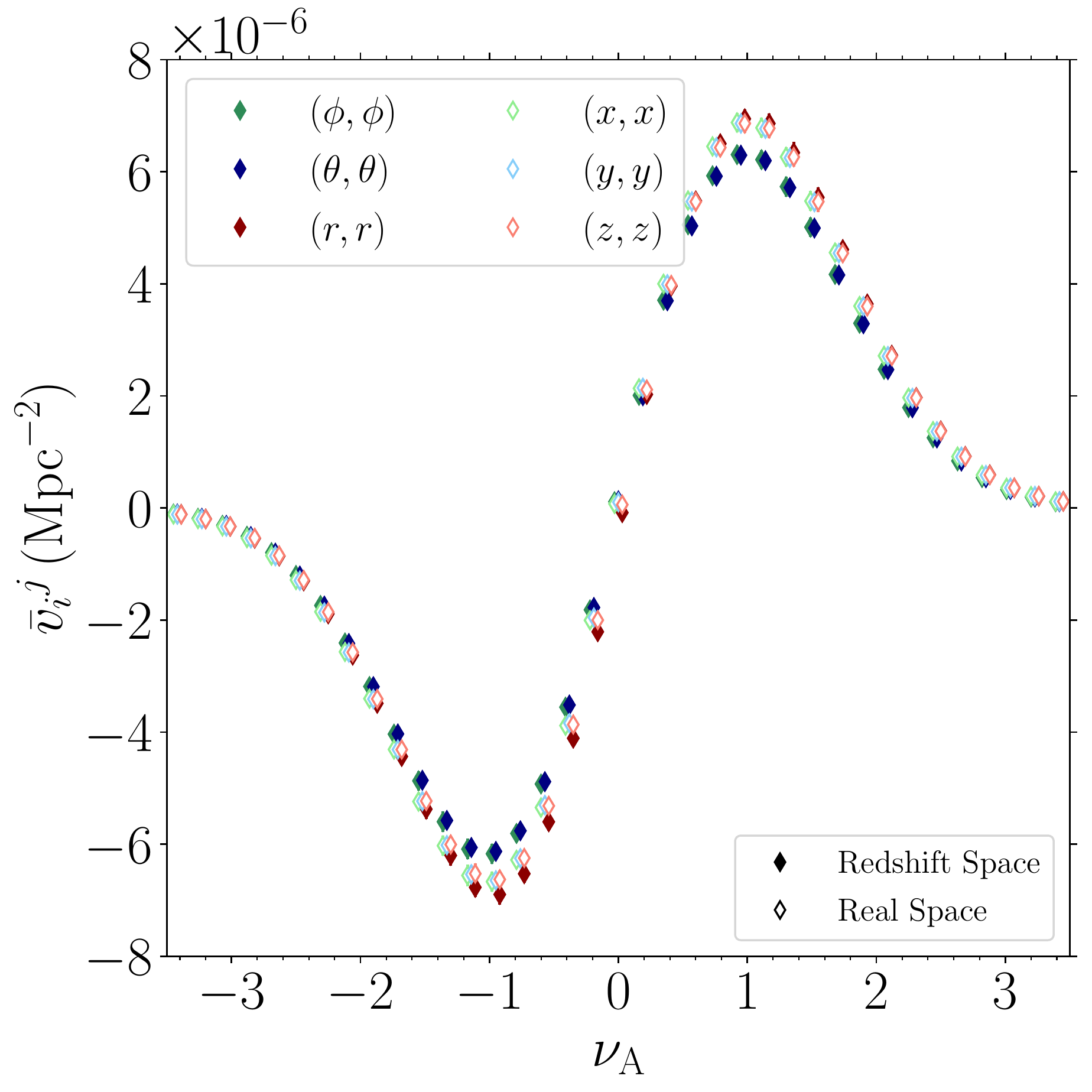} 
  \caption{Components of the the Minkowski tensor $\bar{v}_{i}{}^{j}$ extracted from the Quijote simulations in real space (open diamonds), and spherical redshift space (filled diamonds) as a function of normalised threshold $\nu$ (left panel) and rescaled threshold $\nu_{\rm A}$ (right panel). The off-diagonal elements are consistent with zero and not plotted. The fields have been smoothed with scale $R_{G} = 20 \, {\rm Mpc}$.}
  \label{fig:app1}
\end{figure}

In Figure \ref{fig:app1} we present the components of the Minkowski tensor $\bar{v}_{i}{}^{j}$ from the Quijote $z=0$ snapshot boxes smoothed with scale $R_{\rm G} = 20 \, {\rm Mpc}$ as a function of $\nu$ (left panel) and $\nu_{\rm A}$ (right panel). The color scheme matches Figure \ref{fig:2a} in the main body of the text. All off-diagonal components are consistent with zero and not plotted. The redshift space distortion signal is present, with the $(\theta,\theta)$, $(\phi,\phi)$ components systematically lower in amplitude compared to the real space statistics (cf light hollow diamonds). The radial component (red filled diamonds) is only marginally higher than the isotropic components -- this is due to the same Finger-of-God effect observed in the main body of the text (see Section \ref{sec:ngrf}). 

\begin{figure}[htb]
  \centering 
 \includegraphics[width=0.45\textwidth]{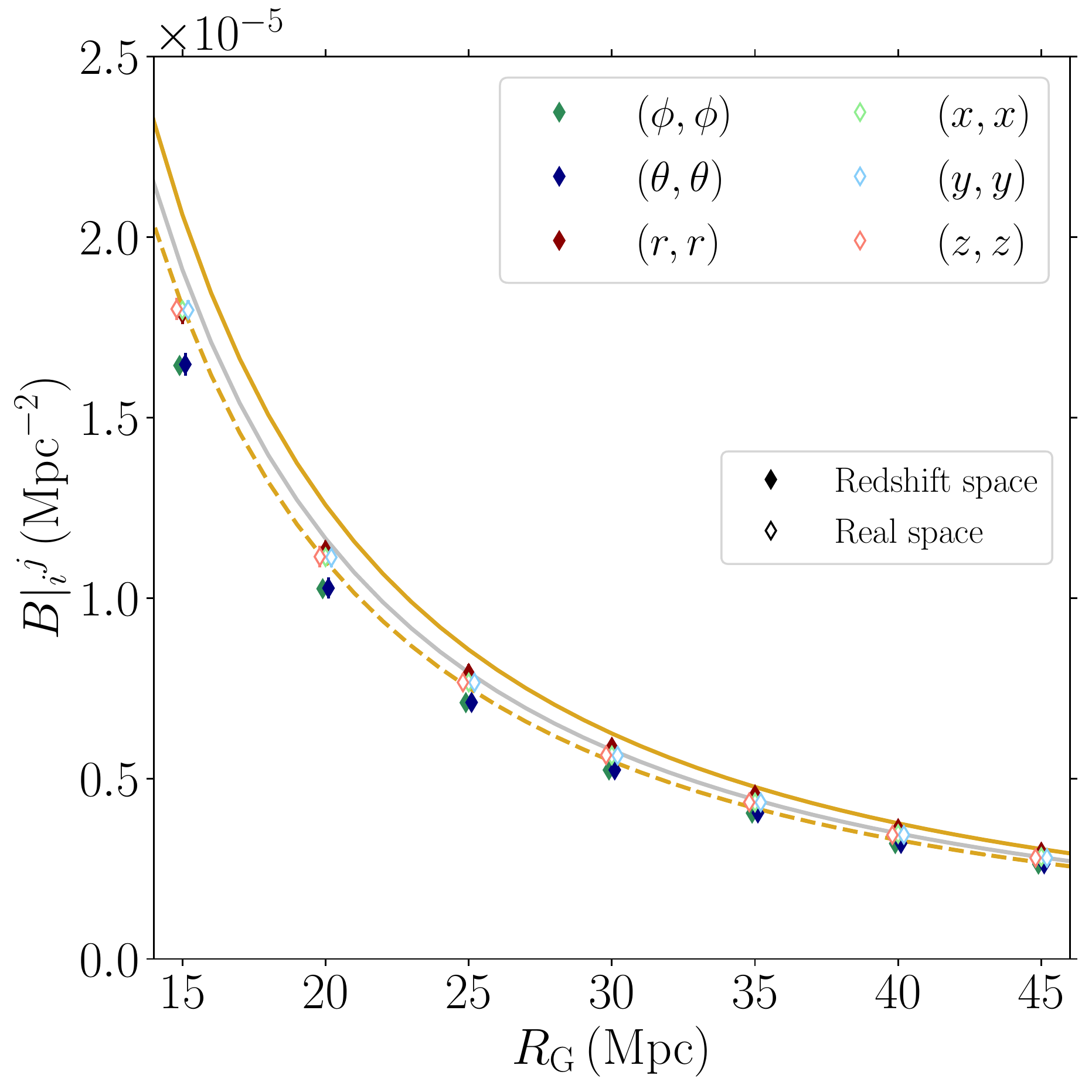} 
\includegraphics[width=0.45\textwidth]{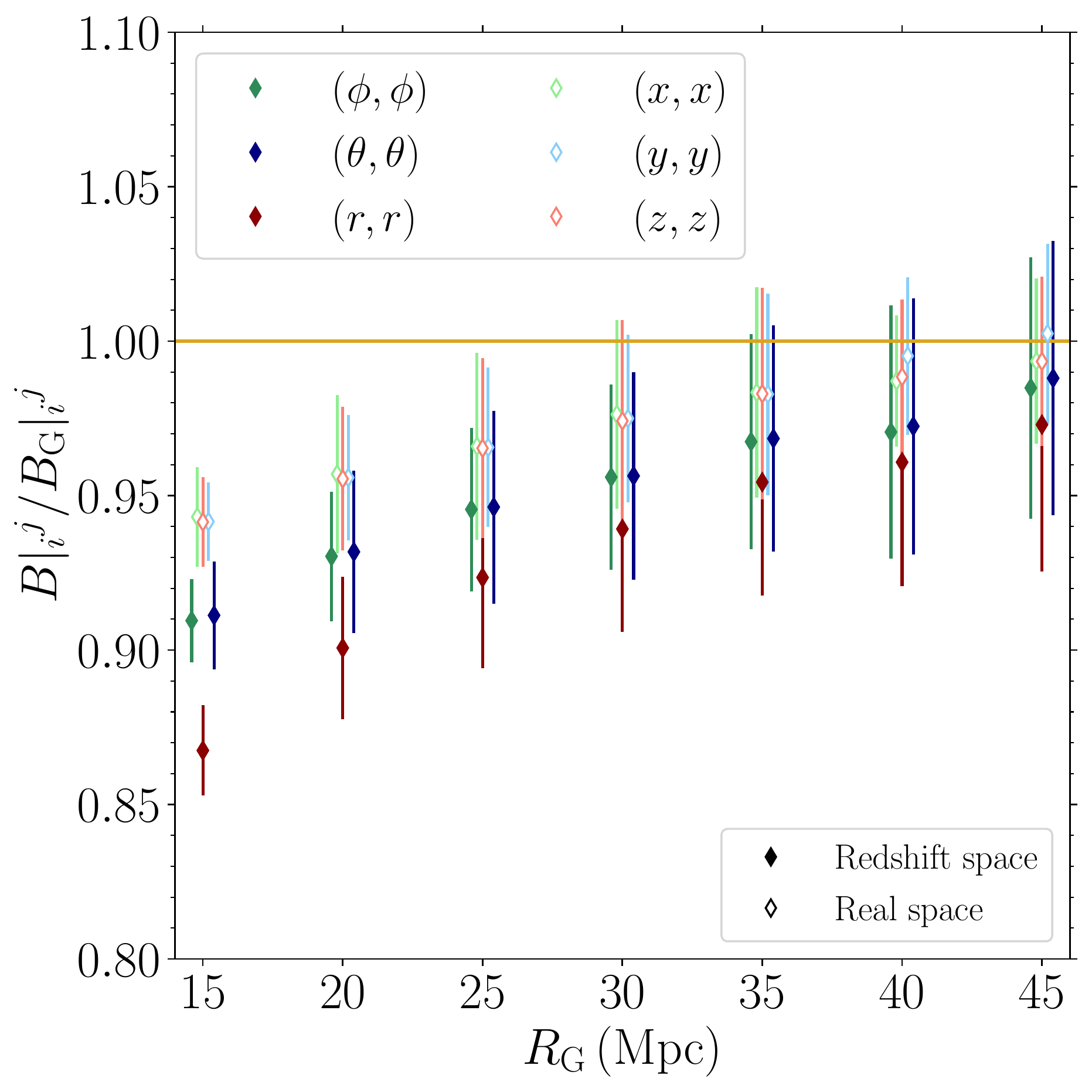} \\ 
 \includegraphics[width=0.45\textwidth]{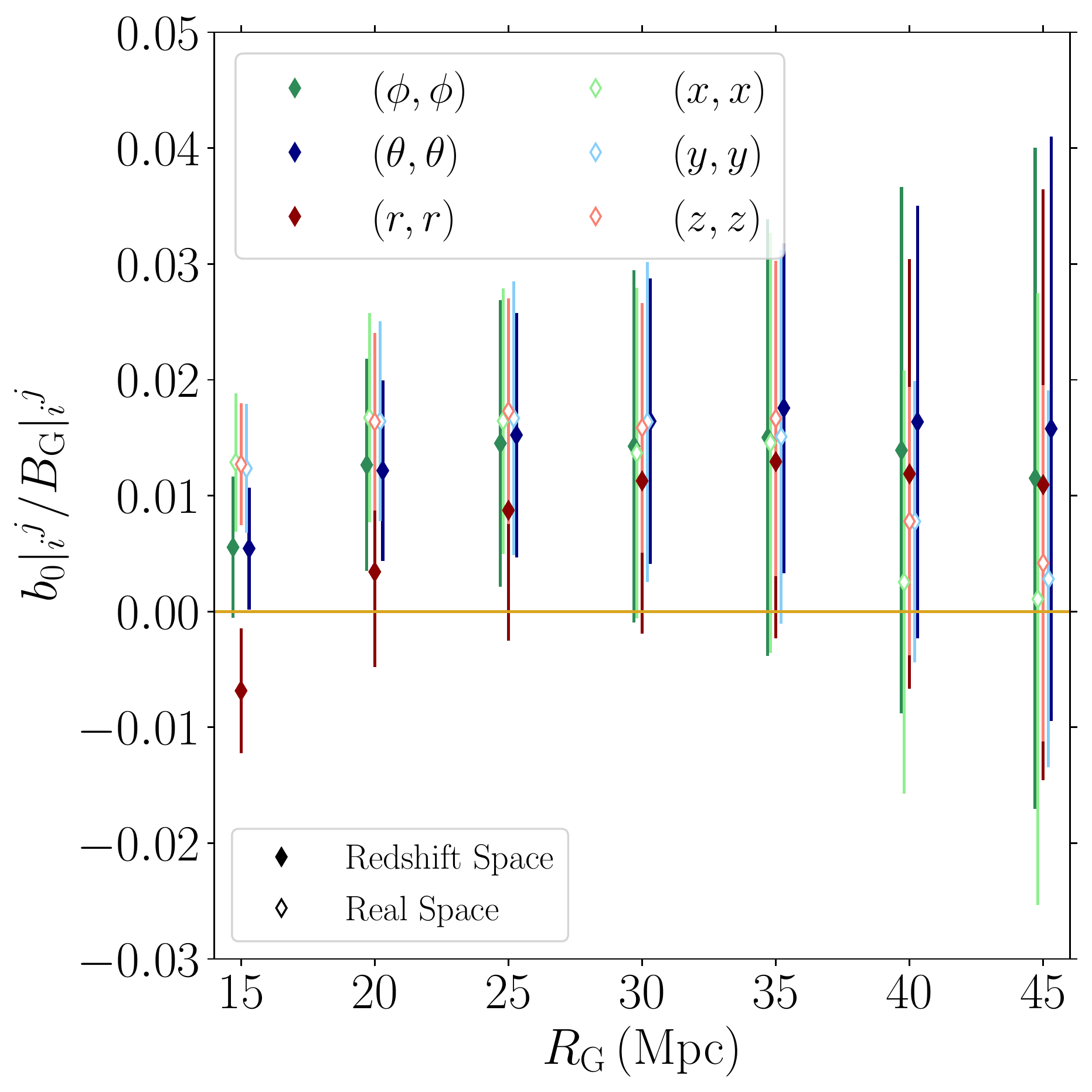} 
\includegraphics[width=0.45\textwidth]{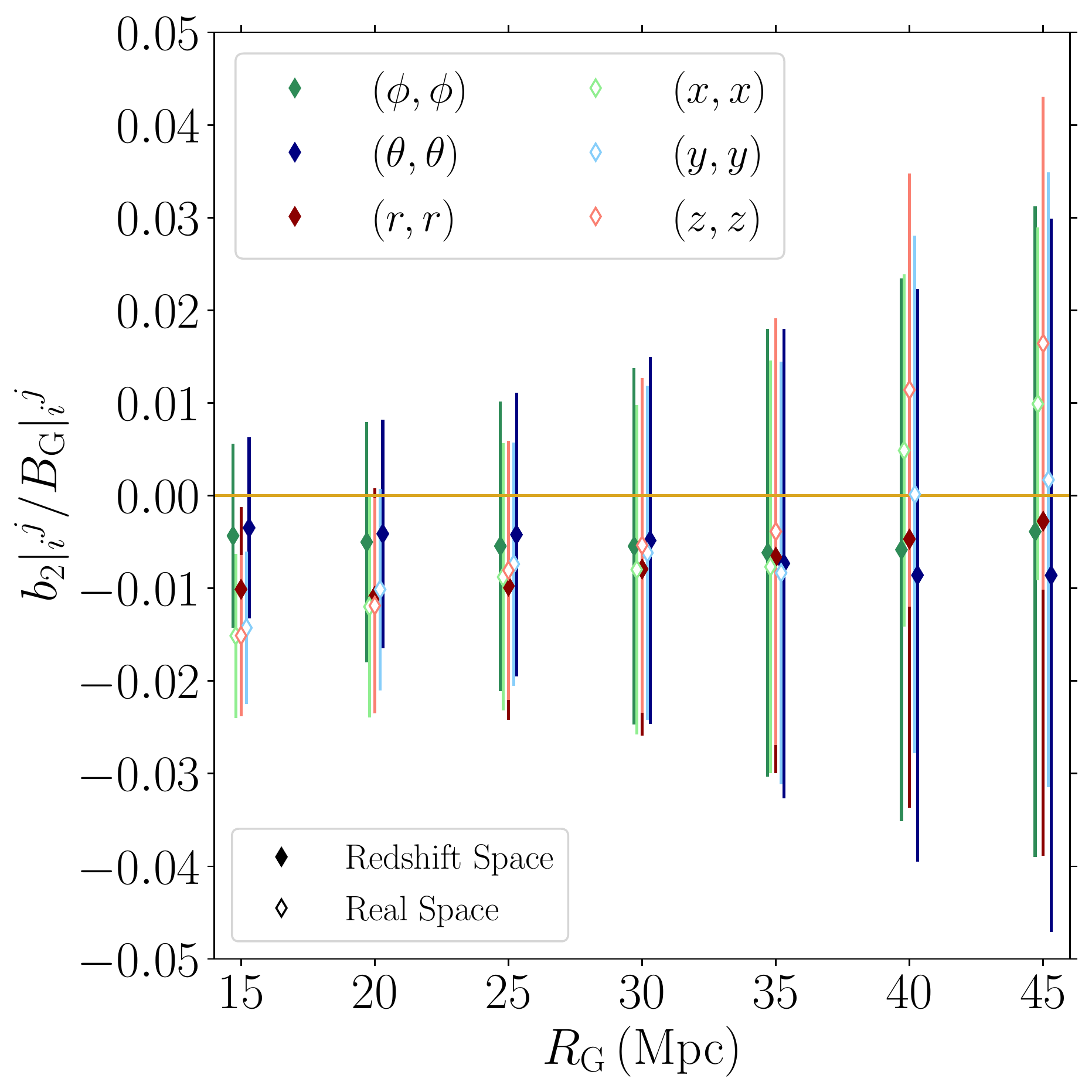} 
  \caption{$B|_{i}{}^{j}$ (top left), $b_{0}|_{i}{}^{j}$ (lower left) and $b_{2}|_{i}{}^{j}$ (lower right) quantities as defined in equations (\ref{eqn:Bij}), (\ref{eqn:b1ij}) and (\ref{eqn:b2ij}) measured from $N_{real}$=50 Quijote simulations at $z=0$, in real and redshift space. The solid/dashed gold lines in the top left panel represent the expectation values of the radial/angular components for a Gaussian field in the plane parallel limit and the silver line is the isotropic Gaussian expectation value. The top right panel shows the fractional difference between the Quijote measurements of $B|_{i}{}^{j}$ and the Gaussian limit of this quantity in real and redshift space.}
  \label{fig:app2}
\end{figure}

In Figure \ref{fig:app2} we present the amplitude of $\bar{v}_{i}{}^{j}$ defined as the coefficient of the $H_{1}$ coefficient --

\begin{equation} B|_{i}{}^{j} = {1 \over \sqrt{2\pi}} \int_{-\nu_{{\rm max}}}^{\nu_{\rm max}} \bar{v}_{i}{}^{j}(\nu_{\rm A}) H_{1}(\nu_{\rm A}) d\nu_{\rm A} ,
\label{eqn:Bij}
\end{equation} 

\noindent and the additional Hermite polynomial coefficients

\begin{eqnarray} b_{0}|_{i}{}^{j} &=& {1 \over \sqrt{2\pi}} \int_{-\nu_{{\rm max}}}^{\nu_{{\rm max}}} \bar{v}_{i}{}^{j}(\nu_{\rm A}) H_{0}(\nu_{\rm A}) d\nu_{\rm A} ,
\label{eqn:b1ij}\\
b_{2}|_{i}{}^{j} &=& {1 \over 2\sqrt{2\pi}} \int_{-\nu_{{\rm max}}}^{\nu_{{\rm max}}} \bar{v}_{i}{}^{j}(\nu_{\rm A}) H_{2}(\nu_{\rm A}) d\nu_{\rm A} .
\label{eqn:b2ij}
\end{eqnarray} 

\noindent 

The colour scheme is the same as in Figure $\ref{fig:2}$. Qualitatively we observe the same behaviour for $W^{0,2}_{2}$ as $W^{0,2}_{1}$, but here it is more pronounced. Both the isotropic and spherically redshift space distorted fields are significantly affected by the non-Gaussianity of the Quijote fields for scales $R_{\rm G} \leq 35 \, {\rm Mpc}$, and the $(r,r)$ component in redshift space most significantly departs from the Gaussian limit (cf top panels, red diamonds/error bars). The redshift space Gaussian and plane parallel limit of the amplitudes are given by

\begin{eqnarray}\label{eq:ampB1} & & {}^{\rm pp}B_{G}|_{x}{}^{x} = {3 \over 2}\sqrt{\pi \over 2} A_{0}^{2}\left[ {(\lambda^{2}-2)(\lambda^{2}-1)^{1/2} + \lambda^{4}\tan^{-1}\sqrt{\lambda^{2}-1} \over (\lambda^{2}-1)^{3/2}  }  \right]  , \\
\label{eq:ampB2} & & {}^{\rm pp}B_{G}|_{y}{}^{y} = {}^{\rm pp}B_{G}|_{x}{}^{x}  , \\
\label{eq:ampB3} & & {}^{\rm pp}B_{G}|_{z}{}^{z} =  3 \sqrt{\pi \over 2}A_{0}^{2}\left[{\lambda^{2} \left[ \left(\lambda^{2}-1\right)^{1/2} + (\lambda^{2}-2) \tan^{-1}\sqrt{\lambda^{2}-1}  \right] \over (\lambda^{2}-1)^{3/2} }\right] ,
\end{eqnarray}

\noindent and the isotropic Gaussian limit is 

\begin{equation}\label{eq:Bgau} {}^{\rm re}B_{\rm G}|_{x}{}^{x} = {}^{\rm re}B_{\rm G}|_{y}{}^{y} = {}^{\rm re}B_{\rm G}|_{z}{}^{z} = {\sigma_{1}^{2} \over 27\pi \sqrt{2\pi} \sigma_{0}^{2}} .
\end{equation} 
The non-Gaussian coefficients $b_{0}|_{i}{}^{j}$ and $b_{2}|_{i}{}^{j}$ remain small even on relatively small scales $R_{\rm G} \gtrsim 15 \, {\rm Mpc}$.

\section{Rotation of Basis Vectors Relative to a Great Arc}
\label{sec:appen3}

In the main body of the paper we constructed an algorithm to describe the rotation of a vector under geodesic transport to a different location on the two-sphere. In this appendix we present the rotation of the spherical basis vectors explicitly using a simple geometric prescription. Starting with the unit sphere, we select two points on the sphere defined with $\mathbb{R}^{3}$ unit vectors $\hat{u}$ and $\hat{v}$. To parameterize the great arc that passes through these two points we introduce the vectors $\hat{m} = \hat{u} \times \hat{v}/|\hat{u}\times \hat{v}|$ and $\hat{n} = \hat{m} \times \hat{u}/|\hat{m}\times \hat{u}|$. The unit vectors $\hat{u}$, $\hat{m}$ and $\hat{n}$ are mutually orthogonal, and $\hat{u}$, $\hat{n}$ form a basis in the plane in which the great circle is defined. Any position on the great arc can then be represented parametrically with the vector 

\begin{equation} \hat{R} = \hat{u}\cos t + \hat{n} \sin t \end{equation}

\noindent for $0 < t \leq 2\pi$. The tangent vector to the great arc is 

\begin{equation} \hat{T} = -\hat{u} \sin t + \hat{n}\cos t \end{equation}

\noindent Each point on the great arc, specified by the vector $\hat{R}$, can be described using the angles $\theta,\phi$ in a spherical coordinate system, and we can then define the spherical basis vectors in the usual way 

\begin{eqnarray} & & {\bf e}_{r} = \sin\theta \cos\phi \, {\bf e}_{x} + \sin\theta \sin\phi \, {\bf e}_{y} + \cos\theta \, {\bf e}_{z} \\
& & {\bf e}_{\theta} = \cos\theta \cos\phi \, {\bf e}_{x} + \cos\theta \sin\phi \, {\bf e}_{y} - \sin\theta \, {\bf e}_{z}  \\
& & {\bf e}_{\phi} = - \sin\phi \, {\bf e}_{x} + \cos\phi \, {\bf e}_{y}  \\
\end{eqnarray} 

The dot products $\hat{T} . {\bf e}_{\theta}$, $\hat{T} . {\bf e}_{\phi}$ then represent the angle rotation of the spherical basis vectors relative to the great arc tangent along the path. This is the rotation that is accounted for in the main body of the text, when summing vector fields at different locations on the manifold. Parallel transport preserves the orientation of a tangent space relative to $\hat{T}$, so after geodesic transport the components of a vector in the basis ${\bf e}_{\theta}$, ${\bf e}_{\phi}$ are rotated. Conversely the dot product $\hat{T} . {\bf e}_{r}$ is always zero and components of a vector parallel to ${\bf e}_{r}$ are preserved. If the great arc lies on the equator of the sphere then the basis vectors do not rotate with respect to the great arc tangent vector. 

\begin{figure}[htb]
  \centering 
 \includegraphics[width=0.45\textwidth]{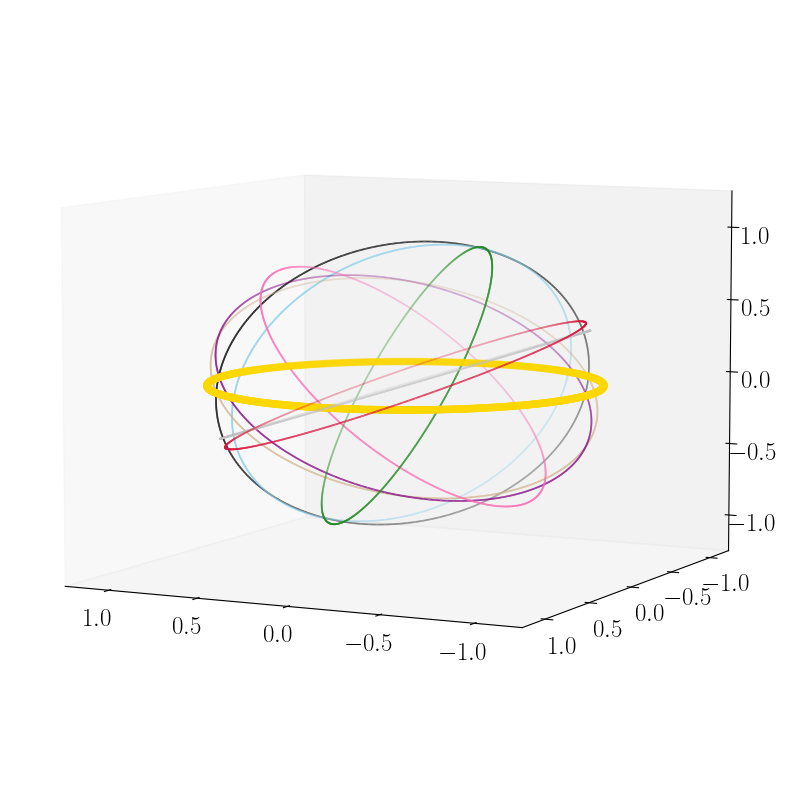} 
\includegraphics[width=0.45\textwidth]{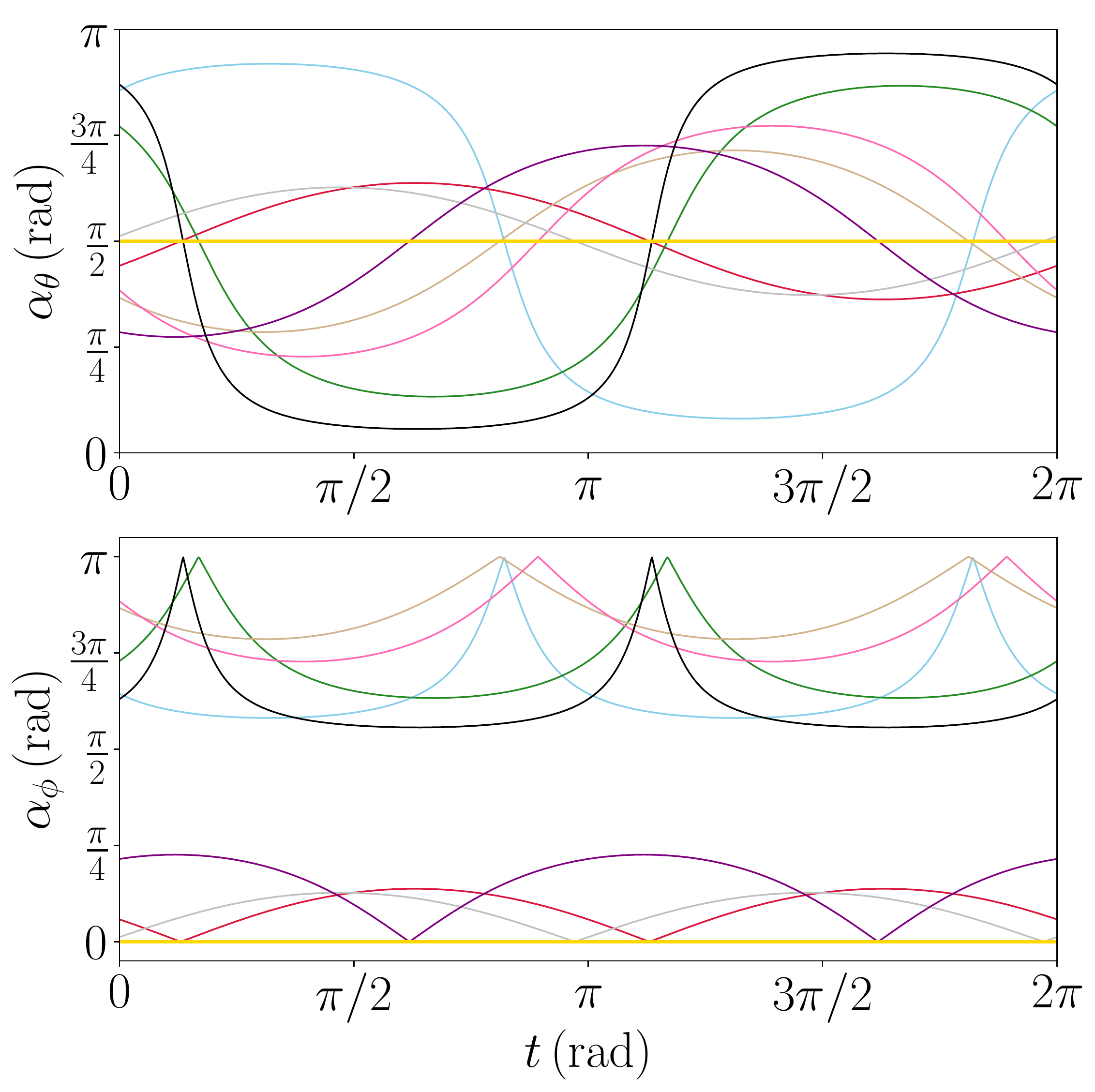}
  \caption{[Left panel] A collection of randomly selected great arcs on the unit sphere. The thick gold line is the great arc that coincides with the equator in the coordinate system adopted. [Right panel] The angle between ${\bf e}_{\theta}$ and $\hat{T}$ (top panel) and ${\bf e}_{\phi}$ and $\hat{T}$ (bottom panel), as a function of great arc parameter $t$. The colors match the great arcs in the left panel. }
  \label{fig:app3a}
\end{figure}

We present $N=10$ great arcs defined by selecting $\hat{u}$, $\hat{v}$ randomly in the left panel of Figure \ref{fig:app3a}. The thick gold arc lies on the equator. The corresponding rotation angles $\alpha_{\theta} = \cos^{-1}(\hat{T}.{\bf e}_{\theta})$ and $\alpha_{\phi} = \cos^{-1}(\hat{T}.{\bf e}_{\phi})$, as a function of the arc parameter $t$, are presented in the right panel of Figure \ref{fig:app3a}. Only when the great arc aligns with the coordinate basis is there no relative rotation of the tangent space (cf gold lines). There are two points on each great arc at which the path is perpendicular to ${\bf e}_{\theta}$ and hence $\hat{T}$ either aligns or anti-aligns with ${\bf e}_{\phi}$, depending on the direction of the arc tangent vector. This is the origin of the dichotomy observed in the $\alpha_{\phi}$ panel. Note that the vectors return to their original orientation if transported along the entire great arc.

%%%%%%%%%%%%%%%%%%%%%%%%%%%%%%%%%%%%%%%%%%%%%%%%%%%%%%%%%%%%%%%%%%%%%%%%%%%%%%%%%%%%
\section{Useful Relations} 
\label{sec:appen2}

We provide some useful identities regarding the spherical Bessel functions and other functions that are used in the paper. Some of these can be found in standard textbooks \citep{mabramowitz64:handbook} --

\begin{eqnarray}
\label{app:b2} & & \sum_{m=-\ell}^{\ell} Y_{\ell m}(\hat{{\bf s}}_{1}) Y_{\ell m}^{*}(\hat{{\bf s}}_{2}) = {2 \ell + 1 \over 4\pi}{\cal L}_{\ell}(\hat{{\bf s}}_{1} . \hat{{\bf s}}_{2}), \\ 
\label{app:b3} & & \sum_{m=-\ell}^{\ell} {\partial Y_{\ell m}(\theta,\phi) \over \partial \phi} {\partial Y_{\ell m}^{*}(\theta,\phi) \over \partial \phi} = {(2 \ell + 1)\ell (\ell + 1 )  \over 8\pi}\sin^{2}\theta, \\ 
\label{app:b4} & & \sum_{m = -\ell}^{\ell} {\partial Y^{m *}_{\ell}(\theta,\phi) \over \partial \theta} {\partial Y^{m }_{\ell}(\theta,\phi) \over \partial \theta} = {(2 \ell + 1) \ell (\ell+1) \over 8\pi}, \\
\label{app:b5} & & \sum_{m = -\ell}^{\ell} {\partial^2 Y^{m *}_{\ell}(\theta,\phi) \over \partial \phi^2} {\partial^2 Y^{m }_{\ell}(\theta,\phi) \over \partial \phi^2} = {(2 \ell + 1) \ell (\ell+1) \over 8\pi}\sin^2\theta\left[ \frac{3\sin^2\theta}{4}\left(\ell(\ell+1) + \left(1-\frac{3\sin^2\theta}{2}\right) \right) \right], \\
\label{app:b6} & & \sum_{m = -\ell}^{\ell} {\partial^2 Y^{m *}_{\ell}(\theta,\phi) \over \partial \theta^2} {\partial^2 Y^{m }_{\ell}(\theta,\phi) \over \partial \theta^2} = {(2 \ell + 1) \ell (\ell+1) \over 8\pi}\left[ \frac{3}{4}\ell(\ell+1) - \frac{1}{2} \right].
\end{eqnarray} 

\noindent These can be derived using the general result ; 

\begin{equation}\label{eq:n1} P_{\ell}(\cos\gamma) = {4\pi \over 2\ell + 1} \sum_{m = -\ell}^{\ell} Y^{m *}_{\ell}(\theta',\phi') Y^{m}_{\ell}(\theta,\phi), \end{equation}

\noindent where

\begin{equation}\label{eq:n2} \cos\gamma = \cos\theta \cos\theta' + \sin\theta \sin\theta' \cos(\phi -\phi'), \end{equation} 

\noindent along with the differential equation that the Legendre polynomial solves -- 

\begin{equation}\label{eq:n3} (1-x^2)P''_{\ell}(x) - 2x P'_{\ell}(x) + \ell (\ell + 1) P_{\ell}(x) = 0, \end{equation} 

\noindent and the normalisation $P_{\ell}(1) = 1$. Taking  derivatives of eq.~($\ref{eq:n1}$) w.r.t. $\phi$, $\phi'$, $\theta$, $\theta'$, and then taking the limit $\theta \to \theta'$, $\phi \to \phi'$ and $x = \cos\gamma \to 1$ yields results such as ($\ref{app:b3}, \ref{app:b4}$).

We also have the following relation for the spherical Bessel function of the first kind, $j_{\ell}$,

\begin{equation}\label{eq:rel1} \sum_{\ell=0}^{\infty} (2\ell + 1)\left[j^{(p)}_{\ell}(x)\right]^{2} = {1 \over 2p + 1}, \end{equation} 

\noindent where the $(p)$ superscript denotes the $p^{\rm th}$ derivative of the spherical Bessel function with respect to its argument. Also important are the following relations

\begin{eqnarray}\label{app:b13} & & \sum_{\ell=0}^{\infty} (2\ell+1) \ell (\ell + 1) j^{2}_{\ell}(x) = {2x^{2} \over 3}, \\
\label{app:b14} & &  \sum_{\ell=0}^{\infty} (2\ell + 1) \ell (\ell + 1) \left[j'_{\ell}(x)\right]^{2} = {2 \over 3} + {2 \over 15}x^{2}, \\
\label{app:b15} & & \sum_{\ell=0}^{\infty} (2\ell+1) \ell(\ell+1) \left[j''_{\ell}(x) \right]^{2} = {8 \over 15} +  {2x^{2} \over 35} . 
\end{eqnarray} 

\noindent The $j_\ell$ functions satisfy the equation,

\begin{equation}
    \ell\left(\ell+1\right)j_\ell(x) = x^2j_{\ell}''(x) + 2xj_{\ell}'(x) + x^2j_{\ell}(x),
    \label{app:b16}
\end{equation}
which can be differentiated w.r.t. $x$ twice to give,

\begin{equation}
    \ell\left(\ell+1\right)j_{\ell}''(x) = x^2j_{\ell}''''(x) +6xj_{\ell}'''(x) + \left(x^2+6\right)j_{\ell}''(x) + 4xj_{\ell}'(x) + 2j_{\ell}(x). 
    \label{app:b17}
\end{equation}

In Section \ref{sec:num}, when reconstructing the Minkowski tensors numerically, we transform between Cartesian and spherical coordinate systems. We adopt the standard angle conventions such that the conversion from Cartesian to radial gradients is 

\begin{equation}\label{eq:other1} {\partial \Omega \over \partial r} = \sin\theta \cos\phi {\partial \Omega \over \partial x} + \sin\theta \sin\phi {\partial \Omega \over \partial y} + \cos\theta  {\partial \Omega \over \partial z} .
\end{equation}

\noindent To define volume averages in Section \ref{sec:num}, we rotate vectors on the two-sphere. To do so we define ${\bf \hat{m}}$ as the unit vector pointing to a position on the manifold at which the $\delta_{D}(\tilde{\delta}_{[m,n,p]}-\delta_{t}) \neq 0$ and ${\bf \hat{a}}$ as the unit vector pointing to a fiducial point at which we take the volume average of $w_{i}{}^{j}$, then the unit quaternion $q = q_{0} + {\bf q}$ with elements 

\begin{equation} q_{0} = \cos{\theta \over 2} , \qquad {\bf q} =   { {\bf \hat{m}} \times {\bf \hat{a}} \over |{\bf \hat{m}} \times {\bf \hat{a}}|} \sin{\theta \over 2} , \end{equation} 

\noindent is used to rotate the gradient vector sampled at $[m,n,p]$ to $[a,b,c]$, where $\cos\theta = {\bf \hat{m}} . {\bf \hat{a}}$.  The complex conjugate is $q^{*} = q_{0} - {\bf q}$ and the rotation operator acting on an arbitrary vector ${\bf v}$ can be written as 

\begin{equation} q {\bf v} q^{*} = (q_{0}^{2} - |{\bf q}|^{2}){\bf v} + 2 ({\bf q}.{\bf v}){\bf q} + 2 q_{0} ({\bf q} \times {\bf v}). \end{equation}

\end{document}